\documentclass[12pt]{iopart}
\usepackage{epsf}
\begin{document}
\title{The Electronic Nature of High Temperature Cuprate Superconductors}
\author{M. R. Norman$^{1,2}$ and C. P\'{e}pin$^2$}
\address{$^1$Materials Science Division, Argonne National Laboratory, 
Argonne, IL 60439, USA}
\address{$^2$SPhT, L'Orme des Merisiers, CEA-Saclay, 91191 Gif-sur-Yvette,
France}
\ead{norman@anl.gov}
\begin{abstract}
We review the field of high temperature cuprate superconductors, with an
emphasis on the nature of their electronic properties.  After a general
overview of experiment and theory, we concentrate on recent results
obtained by angle resolved photoemission, inelastic neutron scattering,
and optical conductivity, along with various proposed explanations for
these results.  We conclude by reviewing efforts which attempt to
identify the energy savings
involved in the formation of the superconducting ground state.
\end{abstract}
\submitto{\RPP}

\maketitle

\tableofcontents

\section{General Overview}

\subsection{History}

As with all endeavors in science, there is a prehistory involved, and 
cuprates are no exception.  Research in superconductivity was not the 
languishing field that it is often portrayed as being prior to the 
Bednorz-Muller ``revolution'' of 1986.  What had begun to be 
diminished, though, was the hope that a truly high temperature 
superconductor would ever be discovered.  At the time of the 
Bednorz-Muller discovery, the highest 
temperature superconductor known was Nb$_3$Ge (23K).  That
material had been known since 1973, and was not much of an 
improvement over NbN (15K) which had been discovered all the way back in 
1941 \cite{DAHL}.  This pessimistic outlook was best articulated by 
Bernd Matthias in a number of papers which 
still make interesting reading today \cite{MATTHIAS}.  Such pessimism was 
not confined to experiment, as witnessed by the famous paper of Cohen 
and Anderson \cite{COHEN}.  As was well appreciated by that time, the 
A15 materials with highest $T_c$ were on the verge of a structural 
transition, and thus it was anticipated that one could not push $T_c$ 
much higher before the lattice became unstable \cite{TESTARDI}.

Despite this, a number of new classes of superconductors had 
been discovered in the period before 1986, including the ternary magnetic 
superconductors such as ErRh$_4$B$_4$ and HoMo$_6$S$_8$, and 
various uranium based superconductors such as $\alpha$-U and U$_6$Fe,
many of these discovered by Matthias and his various associates.  
Matthias' speculation that something really different was going on in 
f electron superconductors was spectacularly confirmed with the 
discovery by Frank Steglich in 1979 of ``heavy fermion'' superconductivity in 
CeCu$_2$Si$_2$ \cite{FRANK}, followed by the discovery of superconductivity
in UPt$_3$ and UBe$_{13}$ \cite{GREG}.

Heavy fermion superconductivity was one of the main 
research topics in fundamental physics prior to 1986, and its history 
has had some impact on the cuprate field.  Unlike the magnetic 
superconductors such as ErRh$_4$B$_4$ where the magnetic moments are 
confined to the rare earth site and the superconductivity to the 
ligand sites, in heavy fermion superconductors, the f electrons 
themselves become superconducting.  This is known from the extremely 
high effective mass of the superconducting carriers.  More properly, 
the carriers should be thought of as composite objects of conduction 
electron charge and f electron spin \cite{MILLIS}.  The fascinating 
thing about these materials, though, is that their superconducting 
ground states do not appear to have the L=0, S=0 symmetry that Cooper 
pairs exhibit in normal superconductors \cite{RMP91,RMP02}.

As with cuprates, heavy fermion superconductivity had its 
own prehistory as well, that being the field of superfluid $^3$He.  
$^3$He had been speculated in the 1960s to possibly be a paired 
superfluid with non-zero orbital angular momentum, in particular 
L=2 pairing \cite{EMERY}.  The idea was that the hard core repulsion of 
the atoms would prevent L=0 pairing, but that longer range pairs could 
be stabilized by the attractive van der Waals interaction between the 
He atoms.  Subsequently, Layzer and Fay \cite{FAY} showed that for 
nearly ferromagnetic metals, and $^3$He as well, spin dependent
interactions instead could stabilize L=1, S=1 pairs, leading to the 
concept of paramagnon mediated pairing.

At that time, experimentalists were beginning to push $^3$He to low 
temperatures, with the idea of searching for magnetic order under 
pressure.  This was subsequently discovered by Bill Halperin.  But 
along the way, Doug Osheroff found superfluidity, and various 
experiments did indeed confirm the L=1, S=1 nature of the pairs
\cite{3He}.  More interestingly, two paired states were found, the 
so-called A and B phases.  Anderson and Brinkman later proposed 
that the stabilization of the anisotropic A phase relative to the 
isotropic B phase 
could be understood by feedback of the pair formation on the spin 
fluctuation interactions which supposedly gave rise to the pairs to 
begin with \cite{BA}.  Such feedback effects are much in vogue lately 
in regards to spin fluctuation mediated theories of cuprates
\cite{AC}.

What does this imply for the lattice case?  Early theories for heavy 
fermions were indeed based on the $^3$He paradigm, but with the 
discovery of antiferromagnetic correlations by inelastic neutron 
scattering \cite{GABE}, people turned away from these nearly ferromagnetic 
models 
(though they have seen a resurgence of late, with the discovery of 
superconductivity in UGe$_{2}$ \cite{UGe2} and ZrZn$_2$ 
\cite{ZrZn2}).  Rather, theoretical work published in 1986 led 
to the concept of L=2, S=0 pairs in the nearly antiferromagnetic 
case \cite{1986}.  This 
d-wave model is still one of the leading candidates to describe 
superconductivity in UPt$_3$, though a competing model based on 
f-wave pairs has been proposed by Norman\cite{MIKE92} and 
Sauls \cite{JIM94}. 
The problems with determining the pair symmetry in heavy fermions are 
the multiple band nature of the problem (orbital degeneracy), along 
with the effects of non-trivial crystal structures (UPt$_3$ has a non 
symmorphic lattice, for instance) and spin-orbit (which destroys L and 
S as good quantum numbers), not to mention the complications 
of dealing with three dimensions.  (Fortunately none of these problems 
exist in the cuprates.)  One of the interesting observations 
of these early calculations was the prediction that for a simple cubic 
lattice, the d-wave pairs should be of the form $(x^2-y^2) \pm i (3z^2-r^2)$ 
\cite{1986}.  If 
one simply eliminates the third dimension, one obtains the order 
paramater now known to be the pair state of the cuprates.  In some 
sense, the prediction of $d_{x^2-y^2}$ pairing in the cuprates was the 
ultimate one liner.

The above path, though, is not what led to the discovery of cuprate 
superconductors.  The history of this is rather lucidly described in 
Bednorz and Muller's Nobel lecture \cite{RMP88}.  Of particular 
interest to them was the case of doped SrTiO$_3$.  This material had a 
$T_c$ of less than 1K, but as it had such an incredibly low carrier density, 
it shouldn't have been superconducting at all, at least according to 
what people thought at the time.  In fact, the properties of this 
material led to a speculation by Eagles about the possibility of Bose 
condensation \cite{EAGLES}, with pairs existing 
above the superconducting transition temperature, a forerunner 
of the pseudogap 
physics currently being discussed for the cuprates.  Although Binning 
and Bednorz did some work on this material, it never led to much,
so Binning got bored and moved on to the discovery of the scanning 
tunneling microscope, which he later got the Nobel prize for.

After this, Bednorz began to work under Alex Muller, who was also 
interested in the possibility of oxide superconductors.  Alex was 
particularly intrigued by the role that Jahn-Teller effects
played in the perovskite structure; that is, in the distortions of the 
oxygen octahedra surrounding the transition metal ions
which lead to the lifting of the degeneracy of the 3d crystal field levels.
While searching around, they 
became aware of work that Raveau's group had done on LaBaCuO, and in 
the course of reproducing this work, they discovered high temperature 
superconductivity.  The story subsequently circulated was that Raveau 
took his samples off the shelf, and found that they too were 
superconducting.  It is on such twists of fate that careers in 
science are often decided.

After the original discovery, several groups got in the act, and by 
use of pressure, Paul Chu's group was able to drive the initial 
transition temperature of 35K up to 50K.  The real quest, though, was 
to find a related structure with a higher transition temperature, and 
this was rapidly discovered in early 1987, when Chu and collaborators 
found 90K superconductivity in YBCO \cite{CHU}.  The liquid 
air barrier (77K) had finally been breached, and true high temperature 
superconductivity had at last been discovered.  By varying the crystal
structure and again exploiting pressure, transition temperatures 
up to 160K have been achieved, again by Chu's group.  Matthias must have 
been smiling from on high.

\subsection{Crystal Symmetry and Electronic Structure}

\begin{figure}
\centerline{\epsfxsize=0.6\textwidth{\epsfbox{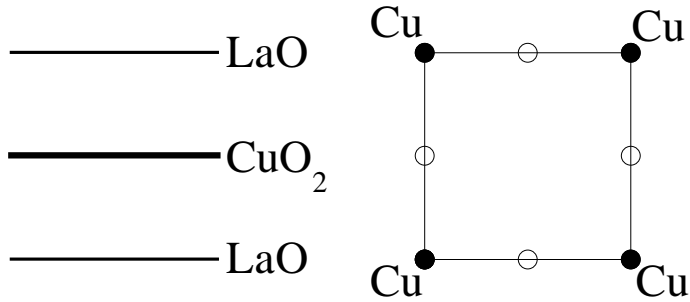}}}
\caption{Crystal structure of $La_2CuO_4$.  Left panel shows the 
layer structure along the c-axis, the right panel the structure of 
the $CuO_{2}$ plane.}
\label{fig1}
\end{figure}

The crystal structures of the cuprates were one of the first things 
elucidated, which is obvious because of the patent rights involved 
(after a very long struggle, Bell Labs eventually won that 
one \cite{BELL}; for an 
illuminating account of those heady days, the reader is referred to 
the book by Hazen \cite{HAZEN}).  Though they come in many variants, 
the basic structure is quite simple (Fig.~\ref{fig1}).  The material consists of 
CuO$_2$ planes, where each Cu ion is four fold coordinated with O ions,
separated by insulating spacer layers.
The exception 
to this is YBCO, which has metallic CuO chain layers as one of the 
spacer layers.  The orginal 
LBCO material had two apical oxygens (the standard
perovskite structure, where the transition metal ion is in the center 
of an octahedron formed from six surrounding oxygen ions), but in 
YBCO, one of these oxygens is absent, and in other structures, 
such as that formed by electron doped cuprates, the apical oxygens are 
totally missing.  What this means is that despite all of these 
complications, the essential structure to worry about are the CuO$_2$ 
planes, which was understood early on, particularly by 
Anderson \cite{RVB}.  Of course, 
the superconducting transition temperature varies a lot from structure to 
structure, and is generally higher the more CuO$_2$ planes per unit 
cell that there are, but after years of study, most researchers have come to the 
conclusion that the main c-axis effect is simply to tune the 
electronic structure of the CuO$_2$ planes.

When considering these planes, one immediately comes across a basic 
fact.  Most transition metal oxides are insulators with
a particular electronic 
structure.  This is due to the fact that the transition metal 3d level 
and oxygen 
2p level are separated by a greater energy than the energy spread of these 
levels from band formation.  The net result is one gets separate 3d 
and 2p energy bands.  The Coulomb repulsion on the transition metal 
site is so large that the 3d band 
``Mott-Hubbardizes'', spliting into upper and lower Hubbard bands 
separated by this energy scale, U (typically 8-10 eV in the solid).  The 
true energy gap then becomes of the charge transfer type,
separating the filled oxygen 2p valence band from the empty 3d 
conduction (i.e., upper Hubbard) band \cite{SAWATZKY}.

\begin{figure}
\centerline{\epsfxsize=0.5\textwidth{\epsfbox{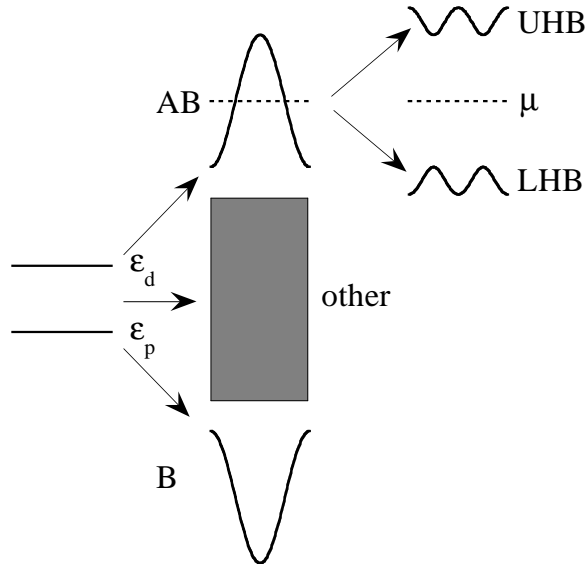}}}
\caption{Electronic structure of the undoped cuprates.  Left panel shows the
atomic Cu d and O p levels, middle panel the band structure of the 
solid (where B is the bonding combination of the atomic levels, AB the 
antibonding one), and right panel the effect of correlations
(Mott-Hubbard gap) on the AB band (LHB and UHB are the lower and upper
Hubbard bands).}
\label{fig2}
\end{figure}

The cuprate case is different, though (Fig.~\ref{fig2}) \cite{PICKETT}.  In the solid,
the Cu ion is in a $d^9$ configuration ($Cu^{++}$) and the O ion in a $p^6$
configuration ($O^{--}$), with the  Cu 3d energy level above but relatively
close to the O 2p energy level.  In the layered perovskites, the tetragonal
environment of the Cu ion leads to the single 3d hole having $d_{x^2-y^2}$
symmetry.  In this case, the dominant energy  is the bonding-antibonding splitting 
involving the
quantum mechanical mixture of the
Cu 3d $x^2-y^2$ orbital and the planar O $2p_x$ and $2p_y$ orbitals (with an
energy of 6 eV).
The net 
result is that in the parent (undoped) structure, one is left with a 
half filled band which is the antibonding combination of these three orbitals, 
with the bonding, non-bonding, and the rest of the Cu and O
orbitals filled.  As Anderson 
speculated early on \cite{RVB},
it is this copper-oxygen antibonding band which ``Mott-Hubbardizes'', forming 
an insulating gap of order 2 eV in the parent compound (the effective 
U being reduced because of the Cu-O orbital admixture).

Therefore, in the end, the complicated electronic structure leads to 
a single 2D energy band near the Fermi energy, which is what makes the 
cuprates so 
attractive from a theoretical perspective \cite{ANDERSON}.
But one can even reduce 
this ``one band Hubbard model'' further by taking the limit of large 
U.  In this case, the upper Hubbard band is projected out (assuming we 
are considering hole doping the insulator), and the effect of U 
becomes virtual, leading to a superexchange interaction between the 
Cu spins, J 
(t$^2$/U, where t is the effective Cu-Cu hopping mediated by intervening
O sites).  This is easily understood by 
noting that two parallel spins are not allowed to occupy the same Cu site 
because of the Pauli exclusion principle, but antiparallel spins can, 
leading to an energy savings of t$^2$/U from second order perturbation 
theory.  This so-called ``t-J'' model is the minimal model for the 
cuprates.  Despite the success of motivating this model from first 
principles calculations \cite{HYBERTSEN}, it is not generally agreed 
upon.  For instance, Varma has advocated that one must consider the full 
three band Hubbard model (one band from each of the three states, Cu 
3d $x^2-y^2$, O 2p$_x$, and O 2p$_y$).  His claim is that upon projection to 
the low energy sector, non-trivial phase factors between the 
three bands become
possible, which can lead to an orbital current state which 
he associates with the pseudogap phase \cite{VARMA}.

\subsection{Phase Diagram}

The original hope of Anderson was that the insulating phase of the 
cuprates would turn out to be a spin liquid \cite{RVB}.
\begin{figure}
\centerline{\epsfxsize=0.5\textwidth{\epsfbox{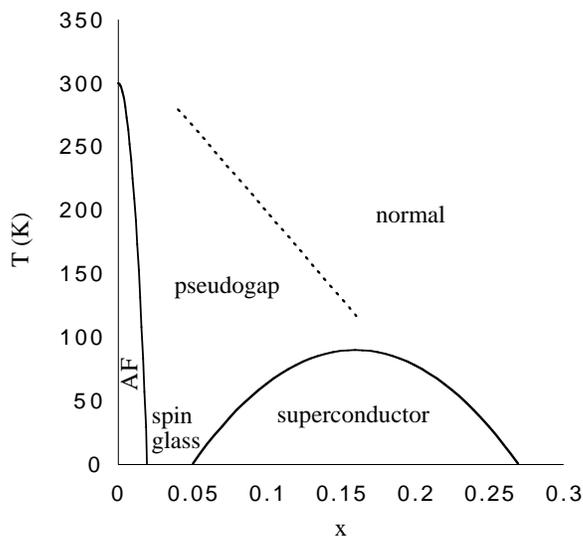}}}
\caption{Phase diagram of the cuprates (x is the hole doping).  AF is 
the antiferromagnetic insulator.  The dotted line is a crossover line 
between the normal metal phase and the pseudogap phase.}
\label{fig3}
\end{figure}
The issue 
here is that most Mott insulators exhibit broken symmetry, such as 
antiferromagnetism.  This means that such insulators can be 
adiabatically continued to an ordinary band insulator with magnetic 
order, as originally proposed by Slater \cite{SLATER}.  The cuprates 
would be another example of this, since the half filled band discussed
above would become one full and one empty band upon magnetic 
ordering due to the unit cell doubling
(in this picture, the lower Hubbard band would correspond to 
the up spin band of the band insulator, the upper Hubbard band to the 
down spin band).  Anderson believed, though, that the Mott phenomenon
should be unrelated to this argument, and that the cuprates 
would be the ideal place to demonstrate this.  Since there is only 
one d hole per Cu site (and thus, S=1/2), and given the 2D nature of 
the material, he felt that quantum fluctuations would be sufficient 
to destroy the order, leading to a spin liquid ground state, which he 
called a resonating valence bond (RVB) state (harking back to the 
theory of benzene rings, where each C-C link resonates between a single bond 
and a double bond state).

As was discovered soon after, though, the undoped phase is indeed 
magnetic, though the moment is reduced by 1/3 from the free ion value 
due to fluctuations \cite{INS}.  On the other hand, magnetic order is rapidly 
destroyed upon hole doping, so in fact the magnetic phase only takes 
up a small sliver (Fig.~\ref{fig3}) of the phase diagram (in the electron doped case, 
though, the magnetism exists over a much larger doping range).  So, 
in that sense, Anderson's intuition was quite good.

For dopings beyond a few percent, the system either enters a messy 
disordered phase exhibiting spin glass behavior (as in LSCO) before
superconducting order sets in, or 
immediately goes to the superconducting phase (as in YBCO).  The 
superconducting transition monotonically rises with doping, reaching a 
maximum at about 16\% doping, after which $T_c$ declines to zero.  The 
net effect is to form a superconducting ``dome'' which extends from 
about 5\% to 25\% doping.

At first sight, the superconducting phase is not so different from 
that of classical superconductors.  We know that it exhibits a zero 
resistance state with a Meissner effect.  Experiments show 
that the superconducting objects are charge 2e, and thus pairs are 
formed.  What is unusual, though, are the short coherence lengths.  
For typical superconductors, the coherence length is quite long, usually 
several hundered $\AA$ or more.  This is in contrast to magnets, which have 
quite short coherence lengths.  Therefore, for most superconductors we 
know, mean field theory works extremely well, as opposed to magnets 
where it almost always fails.  But cuprates exhibit short coherence 
lengths, of order 20$\AA$ in the plane, and a paltry 2$\AA$ between 
planes.  The latter is so short, the cuprates are essentially 
composed of Josephson coupled planes, as has been experimentally 
verified by a number of groups \cite{KLMU}.  Such coupling is necessary, 
of course, since long range superconducting order cannot occur in two 
dimensions (except the Kosterlitz-Thouless phase, whose existence in the 
cuprates is still debated \cite{JOE}).

\begin{figure}
\centerline{\epsfxsize=0.8\textwidth{\epsfbox{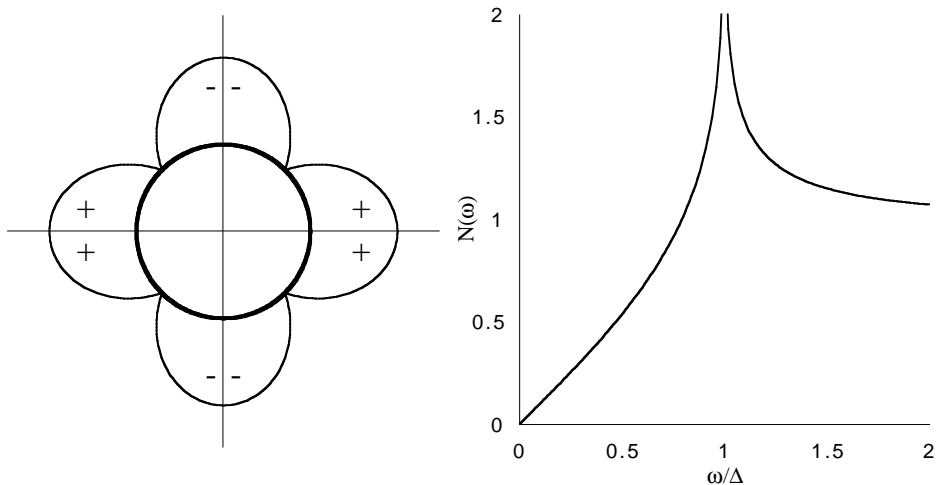}}}
\caption{Angular variation of the d-wave gap around the Fermi surface 
is shown in the left panel (with nodes at $k_x=\pm k_y$).  
The resulting density of states is plotted 
in the right panel.  $\Delta$ is the maximum superconducting energy 
gap.}
\label{fig4}
\end{figure}

Another unusual finding is the symmetry of the order parameter (Fig.~\ref{fig4}).  
For many 
years, it was felt that the order parameter probably had s-wave 
symmetry.
There was no evidence from thermodynamic measurements for nodes in 
the gap as in heavy fermion superconductors, except for an early
report of a non-exponential temperature dependence of the Knight 
shift \cite{TAKIGAWA}.  And the cuprates were 
viewed as quite disordered (doping being achieved by chemical 
substitution), which is known to be pair breaking for unconventional 
superconductors.  This was despite the prediction of d-wave pairing 
from both spin fluctuation models \cite{BICKERS} and RVB theory 
\cite{GROS,KOTLIAR}.  But this begin to change when NMR measurements found 
a $T^3$ variation of the spin lattice relaxation rate, as found in 
heavy fermion superconductors, as opposed to the exponential behavior 
found in s-wave superconductors \cite{MARTINDALE}.
This was soon followed by penetration depth
measurements, where a corresponding linear in T behavior was found
\cite{HARDY}.  At the same time, angle resolved 
photoemission measurements gave direct spectroscopic evidence for 
nodes in the gap \cite{SHEN93}.

Of course, the possibility still remained that the order parameter 
was s-wave, but with a highly anisotropic gap.  These worries were 
put to rest once and for all by phase sensitive measurements.  The 
first of these was by van Harlingen's group.  Following suggestions 
by Leggett, they formed SIS tunneling junctions on the ac and bc faces 
of YBCO, using an ordinary s-wave superconductor as the 
counterelectrode.  These two junctions were then connected, and the 
superconducting phase difference measured from the dependence
of the Josephson critical current on applied magnetic field.
They found exactly the $\pi$ 
phase shift expected for a d-wave state (which differs by a minus 
sign between the two orthogonal a and b directions) \cite{DALE}.
This was soon followed by the tricrystal grain boundary experiments 
of Tsuei and Kirtley \cite{TSUEI}, where three grain boundaries 
at different orientations were brought together at a point.  Thus, about
the tricrystal point, there are three junctions.  Each junction will act as a
``zero" junction or a $\pi$ junction depending on the superconducting
phase difference across the junction.  If the number of such $\pi$ junctions
is odd, then a half integral flux quantum will appear at the tricrystal point.
The advantage of this method is that by varying the crystallographic
orientation of the three grains, the symmetry of the order parameter can
be mapped out in detail.  The net result was that for only those orientations
where d-wave symmetry predicted a half integral flux quantum was one 
observed.  As YBCO is orthorhombic, though, 
there still remained an out (since s-wave and d-wave are the same 
group representation in that crystal structure), but the tricrystal 
experiments were repeated for Tl2201, which has 
tetragonal symmetry, with the same results \cite{TLPH}.  After that, 
there was no question anymore about the order parameter symmetry, and 
for these pioneering efforts, four of the researchers were awarded the 
Buckley prize in 1998.

\begin{figure}
\centerline{\epsfxsize=0.8\textwidth{\epsfbox{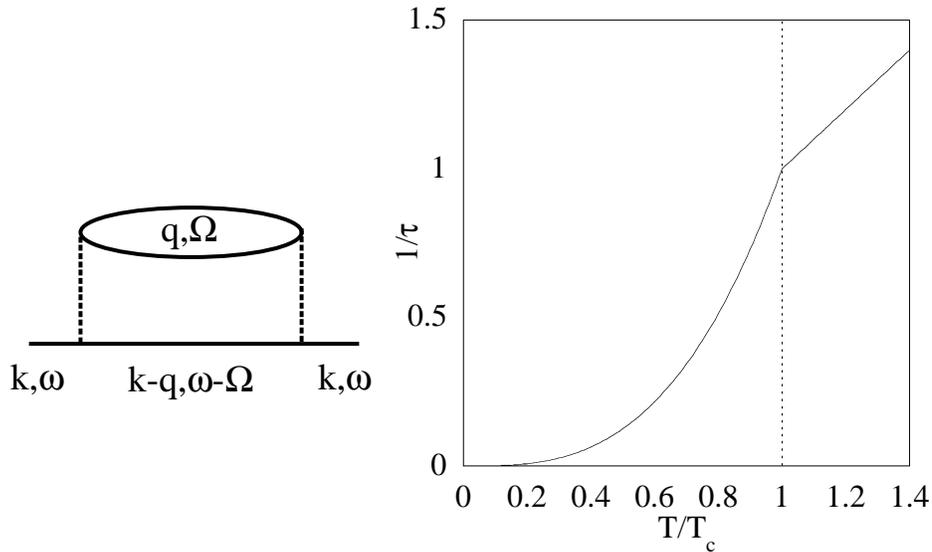}}}
\caption{Left panel: lowest order Feynman diagram for 
electron-electron scattering.  Right panel: resulting temperature 
dependence of the zero energy scattering rate.  $T_{c}$ is the superconducting 
transition temperature (dotted line).}
\label{fig5}
\end{figure}

Perhaps the most unusual finding, though, is the difference of the 
dynamics between the normal and superconducting states.  As will be
discussed below, the normal state of the cuprates (away from 
the overdoped side of the phase diagram) does not appear to be a 
Landau Fermi liquid.  On the other hand, a variety of experiments, 
first microwave conductivity \cite{BONN}, then thermal 
conductivity \cite{ONG}, infrared 
conductivity \cite{PUCH}, and photoemission \cite{ADAM}, revealed that 
the scattering rate of the electrons at low energies
drops precipitously in the superconducting state (right panel, Fig.~\ref{fig5}).  At 
low temperatures in YBCO, mean free paths of the order of microns 
have been inferred for the electrons, as opposed to the very short mean 
free paths found in 
the normal state.  This strong loss in inelastic scattering would be unusual
for an electron-phonon mediated superconductor, since the phonons are not 
gapped in the superconducting state.  The implication, then, is that 
the primary scattering is electron-electron like in character, 
and thus is strongly reduced in the superconducting state since the 
electrons become gapped.  This can be easily seen from the lowest 
order Feynman diagram (left panel, Fig.~\ref{fig5}, showing an electron 
scattering off a particle-hole excitation), since every internal line in the diagram 
is gapped by $\Delta$ (a 
point first noted by Nozieres in his famous book\cite{NOZ}).
This obviously points to an electron-electron 
origin to the pairing as well.

This brings us to a consideration of the rest of the phase diagram.
As discussed above, cuprates should be thought of as doped 
Mott insulators.  What this means is that for low doping, the number 
of carriers is small.  As the superconducting phase is conjugate to 
the number operator, this implies that phase fluctuations could play 
an important role on the underdoped side of the phase diagram.  Again, this was 
realized early on by Anderson, who proposed that the doped holes would 
only be phase coherent 
below a temperature which scaled linearly with 
doping \cite{RVB2}.  This should be contrasted with the ``pairing'' 
scale, which within the RVB model would be maximal for the insulator,
and then drops to zero on the overdoped side when the bandwidth $xt$
of the doped holes
becomes comparable to the superexchange energy $J$ \cite{BZA,GROS}.
\begin{figure}
\centerline{\epsfxsize=0.8\textwidth{\epsfbox{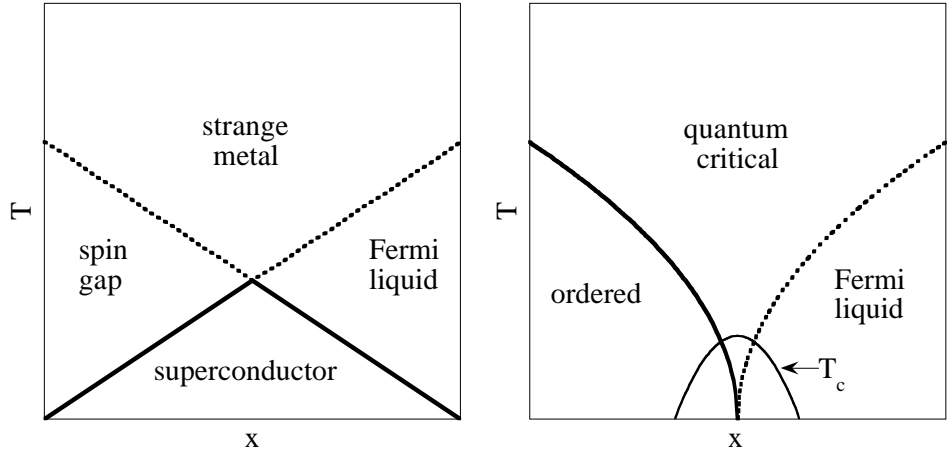}}}
\caption{Two proposed theoretical phase diagrams for the cuprates:  
RVB picture (left panel) \cite{LEE} and the quantum critical scenario (right 
panel).}
\label{fig6}
\end{figure}
These two crossing 
lines led to the 
proposal of a generic ``RVB'' phase diagram (left panel, Fig.~\ref{fig6}) \cite{LEE} 
composed of 
four phases, a superconductor (bottom quadrant), a Fermi liquid 
(right quadrant), a strange metal phase (upper quadrant), 
and a spin gap phase (left quadrant, now known as the 
pseudogap phase).  In this picture, only the superconducting phase 
(which lies below both crossing lines) should be considered as having 
true long range order, otherwise, these ``phase'' lines should be 
considered as crossover lines.

The first of these ``phases'' which was studied in detail was the 
strange metal phase.  Transport measurements revealed that the 
resistivity was dead linear in temperature over a large range, 
the most amazing example of this being in single 
layer Bi2201, where linearity persisted down to 10K (when 
superconductivity finally occurred) \cite{Bi2201}.  No saturation at high 
temperatures was observed as occurs in A15 superconductors.  This 
behavior was further confirmed by a generalized Drude analysis of 
infrared data, which shows a scattering rate linear in energy up to half an 
eV \cite{PUCH}.  These striking observations led Varma and colleagues to 
propose the so-called marginal Fermi liquid phenomenology for the 
strange metal phase \cite{MFL}.  In this model, the electrons are 
assumed to be scattering off a bosonic spectrum which is linear in 
energy up to an energy scale T, then constant afterwards.  Because of 
this, no energy enters the problem except the temperature (modulo an 
ultraviolet cut-off), a phenomenon referred to as quantum critical 
scaling.  This in turn has led to the proposal of an alternate (to the 
``RVB'') phase diagram based on a quantum critical point (right panel,
Fig.~\ref{fig6}).  In such a 
picture, the ordered phase (to the left of the critical point) would 
correspond to the pseudogap phase, its disordered analogue (to the 
right of the critical point) to the Fermi liquid phase, and the 
quantum critical regime (above the critical point) to the strange 
metal phase.  The superconducting ``dome'' surrounds the critical 
point, screening it like an event horizon of a black hole.  We note 
that the Fermi liquid/strange metal boundary is a crossover line in 
both phase diagram scenarios, but the strange metal/pseudogap boundary is a 
crossover line in the RVB model and a true phase line in the quantum 
critical model.  An exception is in certain antiferromagnetic quantum critical
scenarios where the ``phase line" corresponds to short range 2D order \cite{AC}.

\begin{figure}
\centerline{\epsfxsize=0.5\textwidth{\epsfbox{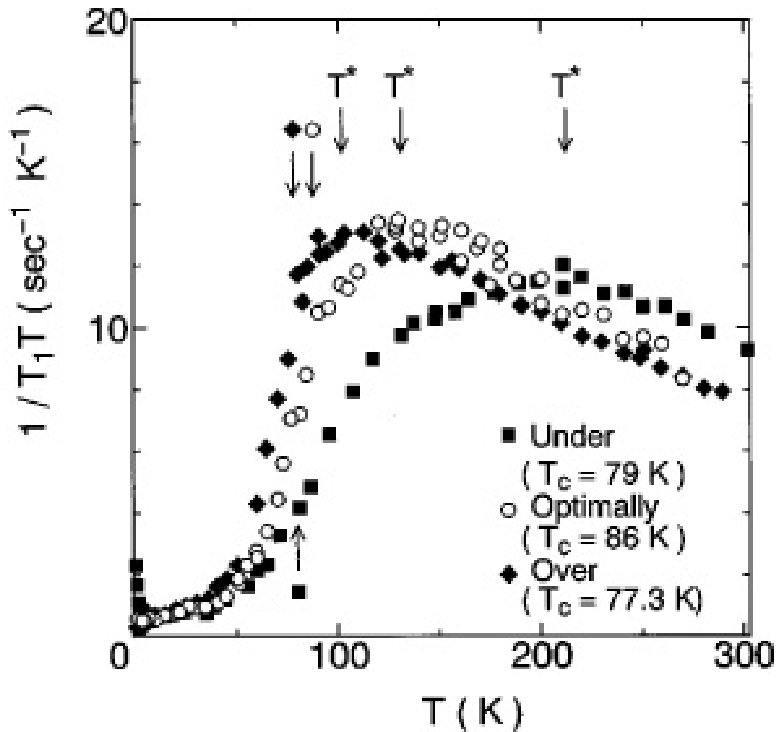}}
\epsfxsize=0.5\textwidth{\epsfbox{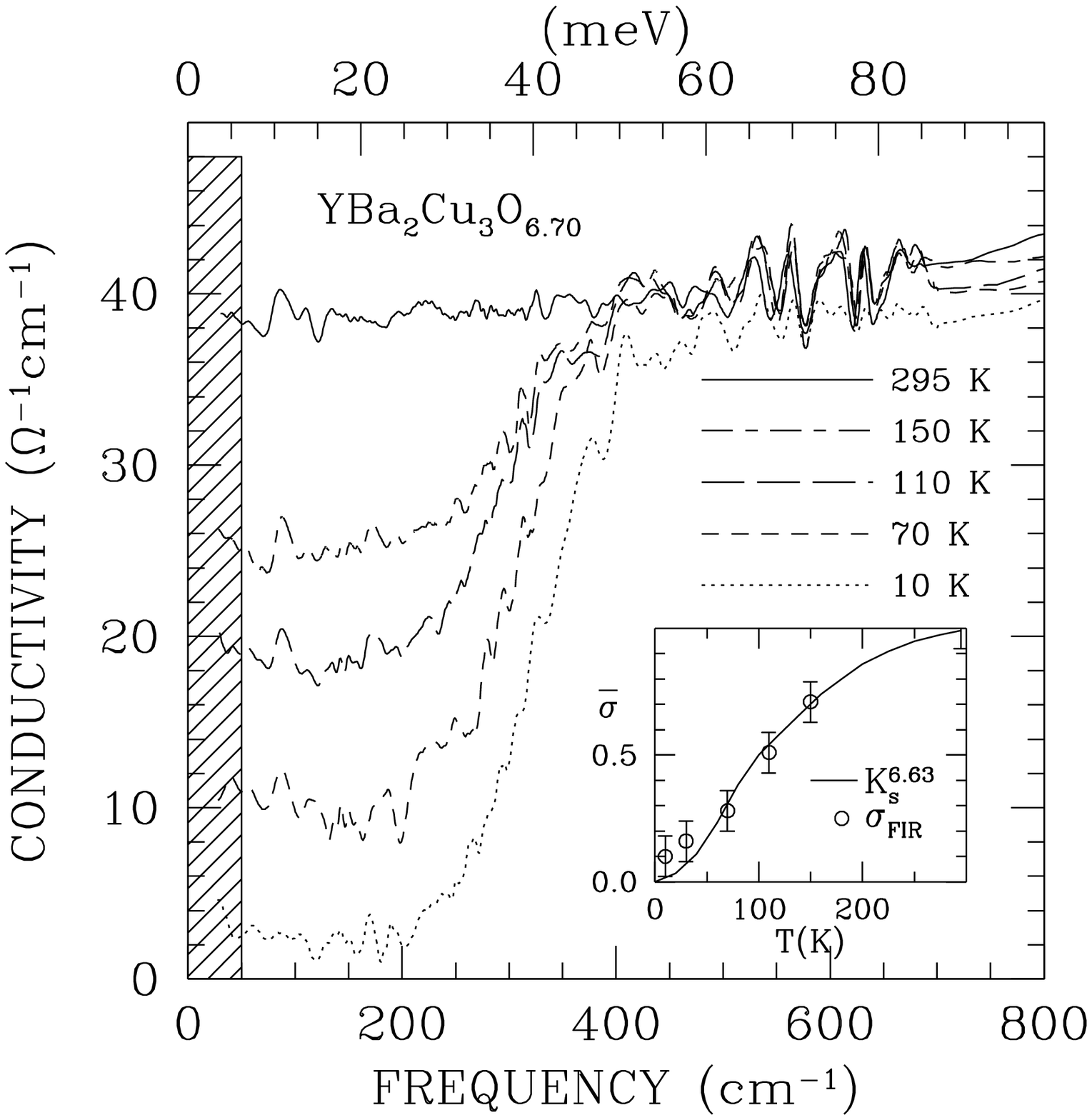}}}
\caption{Experimental evidence for a pseudogap.  Left panel is the NMR 
relaxation rate for various samples of Bi2212 \cite{ISHIDA}, with a 
suppression of $1/T_{1}T$ (spin gap) starting at $T^*$ well above $T_{c}$ 
for underdoped 
samples.  Right panel is the c-axis
conductivity for underdoped YBCO, with a pseudogap which
fills in with temperature \cite{PUCH}.  The inset shows that the 
subgap conductance scales with the Knight shift.}
\label{fig7}
\end{figure}

This brings us to the most controversial aspect of the cuprate field, 
the nature of the pseudogap phase \cite{TIMRPP,VARENNA}.
The first experimental indication 
of such a phase was from NMR measurements by the Bell group, which 
showed that the spin lattice relaxation rate of underdoped cuprates begins to 
decrease well above $T_c$ (left panel, Fig.~\ref{fig7}) \cite{WARREN}.  
A similar decrease is seen in
the Knight shift \cite{ALLOUL}, which measures 
the bulk susceptibility (NMR measuring the zero energy limit of the imaginary 
part of the dynamic susceptibility divided by the energy).  A 
signature of a gap was also evident in infrared measurements, which 
showed a dip in the conductance
separating the low energy Drude peak from the so-called 
mid-infrared bump \cite{COOPER,ROTTER}, with the temperature dependence of 
the conductance near the dip energy  \cite{ROTTER} scaling 
with the spin lattice relaxation rate \cite{TAKI2}.
The resulting sharpening of the Drude
peak leads to a decrease of the planar resistivity in the pseudogap 
phase \cite{RESA}.
But the most dramatic effect was in 
c-axis polarized infrared measurements, which showed a significant gap 
at low energies (no Drude peak), with the temperature dependence of the subgap 
conductance \cite{HOMES} tracking the Knight shift (right panel, Fig.~\ref{fig7})
\cite{TAKI2}.  This 
leads to an insulating up turn below the pseudogap temperature, 
$T^*$, in the c-axis resistivity \cite{RESC}.

What brought the pseudogap effect to the forefront, though, was its 
observation by angle resolved photoemission \cite{MARSHALL,LOESER,DING}.
These experiments found that although the quasiparticle peak in the 
spectral function was destroyed above $T_c$, the spectral gap 
persisted to the higher temperature, $T^*$.  This so-called leading 
edge gap had a similar magnitude and momentum anisotropy as the 
superconducting energy gap, leading to the speculation that this gap 
was a precursor to the superconducting gap, that is, that the 
pseudogap phase represented pairs without long range phase
order \cite{DING}.  This picture was consistent with the NMR and Knight 
shift data, in that pair formation is equivalent to singlet formation, 
and thus the Knight shift and the spin lattice relaxation rate should 
decrease accordingly \cite{MOHIT92}.  The strong coupling limit of this 
picture is simply the RVB physics mentioned above, where the pseudogap 
state corresponds to d-wave pairing of spins.  This picture received 
further support by later ARPES experiments which showed that the 
pseudogap's minimum gap locus in momentum space
coincided with the normal state Fermi 
surface.  That is, the pseudogap is locked to the Fermi surface,
as would be expected for a Q=0 instability \cite{DING97} (in 
superconductors, pairs have zero center of mass momentum).  
Later tunneling experiments were found to be in support of this 
picture as well \cite{FISCHER}.

\begin{figure}
\centerline{\epsfxsize=0.8\textwidth{\epsfbox{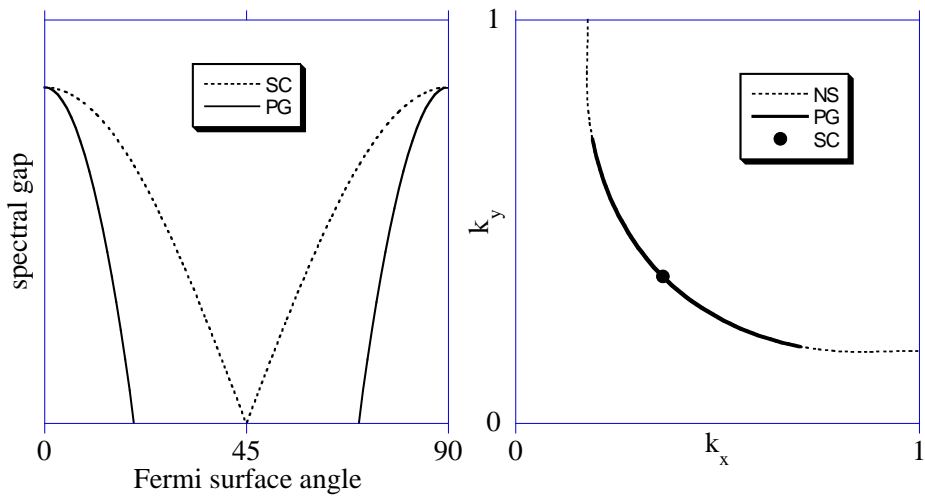}}}
\caption{Momentum anisotropy of the pseudogap from ARPES.  Spectral 
gap around the Fermi surface is plotted in the left panel (SC is the 
superconducting state, PG the pseudogap phase), the locus of gapless 
excitations in the right panel (NS is the normal state Fermi surface, 
PG the Fermi arc of the pseudogap phase, and SC the node of the 
d-wave superconducting state).}
\label{fig8}
\end{figure}

Despite this, the situation, even from ARPES, remains controversial.  
ARPES experiments reveal that the pseudogap turns off at different 
momentum points at different temperatures, leading to the presence of 
temperature dependent Fermi arcs (Fig.~\ref{fig8}) \cite{NAT98}.
These arcs persist to 
low doping; they have even been observed for LSCO at a 3\% doping level,
where the system is on the verge of magnetism \cite{YOSHIDA}.  Some 
authors have taken this as evidence that these arcs represent one side 
of a small hole pocket, the latter picture expected when doping the 
magnetic insulating phase (the idea being that SDW coherence factors 
suppress the intensity on the back side of the pocket).  This magnetic 
precursor scenario is one of the leading alternates to the preformed 
pairs picture.  Despite much searching, though, no ARPES experiment 
to date as ever seen a true hole pocket centered about $(\pi/2,\pi/2)$.

Another explanation has been put forward that the pseudogap represents 
an orbital current phase.  This was implicit in certain treatments of 
the RVB model, which predicted at low dopings the presence of a 
so-called staggered flux phase, which is quantum mechanically 
equivalent to the d-wave pair state in the zero doping 
limit \cite{LEESF}.
This has been generalized to the d density wave state, first 
discussed by Heinz Schulz \cite{SCHULZ}, but popularized by Laughlin 
and colleagues \cite{NAYAK}.  A related picture has been put forth by 
Varma, where his orbital current phase is the result of a non-trivial 
projection of the three band Hubbard model onto the low energy 
sector \cite{VARMA}.  Some experimental evidence for such a state was 
obtained from inelastic neutron scattering which indicated a 
momentum form factor inconsistent with simple Cu spins \cite{MOOK}, but after 
studies by several groups, the feeling is that the observed effect may
represent an impurity phase (always a problem for neutrons given the 
large crystals needed for such measurements).  The latest evidence, 
though, has been given again by ARPES, where Campuzano's group has 
done measurements with circularly polarized light \cite{NAT02}.  What 
they have found is the presence of chiral symmetry breaking below 
$T^*$.  The momentum dependence of the effect, though, is not what is 
expected from the d density wave scenario, though one of the two 
orbital current states proposed by Varma appears to be consistent 
with these observations \cite{VARMA02}.

\begin{figure}
\centerline{\epsfxsize=1.0\textwidth{\epsfbox{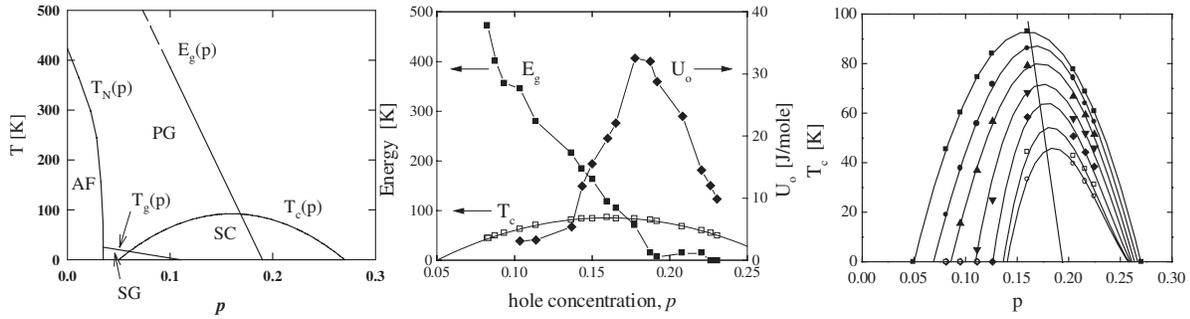}}}
\caption{Left:  Tallon/Loram picture of the phase diagram, with a
quantum critical point at $p$=0.19 where $E_g$, the pseudogap energy scale, 
vanishes ($p$ is the hole doping) .  $SG$ is a spin glass phase.
Middle: Variation with doping of $E_g$ and the superconducting condensation energy
$U_0$ as extracted from specific heat data on YBCO.
Right: Collapse of the superconducting dome about the $E_g$ crossover line
with increasing cobalt doing for Bi2212.  From Ref.~\cite{LORAM}.}
\label{fig9}
\end{figure}

The main debate, though, is whether the pseudogap phase represents a 
state with true long range order (which the neutron and circularly 
polarized ARPES give some evidence for), or simply some precursor 
phase.  If it is the former, then the preformed pairs scenario is 
probably wrong, unless the ordering is some parasitical effect.  Long 
range order is certainly in line with a quantum critical point
scenario.  Additional support for such a scenario comes from
various measurements by Loram and Tallon (Fig.~\ref{fig9}), who claim that the 
pseudogap phase line passes through the superconducting dome and goes 
to zero at some critical point within the dome
(19\% doping) \cite{LORAM}.
One of the strongest points given as evidence 
for their conjecture is that upon impurity doping, 
the superconducting dome appears to collapse about the pseudogap 
phase line (right panel, Fig.~\ref{fig9}).  In their picture, the specific heat data indicate a 
loss of states in the pseudogap phase.  This implies that the 
pseudogap eats up part of the Fermi surface, leaving a smaller part 
available for pairing, thus explaining the collapse of $T_c$ on the 
underdoped side.  This ``mean field'' picture is in total contrast to 
the phase fluctuation picture discussed in the context of the 
precursor pairing scenario \cite{EMERY95}.

Recently, there has been a new measurement which comes out in support 
of the precursor pair scenario.  Ong's group has measured the Nernst 
effect, the coefficient of a higher order transport tensor which is 
very small in normal metals, but is appreciable in 
superconductors because of the presence of vortices.  What they find 
is a sizable Nernst signal on the underdoped side of the phase 
diagram which persists well above $T_c$, though not as high in 
temperature as the ARPES pseudogap \cite{NERNST}.  The only 
explanation that has been put forth for this amazing observation is 
that the pseudogap phase does indeed contain vortices.  This may be 
connected to the results of STM measurements, which reveal that the 
pseudogap forms in the vortex cores in the superconducting state for 
underdoped samples \cite{CORE}.

\subsection{Inhomogeneities}

How one crosses over from the Fermi arc state to the insulator is 
still an unresolved issue.  In one scenario, the chemical potential jumps from the middle 
of the Mott-Hubbard gap to either the lower Hubbard band upon hole 
doping, or the upper Hubbard band upon electron doping.  There is some 
evidence of this from photoemission. In particular, 
the Fermi arc is in a momentum region near the $(\pi/2,\pi/2)$ 
point, and the latter is the top of the valence band in the 
insulator, as seen by photoemission in Sr$_2$CuO$_2$Cl$_2$ \cite{WELLS}.
This picture has been bolstered recently by 
photoemission experiments on the sodium doped version of this 
insulator, which also find a Fermi arc \cite{NADOP}.

The other 
scenario is that upon doping, one creates new states inside the gap.  
Shen's group has given evidence that the latter scenario occurs in 
LSCO, and argues that this is associated with the strong inhomogeneities present in 
that material \cite{RMP03}, though it should be remarked that 3\% doped 
LSCO has the same Fermi arc that is seen in other hole doped cuprates
such as Na doped Sr$_2$CuO$_2$Cl$_2$ \cite{YOSHIDA} and underdoped 
Bi2212 \cite{MARSHALL,NAT98}.

This brings us to the question of stripes.  At low doping, materials can 
be subject to electronic phase separation.  This tendency occurs 
since each doped hole breaks four magnetic bonds, and thus this 
magnetic 
energy loss can be minimized by the holes clumping together.  This
clumping is opposed by the Coulomb repulsion of the holes.  This led 
to the picture that the doped holes, as a compromise,
might form rivers of charge, known 
as stripes \cite{ZAANEN,EMKIV}.

\begin{figure}
\centerline{\epsfxsize=0.8\textwidth{\epsfbox{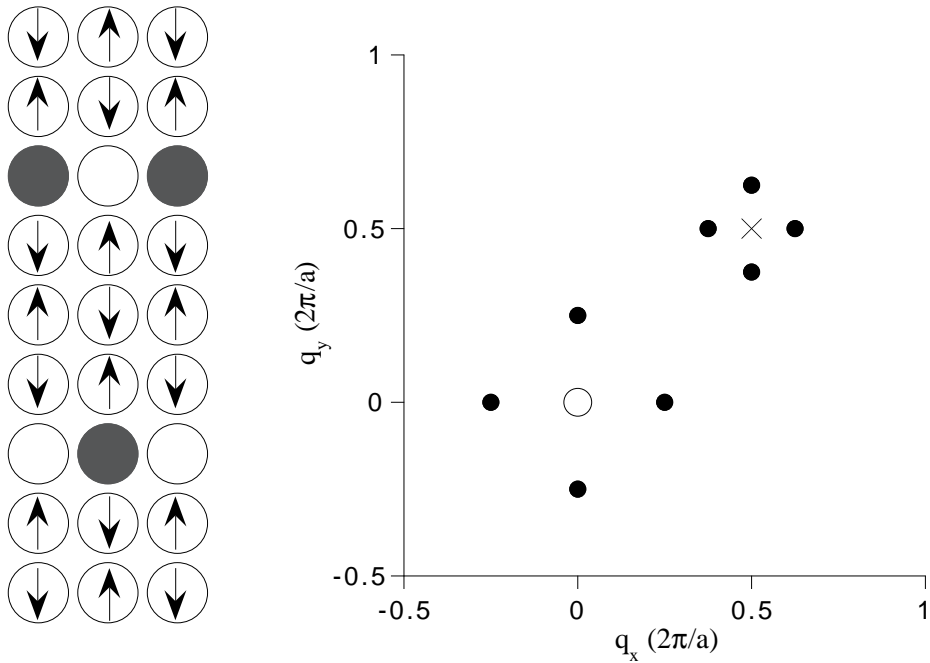}}}
\caption{Stripe picture.  Left panel illustrates stripes for $\delta$=1/8 
doping, arrows represent spins, dark circles doped holes.  Right 
panel plots the resulting neutron scattering peaks (averaged between 
the x and y directions), with charge peaks at $\pm 2\delta$
about (0,0) and spin peaks at $\pm \delta$ about (0.5,0.5).
Adapted from Ref.~\cite{JOHN95}.}
\label{fig10}
\end{figure}

The first evidence for such stripes was given by Tranquada and 
co-workers 
using neutron scattering (Fig.~\ref{fig10}) \cite{JOHN95}.  To understand their result, 
it should be noted that LSCO has a peculiarity in its phase diagram.  
LSCO is normally orthorhombic, the tetragonal phase existing 
either at high temperatures or under pressure.  
Near 1/8 hole doping, though, LSCO has a tendency to distort from its normal 
orthorhombic phase to another distorted phase known as low 
temperature tetragonal, which differs from the high temperature 
tetragonal phase mentioned above \cite{AXE}.
This LTT phase is 
stabilized by neodynium doping.  What Tranquada and co-workers found was that the 
LTT phase exhibited long range ordering consistent with the 
formation of a 1D density wave state.  Both charge and spin ordering 
occurs, but the former sets in at a higher temperature.  This 
stripe formation is consistent with later photoemission \cite{SHENND} and 
transport \cite{ANDO} measurements, which again indicate 1D behavior.

What remains controversial, though, is whether stripes exist only
for this anomalous Nd-doped LSCO compound.  Strong incommensurate 
magnetic spots are seen by inelastic neutron scattering for various dopings 
in LSCO \cite{NS-LSCO} and YBCO \cite{NS-YBCO}, and have been taken as 
evidence for the existence of dynamic stripes \cite{MOOK02}, given their 
resemblance to the static spot pattern of Nd-doped LSCO \cite{JOHN95}.  On the 
other hand, these spot patterns can also be reproduced from standard 
linear response calculations based on the known Fermi 
surface geometry \cite{NORM00}.

What is clear, though, is that there is a definite tendency for 
underdoped materials to exhibit electronic inhomogeneity.  The most 
dramatic example of this has been recently provided by STM studies on 
underdoped Bi2212 by Davis' group \cite{PAN,LANG}.  What they find is the 
existence 
of large gap regions (which have spectra reminiscent of the pseuodgap 
phase) imbedded in smaller gap regions, with the relative fraction of 
the larger gap regions increasing with underdoping, similar to earlier 
results by Roditchev's group \cite{CREN}.  The large gap 
domains have a size of order 30$\AA$.  This granular picture would 
certainly suggest that the pseudogap phase is not a simple precursor 
to the superconducting phase as has been asserted by previous ARPES 
and STM studies.

More recently, the same group has seen a charge density wave state 
associated with the vortex cores which they inferred from Fourier 
transformation of the real space STM spectra \cite{HOFFMAN}.  An 
anisotropy of the pattern gave some indication that this might involve 
stripe formation.  But even more recently, the same group has done a 
Fourier transformation of zero field data \cite{HOFF2}.  They find a weaker 
inhomogeneity in this case, but more interestingly, the Fourier peaks
dispersed with energy.  Although the interpretation of such Fourier 
transforms remain controversial (Kapitulnik's
group \cite{HOWALD} sees similar patterns which they attribute to 
stripe formation), the latest 
results \cite{MCELROY} are consistent with Friedel oscillations from 
impurities whose momentum wavevectors can be used to map out 
the Fermi surface and gap anisotropy.  The results are consistent with 
previous ARPES studies, and have been taken as support of the
interpretation of incommensurability in neutron scattering as due to 
the Fermi surface geometry, as opposed to stripes.

\subsection{Electron Doped Materials}

Electron doped materials have been less studied, mainly due to 
metallurgical problems.  What has been learned, though, is 
that the magnetic phase extends much further in doping than on the hole 
doped side \cite{TOKURA,LUKE90}.  The superconducting phase has a 
lower $T_c$ 
than on the hole doped side, probably for the same reason.  A pseudogap 
phase is observed which appears to be a precursor to the magnetic 
phase in that they exist over the same doping range \cite{ONOSE}, though 
it should be remarked that the pseudogap 
seen is not the leading edge gap (discussed above in the context of 
ARPES), but rather the ``high energy pseudogap'' to be discussed later 
on in the ARPES section.

The symmetry of the superconducting order parameter is still somewhat 
controversial in these materials.  Earlier penetration depth 
measurements \cite{WU93} and point contact tunneling \cite{JOHN90}
showed behavior expected for s-wave pairing, but more recent 
penetration depth measurements have found a power law temperature 
dependence consistent with a disordered d-wave 
state \cite{NCCO}.  Recent ARPES measurements are also consistent 
with d-wave symmetry \cite{SATO,PETER1}, but these experiments are near 
the resolution 
limit because of the small energy gap.  It should be mentioned that 
the tri-crystal experiments mentioned above in the context of hole 
doped superconductors have been performed for the electron-doped case as 
well, and again find a half integral flux quantum in geometries predicted 
by d-wave symmetry \cite{TSUEI2}.  Based on these developments, it is 
fairly certain that the pairing symmetry is d-wave in these systems.

More recently, Raman studies by 
Blumberg and co-workers \cite{GIRSH} find evidence that the 
superconducting gap maximum 
is displaced away from the Fermi surface crossing
along $(\pi,0)-(\pi,\pi)$ (as expected for a d-wave gap based on near 
neighbor pairs)
to the ``hot spots'' (where the Fermi surface crosses the 
magnetic Brillouin zone boundary).  This result is consistent with spin 
fluctuation mediated pairing if the magnetic correlation length is
long (not surprising, given the persistence of long range magnetic order 
over a larger part of the electron doped phase diagram).  
It should be noted that ARPES sees
an intensity suppression at these ``hot spots'' \cite{PETER1} associated 
with the formation of the ``high energy pseudogap'' \cite{PETER2} 
mentioned above.  In 
addition, at low dopings, low energy spectral weight is found around 
the $(\pi,0)$ point \cite{PETER3}, as opposed to the $(\pi/2,\pi/2)$ 
point characteristic of the hole-doped material.  This electron-hole 
asymmetry is what is expected if the chemical potential jumps to the 
upper Hubbard band upon electron doping.

\section{Theory}

\subsection{BCS}

The first microscopic theory of superconductivity, the much 
celebrated BCS theory \cite{BCS}, took many years to come about \cite{SUPER}.
The reason was 
that the machinery needed to construct a proper many-body theory of 
electrons did not emerge until the 1950s.  The motivation behind the 
theory was the isotope experiments of 1950 and their simultaneous 
prediction by Frohlich based on the electron-phonon interaction.  
Bardeen also understood the critical role that the concept of an energy gap 
would play in the ultimate theory.  Once Leon Cooper, an 
expert in many-body theory, joined Bardeen's group as a postdoc, 
progress was rapidly made.

\begin{figure}
\centerline{\epsfxsize=0.6\textwidth{\epsfbox{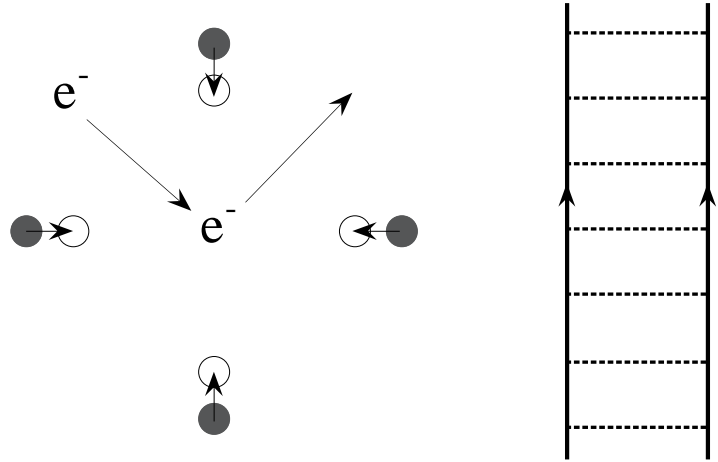}}}
\caption{Electron-phonon interaction leads to attraction (left 
panel).  Arrows joining circles show displaced ions; the time scale 
of these ions for relaxation back is slow compared to the electron 
dynamics.  Right panel is the ladder sum for repeated electron-phonon 
scattering which leads to an electron pairing instability.}
\label{fig11}
\end{figure}

What was known by that time was that the electron-phonon interaction 
could provide attraction among the electrons.  The way this 
works is as follows (left panel, Fig.~\ref{fig11}).  Positive ions are attracted to an electron 
because of the Coulomb interaction.  But, the ion dynamics are slow 
because of their heavy mass.  Thus, once the electron moves away, 
another electron can move into this ``ionic hole'' before the ions 
have a chance to relax back.  This provides attraction at the same 
point in space which can lead to pair 
formation.  The interaction is retarded in time, though, which is what 
ultimately puts a limit on the transition temperature (energy, and 
thus temperature, being conjugate to time).

Such a ``pair'' theory had been proposed in the past, but the physics 
of conventional superconductors does not resemble simple Bose 
condensation, despite the fact that a pair of fermions behaves quantum 
mechanically like a 
boson.  The key discovery by Cooper was the concept of Cooper 
pairs.  The idea is that one is dealing with a degenerate system, 
that is, a filled Fermi sea.  The problem Cooper considered was 
two electrons sitting in unoccupied states above the Fermi sea.  As 
the temperature is lowered, the particle-particle response function 
diverges logarithmically because the 
Fermi distribution function of the electrons approaches a step 
function.  This divergence is strongest when the two electrons are in
time reversed 
states (that is, a state k and a state -k, thus the center of mass 
momentum of the pair is Q=0).
Note the difference from the particle-hole response at Q=0, which 
simply measures the density of states at the Fermi energy.

This logarithmic divergence is cut-off at some ultraviolet energy 
scale, which in the electron-phonon problem is the Debye energy, 
$\omega_D$.  This 
leads to a bare response function that goes as $\chi_0=N\ln(\omega_D/T)$ 
where $N$ is the density of states.
Now 
one sums a ladder series (repeated scattering of the two electrons, right
panel, Fig.~\ref{fig11}),
which leads to an expression for the 
full response function of $\chi=\chi_{0}/(1-V\chi_0)$ where $V$ is 
the interaction.  For positive 
$V$ (attractive in our sign notation), the denominator will have a 
pole when $T_{div}=\omega_D e^{-1/\lambda}$ with $\lambda=NV$.  This is 
the famous Cooper pair divergence.

This doesn't answer the question of what the ground state is.  This 
problem was solved by Schrieffer, Bardeen's graduate student (thus 
BCS).  Based on Cooper's solution, he guessed the many-body ground 
state at T=0 (a rare accomplishment, the other well known example of 
this was Laughlin's guess for the fractional quantum Hall state).  It 
is of the form $\prod_k (u_k + v_k c^{\dag}_k c^{\dag}_{-k} |0>$.  
Here $|0>$ is the vacuum and $u,v$ are ``coherence'' 
factors (the sum of whose squares equals one).  What can be seen here 
is that the BCS ground state is a superposition of states where the 
pair $k,-k$ is either occupied or filled.  Solving the variational 
problem (equivalent to replacing the product of the two creation 
operators by a c number), BCS found that
$u_k^2,v_k^2 = 1/2(1 \pm \epsilon_k/E_k)$ where 
$E_k=\sqrt{\epsilon_k^2+\Delta_k^2}$ and $\Delta_k$, gotten from 
solving an integral equation (the so-called gap equation), has the same
form as $T_{div}$ above.

\begin{figure}
\centerline{\epsfxsize=0.4\textwidth{\epsfbox{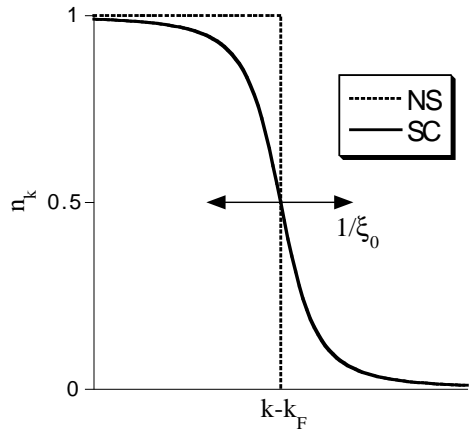}}}
\caption{Momentum distribution function for the normal state (NS) and 
superconducting state (SC).  $\xi_0$ is the BCS correlation length.}
\label{fig12}
\end{figure}

Note that the Fermi distribution function has been replaced by 
$v_k^2$, thus leading to particle-hole mixing (Fig.~\ref{fig12}).
In essence, the BCS 
instability is a consequence of Nature's abhorence of singularities 
(in this case, the step function behavior of the Fermi function at 
T=0).  This distribution is smeared over a momentum range of $\sim 
\Delta/v_F$ where $v_F$ is the Fermi velocity, thus defining the 
inverse correlation length.  Excitations from the 
ground state can be formed by breaking up a pair.  These excitations 
are of the form $\gamma^{\dag}_k = u_k c^{\dag}_k - v_k c_{-k}$ and have an 
energy $E_k$.  That is, on the Fermi surface, the quantity $\Delta$, 
which is related to the order parameter, is 
nothing more than the spectral gap, and thus the energy gap emerges 
quite naturally from the theory.

What allows this conceptually simple picture to work is Migdal's 
theorem \cite{BOB}.  It states that the single particle self-energy 
can be treated to lowest order, since $\omega_D/E_F$ is a small 
expansion parameter (where $E_F$ is the Fermi energy).  Thus, the only 
diagram series which has to be summed is the particle-particle ladder 
mentioned above.  Such a theorem obviously does not apply if the 
pairing is due to electron-electron interactions.

\subsection{Spin Fluctuation Models}

As mentioned above, the electron-phonon attraction is local in space 
and retarded in time.  This leads to L=0 pairs.  By fermion 
antisymmetry, this requires that the pair state be a spin singlet.  
For the case of electron-electron interactions, L=0 pairs are usually 
not favored (because of the direct Coulomb repulsion between the 
electrons).  In fact, one might wonder how one can ever get an
``attractive'' interaction in this case.

Let us start with the nearly ferromagnetic case \cite{FAY}.
\begin{figure}
\centerline{\epsfxsize=0.3\textwidth{\epsfbox{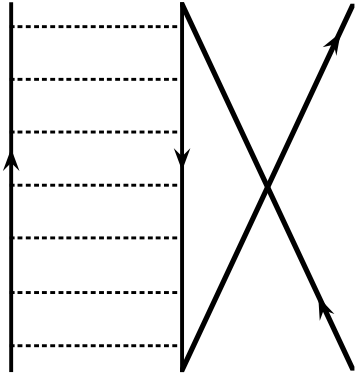}}}
\caption{Particle-particle diagram for the spin fluctuation case.  
Note the particle-hole ladder sum buried inside this diagram.}
\label{fig13}
\end{figure}
The 
particle-particle ladder sum in this case involves exchanging the ends 
of one of the particle lines, thus representing a particle-hole ladder sum 
buried inside of a particle-particle one (Fig.~\ref{fig13}).  Thus, the diverging 
particle-hole response (representing a ferromagnetic instability) drives a 
diverging particle-particle response.  For S=1 pairs, this is 
attractive.  In essence, the bare triplet interaction is zero (due to 
the Pauli exclusion principle) and the induced interaction is 
attractive (representing the tendency for an up spin electron to have 
another up spin electron nearby).  By fermion antisymmetry, the L 
state must be odd, thus allowing the two electrons in the pair to avoid 
coming too close to one another (thus minimizing the direct Coulomb 
repulsion).

For heavy fermions and cuprates, though, the nearly antiferromagnetic 
case is of more interest.  In that case, one again wants to avoid the 
direct Coulomb repulsion, but now a spin up electron wants to have a 
spin down electron nearby.  In the absence of spin-orbit, this implies 
S=0, L=2 pairs.  For strong spin-orbit, $S=1,S_z=0$ pairs can be 
stable as well, which is the basis of the ``f-wave'' scenario postulated by 
Norman in the case of heavy fermions \cite{MIKE92}, but this is not 
relevant for the cuprate case.

\begin{figure}
\centerline{\epsfxsize=0.8\textwidth{\epsfbox{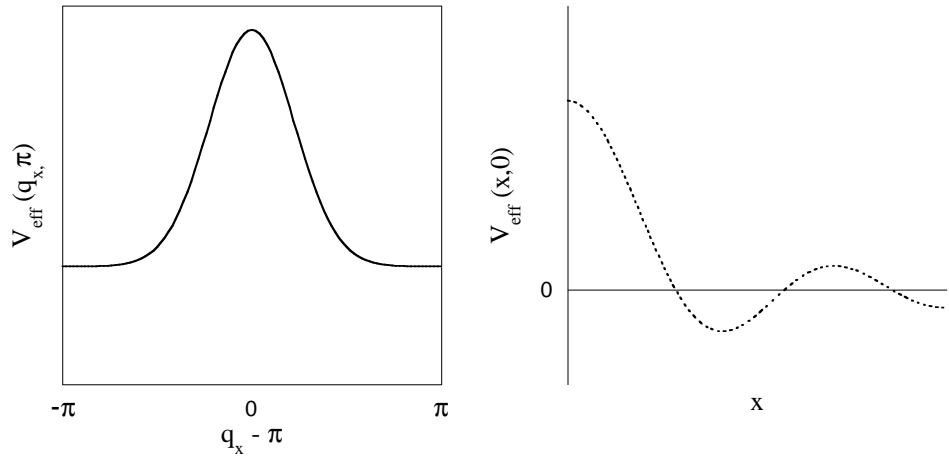}}}
\caption{Effective interaction for spin fluctuation mediated 
pairing (AF case).  Left panel is for momentum space (overall repulsive, peaked 
at $(\pi,\pi)$), right panel for real space (repulsive on-site,
with first attractive minimum at a near neighbor separation).  Adapted from 
Ref.~\cite{DOUG}.}
\label{fig14}
\end{figure}

The above considerations on a 2D square lattice leads to a pair state 
with $d_{x^2-y^2}$ symmetry \cite{1986,BICKERS}.  In fact, the theory 
of this is a bit counterintuitive.  Unlike the nearly ferromagnetic 
case, in the nearly antiferromagnetic case, the interaction is always 
repulsive (in a momentum space representation, left panel, Fig.~\ref{fig14}), 
and in fact is most 
repulsive at $Q=(\pi,\pi)$ where the antiferromagnetic instability would 
occur.  But the d-wave version of $\Delta_k$ changes sign under translation 
by $Q$ 
(it is of the form $\cos(k_x)-\cos(k_y)$), and this sign change 
compensates for the repulsive sign of the interaction when solving the 
integral (gap) equation for $\Delta_k$.  In real space, the picture is more 
clear (right panel, Fig.~\ref{fig14}) \cite{DOUG}.
The on-site interaction is repulsive (which is 
why the overall interaction is repulsive in momentum space), but this 
interaction contains Friedel oscillations, with the first (attractive) 
minimum at a near neighbor separation, representing the tendency of
opposite spin electrons to be on neighboring sites.

Note that retardation does not play the central role in the above arguments
as in the phonon case.  In 
essence, there is no Migdal's theorm in this case \cite{HL}, and so 
one may question such a theory which does not take into account vertex 
corrections.  The justification that is often given is that if one 
considers the electrons and spin fluctuations as separate objects (like 
electrons and phonons), then as the spin fluctuations are ``slow'' 
relative to electrons, one gets something like a Migdal's theorem.  
But this simple argument usually breaks down \cite{HL}, though 
Chubukov has recently made arguments about why an effective Migdal 
theory would apply in the cuprate case \cite{AC}.  Regardless, 
feedback effects definitely have to be considered whenvever 
electron-electron interactions are involved, since the spin 
fluctuation propagator is drastically changed by the introduction of 
the superconducting gap for the electrons \cite{AC}.  The 
classic example of this is the stabilization of the $A$ phase relative to 
the $B$ phase in superfluid $^3He$ \cite{BA}.

\subsection{RVB}

The spin fluctuation theory is essentially a weak coupling approach.  The RVB 
picture mentioned in the introduction is the strong coupling 
version of the spin fluctuation approach \cite{RVB} (though 
Anderson differs on this \cite{ADV}).  The amazing thing was how 
quickly the RVB concept emerged (Anderson first spoke on this 
before the discovery of YBCO).  It has certainly been 
controversial (one well known scientist, whose name will not be 
mentioned here, quipped that RVB actually stood for ``rather vague 
bullshit'').

\begin{figure}
\centerline{\epsfxsize=0.8\textwidth{\epsfbox{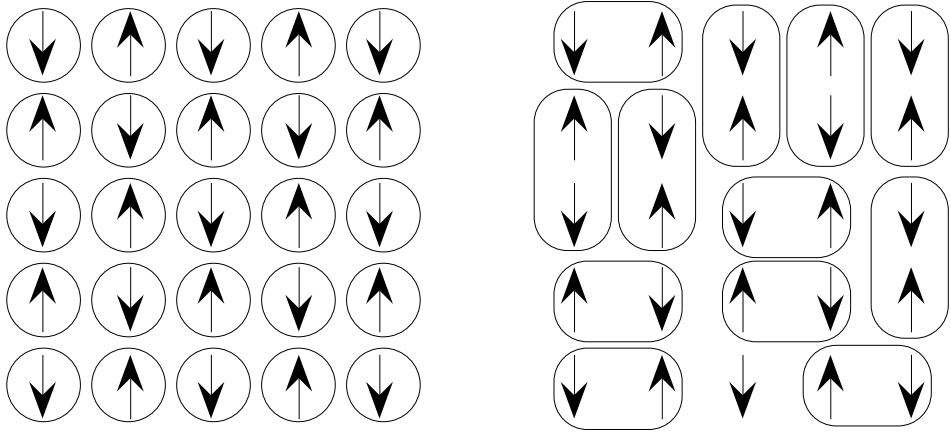}}}
\caption{Neel lattice (left panel) versus RVB (right panel).  
The RVB state is a liquid of spin singlets.}
\label{fig15}
\end{figure}

To understand this approach, consider the undoped insulator, where 
there is one Cu spin per site (the Cu being in a $d^9$ configuration).
The ground state of this system is an antiferromagnet.  The reason is
that if the spins on each site are parallel, then they cannot virtually 
hop because of the Pauli exclusion principle, but they can if the 
spins are antiparallel.  In this case, the virtual hopping leads to 
an energy lowering of $J = t^2/U$ per bond, where $t$ is the effective 
Cu-Cu hopping integral, and $U$ the Coulomb repulsion for double 
occupation.  Mean field theory would then predict that the arrangement 
of spins forms a Neel lattice of alternating up and down spins (left panel,
Fig.~\ref{fig15}).

Anderson, though, suggested that the Cu case was special, since the 
spin of the single d hole was only 1/2.  Because of this, he 
anticipated that quantum fluctuations would melt the Neel lattice, 
leading to a spin liquid ground state, a fluid of singlet pairs of 
spins (right panel, Fig.~\ref{fig15}) \cite{RVB}.  
The ``RVB'' notation comes from the fact that a given spin 
could be taken as paired with any one of its four neighbors, thus 
each bond fluctuates from being paired to not paired.  This is 
analogous to benzene rings, where each C-C link fluctuates between a 
single bond and a double bond.
Although it was discovered soon after \cite{INS} that the undoped 
insulator is indeed a Neel lattice, this lattice does indeed ``melt'' 
with only a few percent of doped holes.  This is rather easy to understand,
since the kinetic energy of the doped holes is frustrated in the Neel state.

\begin{figure}
\centerline{\epsfxsize=1.0\textwidth{\epsfbox{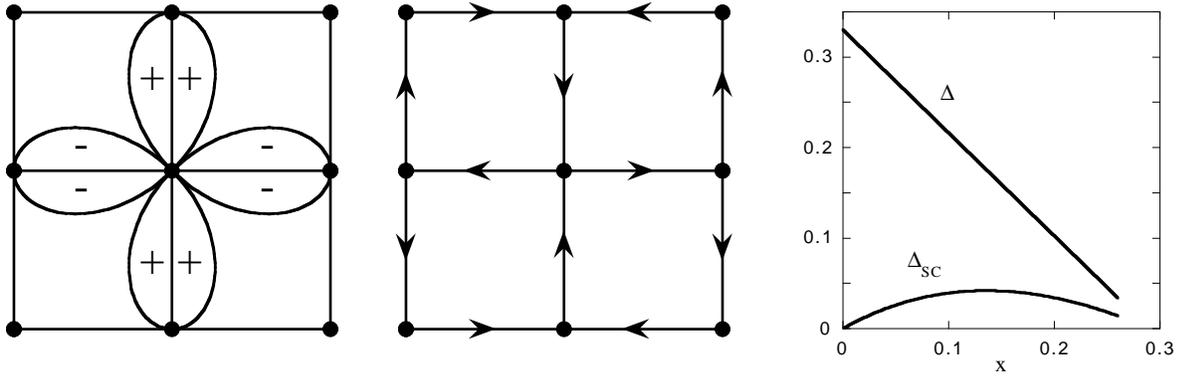}}}
\caption{Two RVB states which are equivalent at half filling.  The 
left panel is a d-wave pairing of spins, the middle panel a $\pi$ flux 
state.  Dots are Cu ions, and arrows are bond currents.  Right:
Variation of the RVB gap parameter, $\Delta$, and the superconducting
order paramter, $\Delta_{SC}$, with doping \cite{GROS}.}
\label{fig16}
\end{figure}

After Anderson's original conjecture, it was realized that at the mean 
field level, there were a number of possible ground states for such a 
spin fluid.  For the undoped case, these states are quantum
mechanically equivalent, since
the presence of a spin up state is equivalent to the absence of 
a spin down state, an effective SU(2) symmetry \cite{SU2}.
That is, various mean field decouplings of the 
Hamiltonian are equivalent since
$<c^{\dag}_{\uparrow}c^{\dag}_{\downarrow}> \equiv 
<c^{\dag}_{\uparrow}c_{\uparrow}>$.
The first decomposition is equivalent to a BCS pairing of spins, 
the second to a bond current state (left panels, Fig.~\ref{fig16}).  
For the spin pairing case, the 
d-wave state is favored since it
does the best job of localizing the two spins on neighboring 
sites.
Its bond current equivalent is the $\pi$ flux phase state, 
where the bond currents flow around an elementary plaquette (square
formed from four Cu-Cu bonds), yielding 
a net phase of $\pi$ per plaquette.

Upon doping, this SU(2) symmetry is broken to U(1).  At the mean 
field level, the d-wave state has the lowest energy \cite{KOTLIAR,GROS}, as
it minimizes the kinetic energy of the doped holes.  The 
variational $\Delta$ associated with the d-wave state is maximal for 
the undoped case and decreases linearly with doping
(right panel, Fig.~\ref{fig16}).
To understand the implications of this for superconductivity, one 
should note that the above variational parameter applies to the 
spins.  But only the doped holes carry the current.  As their density 
increases linearly with doping, then the superfluid density of the real
electrons (a product of spin and charge) varies 
linearly with doping, despite the fact that the variational parameter 
does not.  That is, there is a complete decoupling of $\Delta$ (the
excitation gap) from the order parameter (the superfluid density),
unlike in BCS theory.  These simple 
mean field considerations have been confirmed by recent variational Monte 
Carlo calculations of a ``projected'' d-wave BCS pair state (where double 
occupied states are projected out) \cite{PARM}.  Such projected 
states can also be shown to contain orbital current correlations \cite{ORB}.

Although a finite temperature generalization of RVB theory is 
non-trivial, the overall phase diagram can be easily appreciated by 
noting that the ``pairing'' temperature scale, $T_{RVB}$, will be 
proportional to $\Delta$, and that the phase coherence temperature of 
the doped holes will be proportional to the doping.  The net result 
are two crossing lines with doping, with the spin gap phase in the left
quadrant, 
the Fermi liquid phase in the right quandrant, the superconducting 
phase in the bottom quadrant (below both lines), and the strange metal 
phase in the upper quadrant (see left panel, Fig.~\ref{fig6}) \cite{LEE}.

One of the most important concepts to be introduced by RVB theory is 
the concept of spin-charge separation \cite{ANDERSON}.  This idea can 
be most easily appreciated by the RVB explanation of the spin gap 
phase \cite{PLEE}.  Since only the spins are paired, then strong effects are 
expected for spin probes but only weak effects for charge 
probes.  This is consistent with planar properties 
of the pseudogap phase, which show a strong spin gap in NMR (left 
panel, Fig.~\ref{fig7}), but only a weak gap-like depression
in the in-plane infrared conductivity.  In fact, the drop in 
in-plane resistance in the pseudogap phase
is easily understood since when the spins pair up and become gapped,
there are less states for the doped holes to scatter off of.  On the other hand,
spin-charge separation, being a 2D effect, only occurs within a plane.  As
spins and charges must thus recombine into real physical electrons to 
tunnel from plane to plane, then large effects are expected in any 
experiment measuring a c-axis current.  This is consistent with 
experiment, since a hard gap is seen in c-axis infrared conductivity 
(right panel, Fig.~\ref{fig7}), 
as well as tunneling and photoemission.

Going beyond mean field considerations, though, has proven to be 
difficult in the RVB scheme.  The most promising approach is to use 
gauge theory to treat the various SU(2) and U(1) symmetries of the 
model \cite{PLEE}.  The resulting gauge field fluctuations coupling the 
``spinons'' and 
``holons'', though, are extremely strong, leading to an uncontrolled 
theory, as expected given the strong coupling nature of the problem.
Still, these calculations have given a number of important insights 
into understanding various excited state properties of the cuprates,
particularly in the spin gap phase.

\subsection{Marginal Fermi Liquid}

\begin{figure}
\centerline{\epsfxsize=0.8\textwidth{\epsfbox{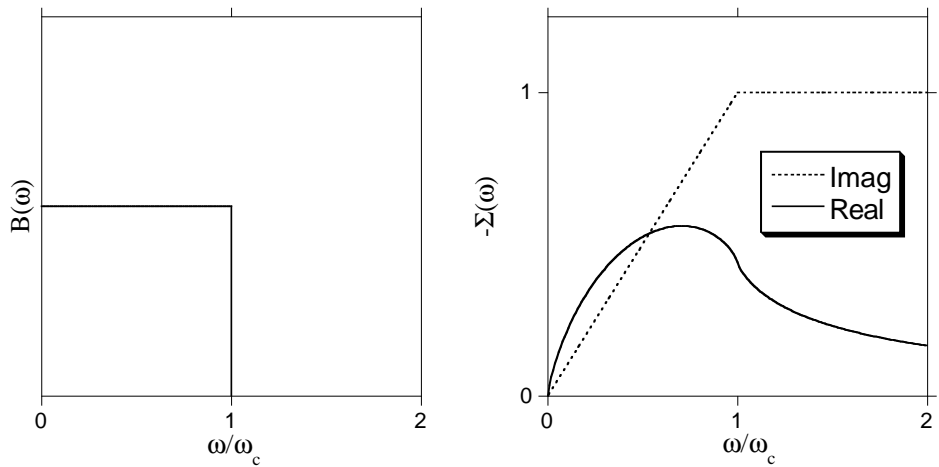}}}
\caption{Bosonic spectrum which yields a marginal Fermi liquid (left 
panel).  Resulting real and imaginary parts of the electron 
self-energy (right panel).  $\omega_{c}$ is the ultraviolet cut-off.  T=0.
Adapted from Ref.\cite{MFL}.}
\label{fig17}
\end{figure}

The striking linear temperature dependence of the in-plane 
resistivity led Varma and co-workers to propose a marginal Fermi 
liquid phenomenology \cite{MFL} to explain many of the anomalous behaviors in 
cuprates.  Their idea was that the electrons are interacting with a 
spectrum of bosonic excitations which has the following form
(left panel, Fig.~\ref{fig17})
\begin{equation}
B(\omega) \propto min(\omega/T,1)
\end{equation}
That is, the bosonic spectrum has no other energy scale present 
besides the temperature, that is, it exhibits quantum critical scaling.
Such a spectrum, though, does not yield a 
convergent fermion self-energy, necessitating the presence of an ultraviolet 
cut-off, $\omega_c$, in the theory.  At zero temperature, the 
resulting self-energy is (right panel, Fig.~\ref{fig17})
\begin{eqnarray}
Im\Sigma \propto \omega \nonumber \\
Re\Sigma \propto \omega \ln \frac{\omega}{\omega_c}
\end{eqnarray}
This result, obtained by a convolution of $B$ and $ImG$ (where $G$ is 
the bare fermion Greens function) can be most easily appreciated by 
noting that for electrons interacting with an Einstein mode, 
$B(\omega)=\delta(\omega-\omega_0)$, $Im\Sigma$ is a step function, 
($Im\Sigma=0, \omega < \omega_0$; $Im\Sigma \propto 1, \omega > 
\omega_0$).  For an array of $\delta$ functions for $B$, $Im\Sigma$ 
becomes a ramp of steps, which in the limit as the energy spacing of the 
$\delta$ functions goes to zero becomes a linear $\omega$ 
behavior.

The ``marginal'' notation comes from the fact that the quantity $1 - 
dRe\Sigma/d\omega$ is logarithmically divergent as $\omega$ 
approaches 0.  As a result, the momentum distribution function, $n(k)$, no 
longer has a step discontinuity at $k_F$ as in a Fermi liquid, but 
rather an inflection behavior.  This logarithm is cut-off by the 
temperature, and a standard calculation of the longitudinal 
conductivity leads to a linear temperature dependence of the 
resistivity \cite{MFL}.

Although an attractive phenomenology for thinking about various 
properties of the cuprates, the deficiency of the model is that there 
is no explicit momentum dependence, leading to the question of
where d-wave pairing would come from.  In later work, Varma has 
claimed that d-wave pairing could arise from vertex corrections 
\cite{VARMA}, but certainly the underlying microscopics behind this 
very successful idea remain somewhat unclear at the present time.

These problems have led to a proposal of another phenomenology to 
explain transport data, the ``cold spots'' model of Ioffe and Millis 
\cite{COLD}.  In this picture, there is a Fermi liquid like 
scattering rate, but it is confined to the vicinity of the d-wave 
nodes.  Such a model can reproduce the linear $T$ resistivity, but this 
nice idea now seems to be ruled out by recent photoemission data
which find that even at the d-wave node, $Im\Sigma$ has 
the linear $\omega$ behavior \cite{VALLA} predicted by the marginal 
Fermi liquid phenomenology \cite{MFL}.

The MFL phenomenology can be easily extended to the superconducting 
state.  Since the $B(\omega)$ spectrum is considered to be electronic 
in origin, then it will acquire a $2\Delta$ gap in the 
superconducting state (see bubble in the left panel, Fig.~\ref{fig5}), 
thus being able to account for the scattering 
rate gap seen in various measurements \cite{KURODA,LITTLEW}.  Such a
``gapped marginal Fermi liquid'', though, cannot account for all 
observations, and has to be supplemented by collective effects 
\cite{NDING}, which we now discuss.

\subsection{SO(5)}

Inelastic neutron scattering measurements by the group of
Rossat-Mignod \cite{ROSSAT} revealed the presence of a
narrow (in energy) resonance in
the superconducting state of YBCO in a small region of momentum centered at 
the $(\pi,\pi)$ wavevector \cite{ROSSAT}.  Subsequent polarized measurements by
the Oak Ridge group \cite{MOOK93} verified that the resonance was 
magnetic in character.  Since the BCS ground state involves S=0 pairs 
with zero center-of-mass momentum, this implies that this 
excited state must involve S=1 pairs with center-of-mass momentum 
$Q=(\pi,\pi)$.  To see this, note that because of the energy gap, only pair 
creation 
processes are present in the particle-hole response at T=0, these processes 
being possible because of particle-hole mixing \cite{SCALAPINO}.  
Since the magnetic signal detected by neutrons involves a spin flip process, 
then the excited pair must have spin one as the ground state is spin
zero \cite{DEMLER}.  Fermion antisymmetry
then implies that the excited pair has odd L.  This is evident as well, since
the d-wave gap function $\cos(k_x)-\cos(k_y)$  translated by $Q/2$
becomes $\sin(k_x)-\sin(k_y)$ \cite{DEMLER}.

At first sight, such a triplet collective mode is a surprise, since they have 
never been found in classic superconductors.  But in the d-wave case, 
since the order parameter changes sign under translation by $Q$, then 
the BCS coherence factor for the pair creation process takes its maximal 
value on the 
Fermi surface, as opposed to the s-wave case where it is zero 
\cite{FONG95}.
\begin{figure}
\centerline{\epsfxsize=0.5\textwidth{\epsfbox{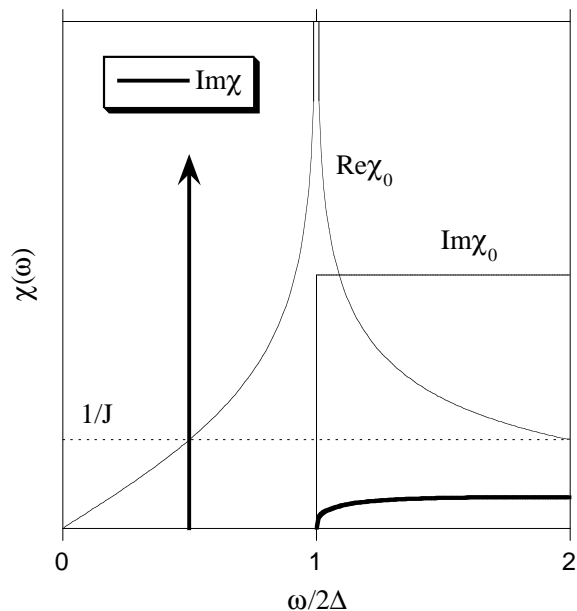}}}
\caption{Real and imaginary parts of the bare bubble for a d-wave 
superconductor.  Intersection of real part with 1/J, where J is the 
superexchange energy, marks the location of the pole in the RPA 
response function (arrow).}
\label{fig18}
\end{figure}
The net result (Fig.~\ref{fig18}) is that the imaginary part of the bare 
particle-hole response, $\chi_0$, has a step function jump from zero to a finite 
value at a threshold of $2\Delta_{hs}$ \cite{NORM00,AC}, where 
$\Delta_{hs}$ is the value of the superconducting gap at the ``hot 
spots'' (points on the Fermi surface connected by $Q$).  By 
Kramers-Kronig, $Re\chi_0$ will then have a 
logarithmic divergence at $2\Delta_{hs}$ because of the step in the 
imaginary part.  Thus, the full response function 
($\chi=\chi_{0}/(1-J\chi_0)$, where $J$ is the superexchange energy) 
will always have an undamped pole at some energy less than the 
threshold energy.  That is, linear response theory (RPA) for a d-wave 
superconductor predicts the presence of a spin triplet collective mode 
below the $2\Delta$ continuum edge.

Experiments, though, reveal that the resonance 
energy does not depend on temperature, and its amplitude scales 
with the temperature dependence of the d-wave order parameter 
\cite{FONG96}.  This is not easily understood in the ``RPA'' framework.
This led Demler and Zhang to 
propose that the spin resonance was in fact a particle-particle 
antibound resonance (antibound, because the triplet interaction is
repulsive).  In such a picture, the resonance is always present, but
can only be detected below $T_c$ by neutron scattering because of 
particle-hole mixing, which allows the particle-particle resonance to 
appear in the particle-hole response \cite{DEMLER}.  This idea 
naturally resolves the two puzzles mentioned above.
Although this original idea has been 
put into question on formal grounds by Greiter \cite{GREITER} (in the 
t-J model, the triplet interaction is formally zero), and on 
kinematics grounds by Tchernyshyov \etal \cite{OLEG} (an 
antibound state is inconsistent with photoemission and tunneling, 
which indicate that the resonance has an energy lower than the 
two-particle continuum), the idea led 
Zhang to propose a very interesting SO(5) phenomenology to explain the 
cuprate phase diagram \cite{ZHANG}.

In the SO(5) picture, the ``5'' stands for the three degrees 
of freedom of the Neel order ($N_x,N_y,N_z$), and the two degrees of 
freedom of the superconducting order (real and imaginary parts of 
$\Delta$).  In an imaginary world where these two order parameters 
were degenerate, then the underlying Hamiltonian would have SO(5) 
symmetry.  This group has ten generators, the three components of the 
spin operator, the charge operator, and six new generators which 
rotate the ``superspin'' between the Neel and superconducting 
sectors.  These new generators are nothing more than the spin 
resonance discussed above (a spin triplet pair with complex 
$\Delta$).  This idea provides a new framework for thinking about 
the phase diagram and the various collective excitations of cuprates
\cite{ZHANG}.  Of course, cuprates are doped Mott insulators, and 
this effect is not present in the theory as stands (that is, charge 
fluctuations are suppressed strongly at low doping).  This has led to 
the development of a version of the theory known as 
``projected SO(5)''  where double occupation has been projected
out \cite{PROJ}.  One result of projected SO(5) 
theory is the claim that it explains the ``d-wave-like'' dispersion of 
the valence band seen in the undoped insulator by ARPES \cite{HANKE}.

\subsection{Stripes}

A number of models of correlated electron systems predict the 
presence of phase separation at small doping, with the system 
bifurcating into hole rich (metallic) and hole poor (magnetic 
insulating) regions.  In some models, these regions form a lamellar 
pattern, i.e., one dimensional ``stripes''.

To connect with experiment, it had been known for some time that 
inelastic neutron scattering experiments for LSCO indicated the 
presence of four incommensurate peaks displaced a distance $\delta$ 
from the commensurate wavevector $Q=(\pi,\pi)$ which characterizes 
the magnetic insulator (right panel, Fig.~\ref{fig10}) \cite{NS-LSCO}.  The standard way of thinking 
about these peaks was that they were due to the Fermi surface geometry 
\cite{LEVIN}.

An alternate way, though, was to consider having holes residing in 1D 
stripes, with magnetic domains between these stripes (left panel, 
Fig.~\ref{fig10}).  Even if the 
local ordering within the magnetic domains is commensurate, if the 
stripes represent an antiphase domain wall, then neutron scattering 
will see incommensurate magnetic peaks, with $\delta$ a measure of the 
spacing between the stripes.  Since the hole density is the doping, 
then this predicts that $\delta$ will have a linear variation with 
doping.  This is indeed what is seen in LSCO, and is known as the Yamada plot 
\cite{YAMADA}.  Associated with these magnetic peaks should be charge 
peaks at positions of $2\delta$ relative to the Bragg peaks.  These have 
been observed as well (right panel, Fig.~\ref{fig10}) \cite{JOHN95}.

An attractive feature of the stripes model is its 1D physics 
\cite{EMKIV}.  The jury is still out whether Fermi liquids are 
inherently unstable in 2D \cite{ANDERSON}, but they definitely are in 
1D.  So, such 1D models naturally contain non Fermi liquid normal 
states exhibiting spin-charge separation.  Moreover, in this picture, 
the pseudogap is nothing more than the spin gap associated with the 
magnetic domains.  Pairs of holes from the stripes 
can obtain pairing correlations by virtually hopping into the magnetic 
domains (the fluctuating Neel order in the magnetic domains 
favors antiparallel spins on neighboring Cu sites).
\begin{figure}
\centerline{\epsfxsize=0.8\textwidth{\epsfbox{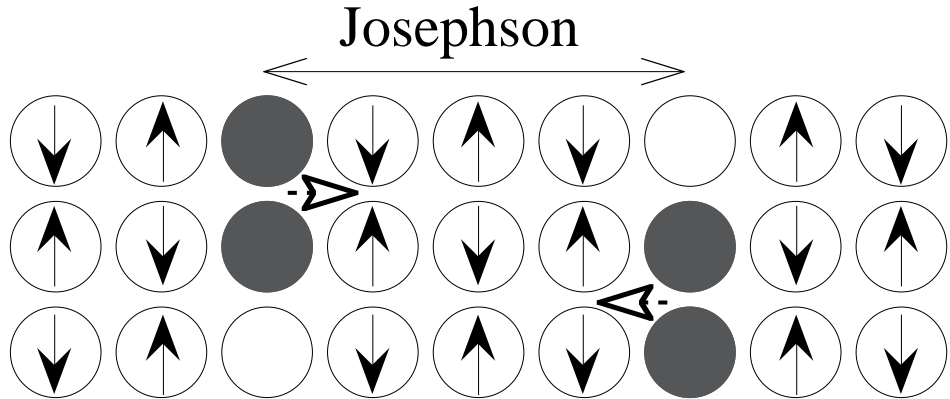}}}
\caption{Stripes model for cuprates.  Pairs of doped holes (dark 
circles) virtually 
hop into AF (spin gap) domains, acquiring spin pairing correlations.  
Josephson coupling of the stripes leads to long range superconducting 
order.  Adapted from Ref.~\cite{EMKIV}.}
\label{fig19}
\end{figure}
Below some 
temperature, the stripes phase coherently lock via Josephson coupling, 
leading to long range (3D) superconducting order (Fig.~\ref{fig19}).  That is, the system 
crosses over from a 1D non-Fermi liquid normal state to a 3D coherent
superconducting state \cite{EMKIV}.

\subsection{Pseudogap}

Most authors agree that the superconducting state is isomorphic to a 
BCS ground state of d-wave pairs.  There is no agreement, however, on 
the nature of the pseudogap phase.  The general hope is that once 
the pseudogap phase is sorted out experimentally, then the number of 
possible theories for cuprates will be drastically reduced.

One general class of theories is that the pseudogap phase represents 
preformed pairs \cite{VARENNA}.  Cuprates are characterized by 
short superconducting coherence lengths, low carrier densities, and 
quasi-two dimensionality.  All of these conditions favor a 
suppression of the transition temperature relative to its mean field 
value due to phase fluctuations.  In the intermediate region between 
these two temperatures, 
preformed pairs are possible.  The cuprates, though, are not in the 
Bose condensation (local pair) limit, in that photoemission still reveals the 
presence of a large Fermi surface (in the Bose limit, the chemical 
potential would actually lie beneath the bottom of the energy band).  
Still, specific heat data clearly reveal the non-mean-field like 
character of the superconducting phase transition, particulary for 
underdoped samples \cite{JUNOD}.

The ``RVB'' picture has subtle differences from that of pre-formed 
pairs.  In this case, the pseudogap phase is a spin gap phase (that 
is, the spins bind into singlets).  As an electron is a product of 
spin and charge, then real electrons acquire an energy gap because of 
the spin gap, though charge excitations confined to the plane do not.  
At low enough temperatures, the doped holes become phase coherent, 
leading to rebinding of spin and charge, and the formation of a true 
superconducting ground state.
The stripes picture is not unrelated to the RVB picture, in that the 
pseudogap is due to the spin gap present in the magnetic insulating 
domains between the stripes.

In the other class of scenarios, the pseudogap is not related to 
superconductivity per se, but rather is competitive with it.  Most of 
these scenarios involve either a charge density wave or spin density 
wave, usually without long range order.  As the spectral gap 
associated with the ordering grows with reduced doping, then more and 
more of the Fermi surface becomes unavailable for pairing, leading to 
an increasing suppression of the superconducting transition 
temperature on the underdoped side.  This is an old idea, going back to 
the A15 superconductors where the martensitic phase transition 
competes with superconductivity \cite{BILBRO}.  These models 
have a certain attractiveness, since the basic physics can 
be appreciated at the mean field level.

\begin{figure}
\centerline{\epsfxsize=0.8\textwidth{\epsfbox{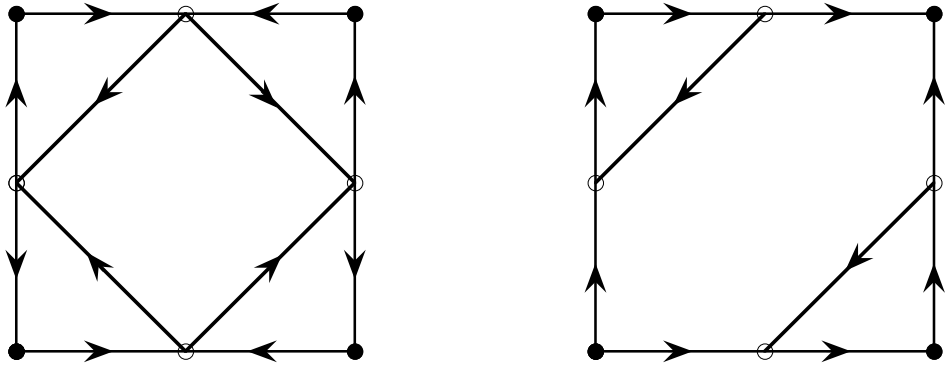}}}
\caption{Orbital currents states proposed by Varma 
(Ref.~\cite{VARMA02}).  Solid dots are Cu ions, open dots O 
ions, and arrows are bond currents.  The right panel has a form factor 
consistent with recent ARPES results \cite{NAT02}.}
\label{fig20}
\end{figure}

The most interesting example of the competitive scenario is that 
of orbital currents.
These involve bond currents either circulating around an elementary 
plaquette of four coppers (middle panel, Fig.~\ref{fig16})
\cite{LEESF,SCHULZ,PLEE,ORB,NAYAK} or within a 
subplaquette involving just the Cu-O bonds (Fig.~\ref{fig20}) \cite{VARMA,VARMA02}.
So far, only two pieces of experimental evidence point to 
such a state.  Inelastic neutron scattering experiments find small 
moment magnetism in underdoped YBCO whose form factor drops off 
more rapidly in momentum space than that associated with Cu spins 
\cite{MOOK}, indicating the presence of a moment extended in real
space.  And recent circularly polarized ARPES experiments reveal 
the presence of time reversal symmetry breaking below $T^*$ 
\cite{NAT02}, whose form factor in momentum space (if interpreted in 
terms of orbital currents) favors the Varma picture \cite{VARMA02}.

One of the most interesting aspects of the competitive scenarios is 
the prediction of a quantum critical point where the pseudogap effect 
disappears.  From experiment, it has been claimed that the $T^*$ line 
passes through the $T_c$ line and vanishes within the superconducting 
dome at a concentration of 19\%, just beyond optimal doping (Fig.~\ref{fig9})
\cite{LORAM}.

\begin{figure}
\centerline{\epsfxsize=0.5\textwidth{\epsfbox{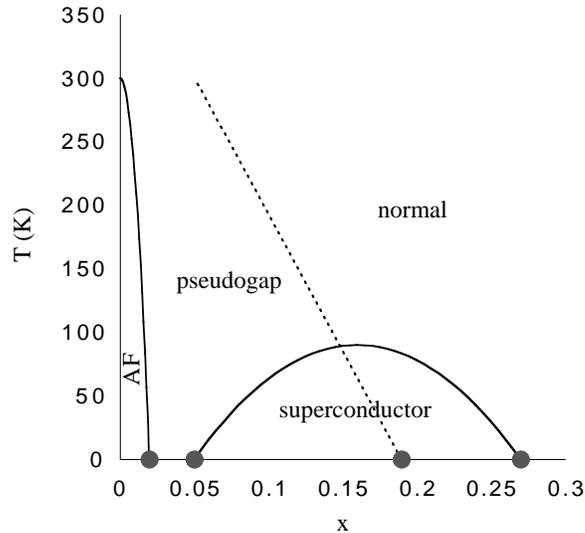}}}
\caption{Four possible quantum critical points (dark circles)
in the cuprate phase 
diagram.  Dotted line is the pseudogap phase line.}
\label{fig21}
\end{figure}

In fact, there are several possible quantum critical points in the 
cuprate phase diagram (Fig.~\ref{fig21}).
Starting from the undoped material, as the 
doping progresses, one first finds the point where the Neel 
temperature vanishes, then the point where superconductivity first 
occurs, then the critical point mentioned above, and finally at higher 
doping the point where superconductivity disappears, for a total of 
four possible quantum critical points.  The last point may correspond 
to where the Fermi surface topology changes from hole-like to 
electron-like (that is, the saddle point in the dispersion at $(\pi,0)$ passes 
through the Fermi energy), as there is some evidence of this from ARPES.

It remains to be seen whether the ``quantum critical paradigm'' with
its emphasis on competing phases is the proper way of thinking about 
cuprates \cite{LAUGH}.  It has certainly led to an enrichment in our 
understanding of these novel materials \cite{SUBIR}.

\section{Photoemission}

\subsection{General Principles}

As emphasized by Anderson, angle resolved 
photoemission has emerged as one of the most important spectroscopic 
probes of cuprate superconductors, in some sense playing the role 
that tunneling spectroscopy played in conventional superconductors 
\cite{ANDERSON}.  Of course, much has 
been discovered about cuprates using tunneling, but Anderson's statement 
was meant to emphasize the fact that photoemission played no role in 
the past, and then all of the sudden stepped up to play a major role 
in the cuprate problem.  For those of us working in the ``old days'', 
these developments have been nothing short of amazing.

To understand how this came about, a few comments are in order about 
photoemission.  This technique has a venerable history; in fact, it 
was for explaining the photoelectric effect, discovered by 
Hertz in 1887, that Einstein got his Nobel 
prize.  Although the concept for doing angle resolved experiments had 
been recognized, the general perception was that not much 
useful would be learned.  This changed in the early 1960s when Spicer 
developed the three-step model for photoemission, showing that in 
principle, important information about the electronic structure could 
be elucidated.  Subsequent experiments by a number of groups, including
Dean Eastman's, were able 
to determine the electronic dispersion of transition metals using 
this technique \cite{HUFNER}.  For these developments, Spicer and 
Eastman received the Buckley Prize in 1980.

The technique involves shining photons on a sample with a specific 
energy.  If the photons have an energy larger than the work function 
of the metal, then electrons will be emitted.  In angle resolved 
mode, an electrostatic detector measures the azimuthal and polar 
angles of the electrons (relative to the surface normal), as well as 
their energy.  Knowing the energy of the photon, then the initial 
energy of the electrons in the crystal can be determined, as well as 
the components of the momentum parallel to the sample surface.  In 
principle, the perpendicular component of the momentum is also 
determined from the energy-momentum relation (the electrons in vacuum 
having an energy which is quadratic in momentum), but there are
subtleties connected with the breaking of the crystal symmetry by 
the surface in this direction.

The actual photocurrent is very complicated, since it is formally a 
three current correlation function (this is the so-called one-step 
model for photoemission).  But in the three-step approximation of 
Spicer, where the initial electron is photoexcited, this 
photoelectron transports through the crystal, then out in the 
vacuum to the detector, the photocurrent (for one band) can be written as
\begin{equation}
I({\bf k},\omega) = c_{\bf k} \int_{\delta {\bf k}} d{\bf k'} \int d\omega'
A({\bf k'},\omega') f(\omega') R(\omega,\omega')
\end{equation}
where $c_{\bf k}$ is the modulus squared of the matrix element of the 
operator ${\bf A}\cdot{\bf p}$ between initial and final states 
(${\bf A}$ is the vector potential, ${\bf p}$ the momentum operator), $A$
the single particle spectral function ($-ImG/\pi$, where $G$ is the 
electron Greens function), $f$ the Fermi-Dirac function, and $R$ the 
energy resolution function (a guassian).  The momentum integration is 
a window ($\delta {\bf k}$)
centered about ${\bf k}$ which represents the finite 
momentum resolution of the spectrometer.  This expression assumes the 
impulse (or sudden) approximation, where the interaction of the 
photoelectron with the photohole is ignored.  Moreover, the 
expression implicitly assumes the 2D limit, which fortunately is 
relevant for the cuprate case, where $k_z$ dispersion effects are 
weak (``bilayer splitting'' is a different matter).

The significance of this expression is obvious.  The single particle 
spectral function is the simplest quantity which emerges from a 
many-body theory of electrons.  We note that $G^{-1}({\bf 
k},\omega) = \omega -\epsilon_{\bf k} -\Sigma({\bf k},\omega)$
where $\epsilon$ is the bare energy and $\Sigma$ the Dyson 
self-energy.  A simple example of this is 
BCS theory, where $\Sigma_{BCS}({\bf k},\omega) = \Delta_{\bf 
k}^2/(\omega+\epsilon_{\bf k})$.  Since the momentum distribution 
function (many-body occupation factor) is given by $n_{\bf k} = \int 
d\omega A({\bf k},\omega) f(\omega)$, then modulo resolution and dipole 
matrix elements, the frequency integral of the ARPES spectrum is 
$n_{\bf k}$ \cite{MOHIT95}.

One problem with photoemission is that it is a surface sensitive 
probe, particularly for the low energy ($\sim$ 20 eV) photons typically used 
to achieve high energy and momentum resolution.  Even in the quasi-2D 
cuprates, this can lead to problems unless a natural cleavage plane 
exists.  This is why most measurements have been done on BSCCO, which 
contains a double BiO spacer layer, with the two BiO layers having 
the biggest interplanar separation in the cuprates (these layers are 
at a separation typical of van der Waals interactions).  This 
provides the best possible cleavage in the cuprates.  The penalty one 
pays is that the BiO layers have planar bonds which are longer than 
the CuO bonds.  The material tries to compensate for this by 
developing a superstructure deformation of the BiO planes.  This can 
lead to ``ghost'' images of the main CuO signal due to diffraction of 
the photoelectrons off the BiO surface layer, which have to be taken 
into account when interpreting ARPES spectra \cite{REVIEW}.

The advantage of ARPES, though, is now obvious.  It is both a 
momentum and frequency resolved probe.  The only other probe 
comparable to this is inelastic neutron scattering, which measures a 
more complicated function (the spin part of the particle-hole response).
Such a 
momentum resolved probe was not essential in classic s-wave
superconductors where momentum dependent effects are not important,
but we know they are essential to consider in the d-wave case.  This 
is an obvious advantange of ARPES over tunneling, though the latter 
still has much better energy resolution, and has the advantage of 
being able to see unoccupied states as well.  On the other hand, 
important spatial information can be obtained by STM, which measures
the local density of states, which is beyond 
the scope of this review.

The true impact of ARPES was recently realized by the development of 
the Scienta detector, which allows collection of data simulataneously 
as a function of momentum and energy.  In angle integrated mode, 
energy resolutions of order 2 meV become possible, allowing even 
conventional superconductors to be studied by photoemission 
\cite{YOKOYA}.  But even though the energy resolution is not as good 
in angle resolved mode (typically 10-20 meV), it is adequate for the
cuprates, and the momentum resolution is 
quite good, of order 0.005 of a reciprocal lattice vector.  This 
precision in momentum allows fine details of the spectral function to 
now be resolved.  It is certainly a far cry from the pre-cuprate era, 
when momentum resolutions were typically 0.1 of a reciprocal lattice 
vector, and energy resolutions were typically 100 meV.

The power of the Scienta detectors can be appreciated by looking at 
the expression for the spectral function
\begin{equation}
A({\bf k},\omega) = \frac{1}{\pi}\frac{Im\Sigma({\bf k},\omega)}
{(\omega-\epsilon_{\bf k} -Re\Sigma({\bf k},\omega))^2+
(Im\Sigma({\bf k},\omega))^2}
\end{equation}
In the past, spectra were typically analyzed at fixed ${\bf k}$ as a 
function of energy (EDC).  This energy lineshape is obviously complicated 
given the 
non-trivial $\omega$ dependence of $\Sigma$.  But at fixed $\omega$
as a function of ${\bf k}$
instead, the momentum lineshape (so-called MDC \cite{VALLA}) is
considerably simpler.  In the normal state near the Fermi energy, we 
can typically linearize $\epsilon_{\bf k}$ in momenta normal to the
Fermi surface.  As long as 
$Re\Sigma$ can also be linearized, then the MDC 
reduces to a Lorentzian, with half width $\Sigma(\omega)/v_{F0}$, 
where $v_{F0}$ is (modulo $dRe\Sigma(k)/dk$) the bare Fermi velocity 
(obtained from 
$\epsilon_{\bf k}$) \cite{ADAM2}.  Given an estimate for the bare 
velocity, then $Re\Sigma$ can be read off from the MDC dispersion, 
and $Im\Sigma$ from the MDC width, though it should be noted that the 
latter is only true if the resolution is accounted for in the analysis.

\subsection{Normal State}

\begin{figure}
\centerline{\epsfxsize=0.8\textwidth{\epsfbox{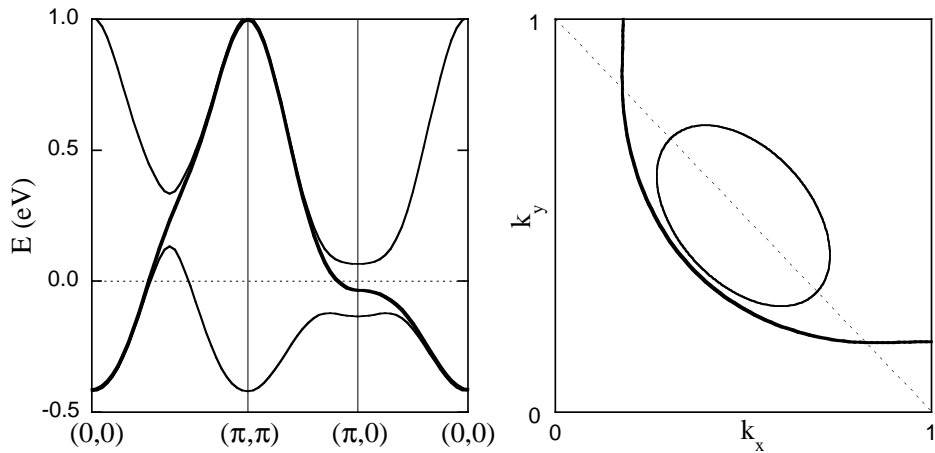}}}
\caption{Left panel is the dispersion in the non magnetic phase (thick
line) and magnetic phase (thin lines), the latter assuming an SDW gap 
of 200 meV.  Resulting Fermi surfaces 
are shown in the right panel, with the dashed line the magnetic zone
boundary.}
\label{fig22}
\end{figure}

The undoped version of the cuprates is a magnetic insulator.  Simple 
considerations lead one to expect that the valence band maximum is at 
the $(\pi/2,\pi/2)$ points of the zone (left panel, Fig.~\ref{fig22}).
This has been beautifully 
confirmed by photoemission measurements \cite{WELLS}.  When hole doping 
such a state, then one might expect to form small hole pockets with 
volume $x$ centered at these points (right panel, Fig.~\ref{fig22}).  This would certainly be
consistent with transport and optics data, which indicate a carrier 
density of $x$.  Somewhat surprisingly, though, this is not what is 
seen by ARPES \cite{JC90}.  Rather, what is seen is a large hole 
surface (right panel, Fig.~\ref{fig22}) with volume $1+x$ centered about the $(\pi,\pi)$ points
(remembering the full Brillouin zone corresponds to 2 filled states
because of spin degeneracy).  This 
is more or less what is predicted by paramagnetic band theory.  To be 
consistent with transport, this would mean that the spectral weight 
would have to scale with $x$.  This was subsequently found to be the 
case from ARPES, first at 
the $(\pi,0)$ points in Bi2212
\cite{JC99,FENG00,DING01}, then most recently along the nodal direction
in LSCO \cite{YOSHIDA}.  In some sense, the Fermi surface disappears 
by losing its spectral weight, much like the Cheshire Cat in Alice in 
Wonderland.

The energy dispersion seen in the doped case involves the presence of 
a saddle point at $(\pi,0)$ on the occupied side which is relatively 
close to the Fermi energy (left panel, Fig.~\ref{fig22}).  
This has led to many theories based on van 
Hove singularities in the density of states.  The dispersion near the 
saddle point is quite flat, especially in the superconducting state, 
which has led to the ``extended'' van Hove singularity concept 
proposed by Abrikosov \cite{ALEX}.  As the hole doping 
increases, the saddle point approaches the Fermi energy.  In the case 
of Bi2201, which can be heavily overdoped, the saddle point appears 
to be almost degenerate with the Fermi energy at a concentration 
where $T_{c}$ is essentially zero \cite{SATOB}.  Beyond this 
concentration, the saddle point is expected to pass through the Fermi 
energy, leading to an electron surface centered at $(0,0)$.  There is
evidence from ARPES that this occurs in LSCO \cite{INO}.  This 
would be consistent with Hall measurements, which see a sign change 
in the Hall number near this concentration \cite{HALL}.

This dispersion, though, should be taken with a very large grain of 
salt.  In particular, no well defined spectral peaks appear in the 
normal state, at least for optimal and underdoped samples.  That is, 
the widths of the peaks are of order their energy separation from the 
Fermi energy.  This is why no van Hove singularity in the 
density of states has been inferred from any experimental measurement.

The momentum and energy dependence of the spectral peaks in the 
normal state was of great interest from the beginning.  The original 
ARPES analysis of Olson \etal \cite{OLSON} found the peak width to scale 
linearly with peak energy.  This is the behavior predicted by the 
marginal Fermi liquid phenomenology \cite{MFL}.
\begin{figure}
\centerline{\epsfxsize=0.5\textwidth{\epsfbox{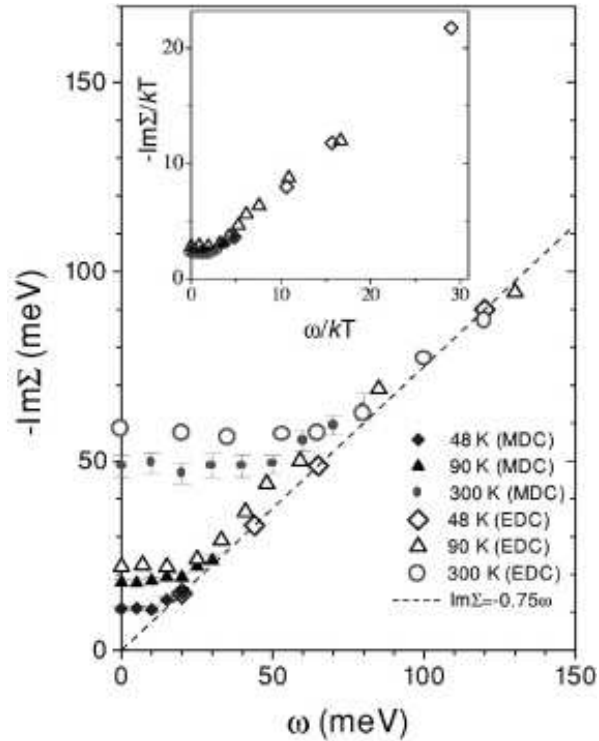}}}
\caption{Imaginary part of the self-energy determined by ARPES along 
the nodal direction in Bi2212, demonstrating marginal behavior (quantum
critical scaling) as a function of 
both $\omega$ and $T$.  From Ref.~\cite{VALLA}.}
\label{fig23}
\end{figure}
This behavior has 
been confirmed to much better precision with high momentum resolution 
data along the nodal direction (Fig.~\ref{fig23}) \cite{VALLA}, which also indicate a 
linear $T$ dependence of the linewidth as well.

The momentum anisotropy of the linewidth, though, is still a matter of 
controversy.  Data for optimal doping indicate that the linewidths get 
broader as the $(\pi,0)$ point is approached.  This is as opposed to 
heavily overdoped Bi2201, where the lineshape is relatively isotropic 
around the Fermi surface.  Such momentum anisotropies are not 
unexpected, given that the d-wave nature of the superconducting order 
parameter implies momentum dependent interactions.
On the other hand, some authors have suggested that the 
intrinsic lineshape might be fairly isotropic, with the observed 
anisotropy in Bi2212 a combination of overlap of features due to the ``ghost'' 
images associated with the superstructure, along with bilayer 
splitting of the energy bands \cite{BOGDANOV}.

\begin{figure}
\centerline{\epsfxsize=0.5\textwidth{\epsfbox{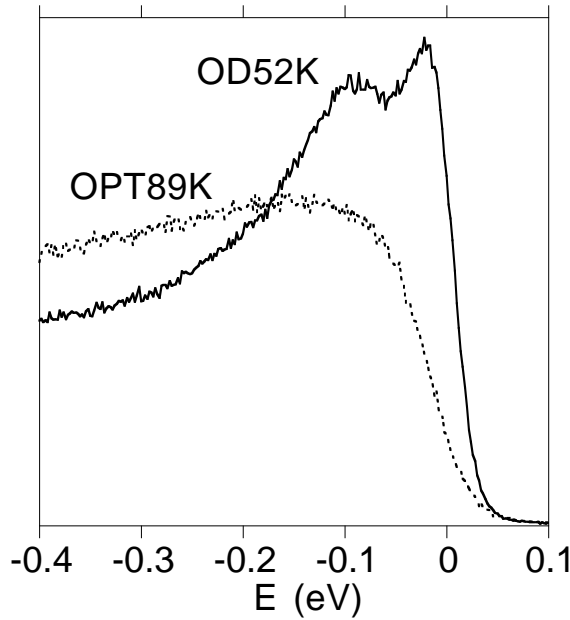}}}
\caption{ARPES spectrum at $(\pi,0)$ for optimal doped and 
overdoped Bi2212 samples in the normal state.  Separate peaks in the 
overdoped case are due to bilayer splitting.  Data from Ref.~\cite{JCNEW}.}
\label{fig24}
\end{figure}

The issue of bilayer splitting has been somewhat controversial, and 
thus deserves some attention.  Most experiments have been done in 
Bi2212, which has two CuO layers in each formula unit, the two layers
being separated by Ca 
ions.  Mixing of the levels on the two planes will lead to an
antibonding and bonding combination, thus bilayer splitting (Fig.~\ref{fig24}).  Band 
theory predicts that the splitting should be sizable (of order 1/4 
eV), with the splitting varying with planar momentum like 
$(\cos k_{x}-\cos k_{y})^{2}$.  Therefore, the effect is largest at 
$(\pi,0)$ \cite{INTLAY}.  As spectral peaks are broad in the normal state, 
then it is quite possible that the two overlapping features would smear 
together into a single feature, making the $(\pi,0)$ 
lineshape look anomalously broad.

Most early reports of bilayer 
splitting were later attributed to the ``ghost'' images associated 
with the superstructure.  The claim was that once this was factored 
out, no bilayer splitting was apparent in the data, at least for an
optimal doped sample \cite{DING96}.  
This was of interest, since several theories of cupratres predicted 
incoherent behavior along the c-axis in the normal state 
\cite{ANDERSON}.

Subsequent measurements at high momentum resolution, though, revealed 
the presence of bilayer splitting in heavily overdoped samples 
\cite{BILAYER}.  The $(\pi,0)$ spectrum is characterized by a 
relatively sharp antibonding peak near the Fermi energy (consistent 
with the more Fermi liquid like character of heavily overdoped 
samples), plus a broader (bonding) peak at higher binding energy  
(its width being larger due to its greater binding energy).
This leads to a peak-shoulder type spectrum, as opposed to the single 
broad spectrum seen for optimal and underdoped samples (Fig.~\ref{fig24}).  The doping
dependence of the bilayer splitting, though, is still a controversial issue.
Recently,
Campuzano's group has presented evidence that bilayer effects
disappear in the ``strange metal" (quantum critical) phase of 
Fig.~\ref{fig6} \cite{JCNEW}.

\subsection{Superconducting State}

There are two remarkable features revealed by ARPES in the superconducting 
state.  First, the opening of an anisotropic superconducting gap in the 
spectrum.  Second, the appearance of a sharp coherent peak below $T_{c}$.

The first observation of a gap was made in 1989 by Yves 
Baer's group \cite{BAER}, and was considered a tour-de-force at the 
time.  Later, angle resolved, measurements did not detect any gap 
anisotropy, probably due to the sample qualities at the time.  This 
all changed in 1993 with the observation by Z-X Shen's group of an 
anisotropic gap consistent with d-wave symmetry \cite{SHEN93}.

\begin{figure}
\centerline{\epsfxsize=0.8\textwidth{\epsfbox{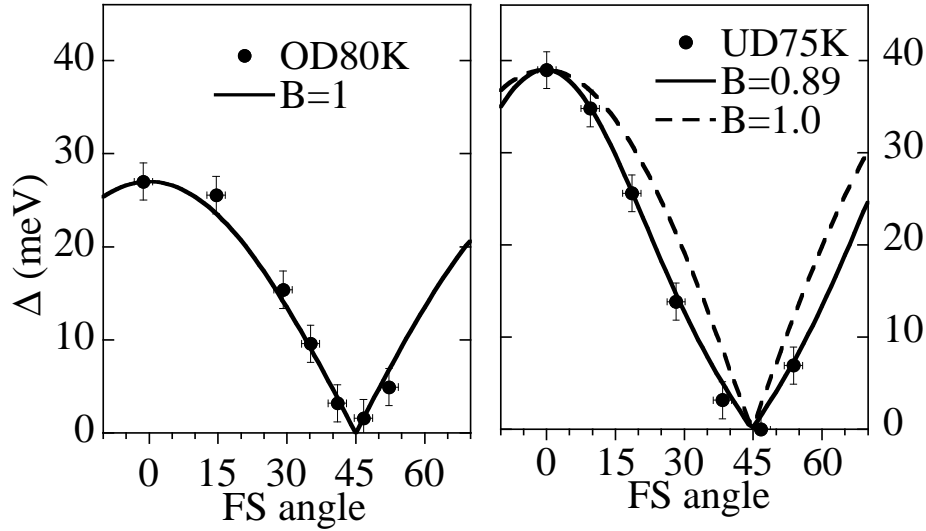}}}
\caption{Doping dependence of the spectral gap in Bi2212 from ARPES.  
Left panel for an overdoped sample, right panel for an underdoped 
one.  B=1 is a $\cos(2\phi)$ dependence of the gap on Fermi surface 
angle, solid line in the right panel includes a contribution from the 
next d-wave 
harmonic, $\cos(6\phi)$.  From Ref.~\cite{MESOT99}.}
\label{fig25}
\end{figure}

Not all groups found this behavior, though.  It was realized 
later that the discrepancies were due to complications caused 
by the ``ghost'' images associated with the Bi2212 superstructure.  Once 
this was appreciated, then it was evident that the best results could be 
obtained in the $Y$ quadrant of the Brillouin zone where these images 
were well separated from the main image.  By fitting the leading edge 
of the spectrum (taking into account the known resolution), precise 
values of the energy gap can be obtained.  The resultant plot of the
gap along the Fermi surface shows a 
clear V-shaped behavior around the node, as expected for d-wave 
symmetry (left panel, Fig.~\ref{fig25}) \cite{GAP96}.  Moreover, the functional dependence on 
momentum is precisely of the form $|\cos k_{x} - \cos k_{y}|$ 
\cite{SHEN93,GAP96} as would be expected for pairs of electrons 
sitting on near neighbor Cu sites.  In fact, along the observed Fermi
surface, the functional dependence is essentially $\cos(2\phi)$, where
$\phi$ is the angle of the line connecting $(\pi,\pi)$ to the 
Fermi surface with the line $(\pi,0)-(\pi,\pi)$.  Subsequent measurements have 
revealed a deviation from this form with increasing underdoping (right panel,
Fig.~\ref{fig25}) \cite{MESOT99}, which can be fit by inclusion of the next 
harmonic in the gap expansion, $\cos(6\phi)$, which is related to
$\cos 2k_x - \cos 2k_y$.  This indicates that the 
pair interaction is becoming longer range in real space as the doping is 
reduced.  The trend is expected, given that the correlation length 
associated with magnetic fluctuations increases with underdoping.  
Recently, the same deviation from the $\cos(2\phi)$ form has been 
inferred from Fourier transformation of STM data \cite{MCELROY}.

\begin{figure}
\centerline{\epsfxsize=0.8\textwidth{\epsfbox{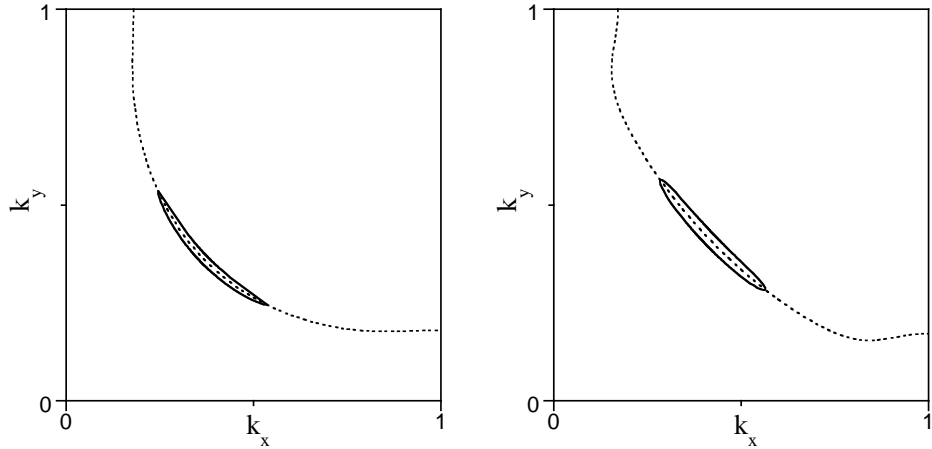}}}
\caption{Left: Fermi surface (dashed line) and a constant energy contour for
quasiparticle excitations in the d-wave superconducting state (solid 
line) based on ARPES data for Bi2212.  Right: Modified version proposed to
explain the incommensurability seen in neutron scattering data in YBCO.
From Ref.~\cite{NORM00}.}
\label{fig26}
\end{figure}

One of the most interesting aspects of the superconducting state is 
that the low energy states have a Dirac-like dispersion, i.e., Dirac cones,
whose constant energy contours are centered 
about the nodes (actually, these contours are banana shaped due to the
Fermi surface curvature, Fig.~\ref{fig26}).
According to ARPES, these 
cones are quite anisotropic, with the ratio of the velocity normal to 
the Fermi surface to that along the Fermi surface (the latter 
the slope of the gap around the node) of 20 
\cite{MESOT99}.  This value has also been inferred from thermal 
conductivity measurements \cite{CHIAO}.  In principle, by comparison 
of this ratio to the value of the linear T coefficient of the
penetration depth, important information can be obtained about the 
electromagnetic coupling of the quasiparticles.  Present results are 
consistent with a linear doping variation of the particular Landau 
interaction parameter involved \cite{MESOT99}, but the associated 
error bars are quite large.  More precise ARPES and 
penetration depth measurements on Bi2212 would be gratifying in this 
regard.  In particular, the doping variation of the gap slope around 
the node has not been studied yet with high resolution detectors.  It 
is of some interest to see whether this quantity scales with $T_c$ on 
the underdoped side of the phase diagram.  Recent thermal conductivity
data indicate that this is not the case \cite{SUTHER}.

\begin{figure}
\centerline{\epsfxsize=0.5\textwidth{\epsfbox{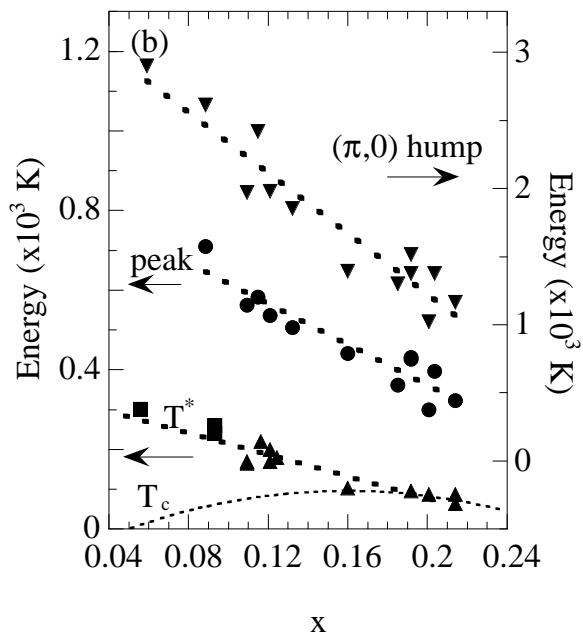}}}
\caption{Spectral peak energy (maximum superconducting energy gap), 
energy of the hump, and pseudogap temperature, $T^{*}$, versus doping, 
x, from ARPES data on Bi2212.  From Ref.~\cite{JC99}.}
\label{fig27}
\end{figure}

What is known, though, is that the maximum superconducting energy gap 
does not scale with $T_c$ on the underdoped side (Fig.~\ref{fig27}) \cite{HARRIS,JC99}.
Instead, this quantity monotonically increases with underdoping, 
scaling with the pseudogap temperature, $T^{*}$.  The same trend has 
been seen by tunneling \cite{JOHNZ}.  The observed behavior would 
be consistent with $T^{*}$ representing some mean field transition 
temperature for pairing.  This doping trend was actually predicted 
many years ago by RVB theory \cite{GROS}, where the spin pairing 
energy scale is decoupled from the phase stiffness energy
associated with the doped holes, since the latter is proportional to $x$
(see Fig.~\ref{fig16}).

\begin{figure}
\centerline{\epsfxsize=0.5\textwidth{\epsfbox{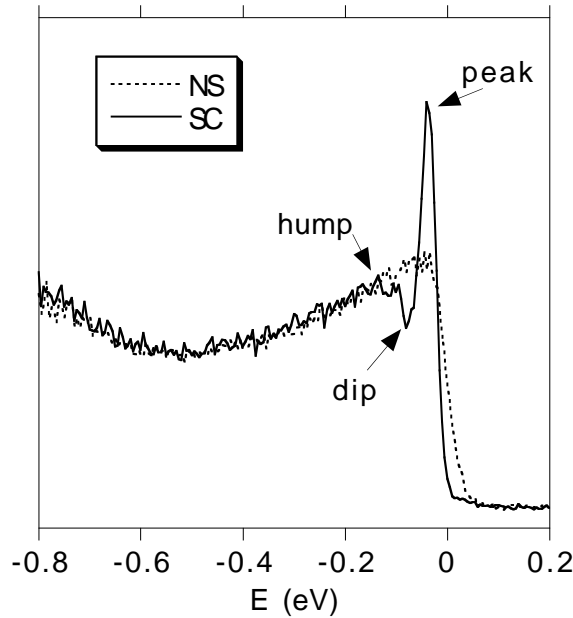}}}
\caption{ARPES spectra at $(\pi,0)$ for an overdoped (87K) Bi2212 sample in 
the normal state (NS) and superconducting state (SC).  Data from 
Ref.~\cite{NORM97}.}
\label{fig28}
\end{figure}

The lineshape changes between normal and superconducting states, though, 
are perhaps the most fascinating aspect of the ARPES data (Fig.~\ref{fig28}).  The 
changes are most spectacular near the $(\pi,0)$ point for optimal and 
underdoped samples.
As the temperature is lowered, the broad peak in 
the normal state develops into a sharp coherent peak separated by a 
spectral dip (near $(\pi,0)$) or a spectral break (near the 
$(\pi,0)-(\pi\pi)$ Fermi surface crossing) from the higher energy 
incoherent part \cite{MOHIT95}.  This behavior is consistent with a 
strong increase in the lifetime of the electrons
as the temperature is lowered below $T_{c}$, as has 
been earlier inferred from microwave conductivity \cite{BONN} and 
thermal conductivity \cite{ONG} experiments.  That is, a gap is being 
opened in the scattering rate, as also derived from infrared 
conductivity measurements \cite{PUCH}.  In ARPES, this can be seen 
very clearly by ``inverting'' the data to directly extract the 
temperature dependence of the electron self-energy \cite{TEMP01}.

An alternate interpretation has been given to the data, however 
\cite{FEDEROV,FENG00,DING01}.  In this picture, a gap 
develops in the incoherent part of the spectrum, with a quasiparticle 
pole appearing inside the gap.  The pole weight monotonically 
increases with decreasing temperature, and it has been suggested that 
this behavior tracks the superfluid density \cite{FENG00}.  In some 
sense, this would imply that the quasiparticle weight was equal to 
the superconducting order parameter.  One particular model which 
is suggestive of this is the Josephson coupling of stripes below 
$T_{c}$ \cite{EMKIV}.

The remarkable spectral changes near $(\pi,0)$ leading to the unusual 
peak-dip-hump lineshape below $T_{c}$ were actually first 
observed 
by tunneling \cite{HUANG}.  When they were subsequently observed by
ARPES, the obvious explanation was 
that they were due to bilayer splitting (the ``hump'' representing the 
bonding band, the ``peak'' the antibonding band).  There are a number 
of arguments against this (including the fact that tunneling 
spectroscopy sees this lineshape for single layer materials like 
Tl2201).  What is clear, though, is that bilayer splitting alone
is not sufficient to explain the lineshape.  In particular, 
the spectral dip represents a depletion of states which fills in as 
the temperature is raised \cite{TEMP01}.  Moreover, the dip energy 
scale appears to exist at the same energy throughout the Brillouin 
zone \cite{ADAM2}.

These considerations have led to many speculations that the spectral dip 
represents some sort of many body effect.  One of the first 
treatments of this problem was by Arnold \etal \cite{ARNOLD}, where 
they applied the McMillan-Rowell ``inversion" procedure \cite{MCMR} to
the data to determine the boson spectral function from the frequency 
dependence of the gap function, $\Delta(\omega)$.  From this analysis, a 
sharp bosonic mode was inferred at about 10 meV.
The problem with this pioneering analysis was that it assumed the data 
represent an isotropic 
density of states proportional to $\frac{\omega}{\sqrt{\omega^2-\Delta^2(\omega)}}$,
with the spectral dip corresponding to a strong frequency variation
of $\Delta(\omega)$.
In this case, the ``normal'' part of 
the self-energy (diagonal in particle-hole space)
drops out, and so all structure in the data can 
be associated with the pairing self-energy (off-diagonal part).
This is not the case if 
the data represent a spectral function.

The data were later analyzed assuming the primary effects 
were due to the normal self-energy (the pairing part being 
treated in a BCS approximation) \cite{NDING}.  In this 
analysis, the spectral dip can be understood as a sharp threshold for 
inelastic scattering.  To understand this, consider the Feynman diagram 
for an electron scattering off particle-hole excitations (left panel, 
Fig.~\ref{fig5}).
\begin{figure}
\centerline{\epsfxsize=1.0\textwidth{\epsfbox{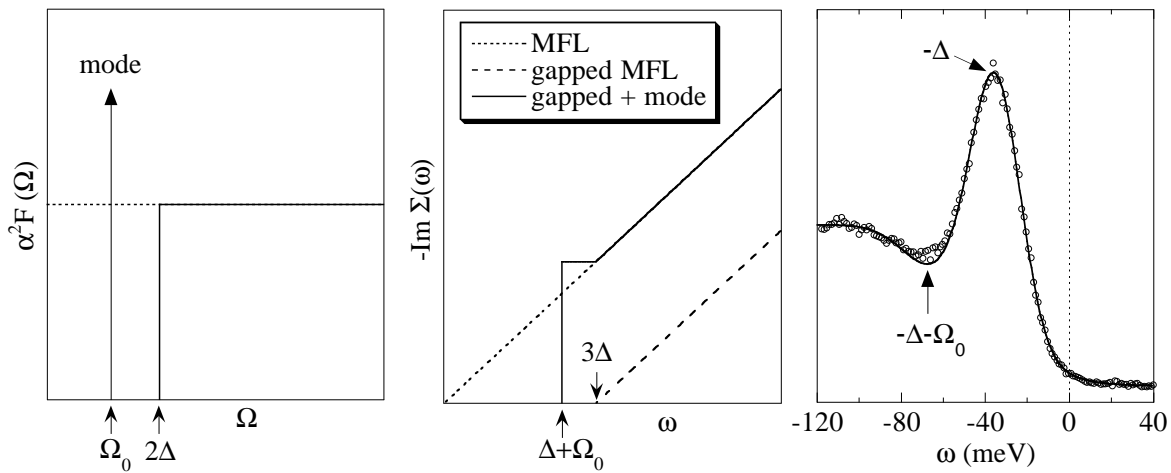}}}
\caption{Left: boson spectrum (``$\alpha^2F$") in the normal state (dashed line, 
corresponding to a marginal Fermi liquid) and superconducting state (solid line, 
corresponding to a gapped marginal Fermi liquid).  The arrow marks the 
possibility of a collective mode with energy $\Omega_0$ inside the 
continuum gap of $2\Delta$ (corresponding to 
a gapped marginal Fermi liquid plus a mode).  Middle: 
Resulting imaginary self-energy for the electrons.
Right:  Superconducting state spectral function from this model
as compared to ARPES data at $(\pi,0)$ for slightly overdoped Bi2212.
The mode energy, $\Omega_0$, equals the spin resonance energy
determined independently from neutron scattering.
Adapted from Ref.~\cite{NDING}.}
\label{fig29}
\end{figure}
In the superconducting state, the particle-hole continuum will have a 
gap of order $2\Delta$ (left panel, Fig.~\ref{fig29}).  If interactions are strong enough that a
bound state (energy $\Omega_{0}$) emerges inside of this continuum gap, 
then $Im\Sigma$ will develop a sharp threshold, as implied by the 
data, at an energy $\Delta+\Omega_{0}$ \cite{NORM97} (middle panel, Fig.~\ref{fig29}).  
In this picture, the energy of the bosonic mode 
will be equal to the energy difference of the dip ($\Delta + 
\Omega_{0}$) and the peak ($\Delta$) \cite{NDING,AC2} (right panel, Fig.~\ref{fig29}).  
Moreover, the resulting spectral function 
will consist of two features:  a broad feature at higher binding
energy whose dispersion roughly tracks the dispersion of the 
single feature in the normal state (hump), and a sharp, weakly dispersive feature 
for smaller
binding energies representing the renormalized quasiparticle branch (peak).  
These two features are separated by the dip energy, which is 
roughly constant in momentum.  This model gives a natural explanation of the 
unusual dispersions associated with the peak and hump \cite{NORM97}.

\begin{figure}
\centerline{\epsfxsize=0.5\textwidth{\epsfbox{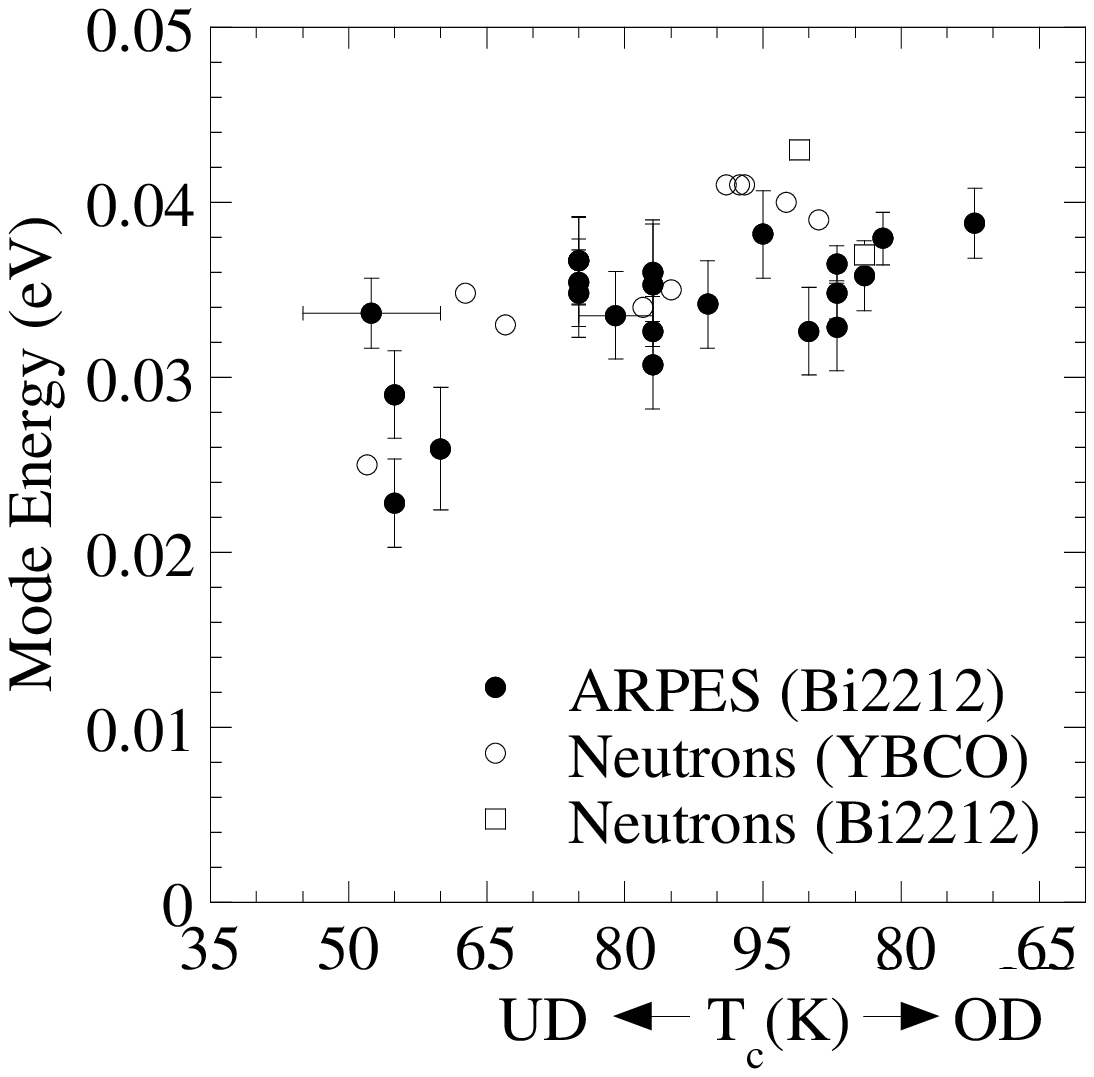}}
\epsfxsize=0.5\textwidth{\epsfbox{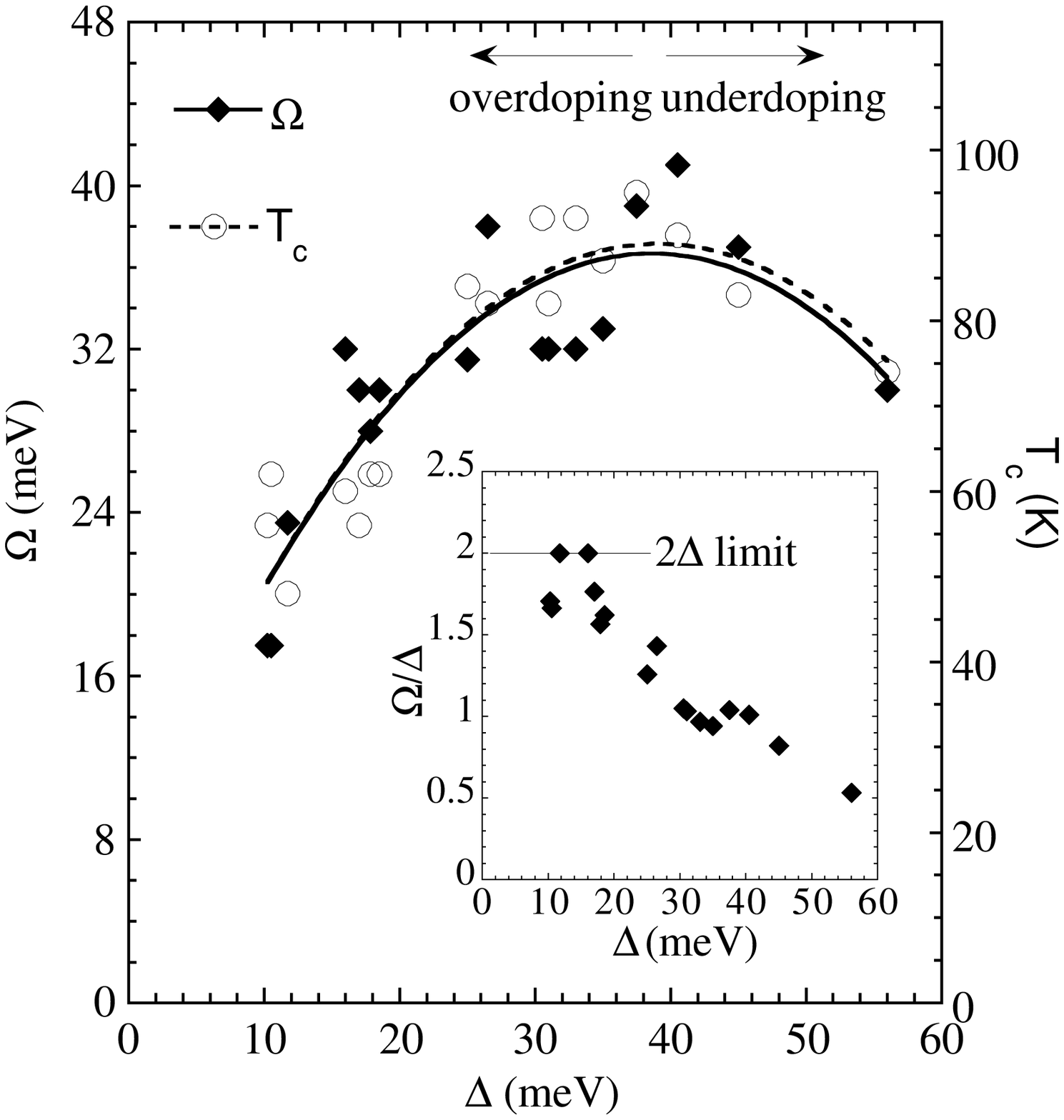}}}
\caption{Left panel:  mode energy inferred from ARPES as compared to 
that determined directly from neutron data, showing the two data sets agree.
From Ref.~\cite{JC99}.  Right panel: mode energy inferred from tunneling data, 
showing scaling with $T_{c}$.  Inset demonstrates that the mode energy 
saturates to $2\Delta$ in the overdoped limit.  From Ref.~\cite{JOHNZ}.}
\label{fig30}
\end{figure}

Moreover, the peak-dip-hump is strongest at the $(\pi,0)$ points.  As 
these points are connected by $(\pi,\pi)$ wavevectors, this would 
imply that the bosonic excitations involved are associated with this 
wavevector \cite{SS}.  This, coupled with the inferred mode energy (40 
meV), points to the spin resonance as the boson \cite{NORM97}.
At the time of this conjecture, the resonance had only been seen in 
YBCO, but later experiments found it in Bi2212 at the energy inferred 
from ARPES \cite{KEIMER}.  Despite this, a criticism offered 
against such an interpretation was that all energy scales from ARPES 
and tunneling appear to increase with underdoping 
(Fig.~\ref{fig27}), but the resonance energy decreases 
\cite{TIMRPP}.  This was answered later by a doping dependent ARPES 
study, which found that
the mode energy inferred from the data had a doping 
dependence which indeed tracks
the resonance energy (left panel, Fig.~\ref{fig30}) \cite{JC99}.
This was confirmed in greater 
detail by tunneling \cite{JOHNZ}, where the energies were tracked over 
a larger doping range with higher energy resolution (right panel, Fig.~\ref{fig30}).  
Not only were 
the inferred mode energy the same as the resonance energy, but the 
mode energy saturated to $2\Delta$ in the overdoped limit as would be 
expected for a collective mode inside a continuum gap (inset, right panel,
Fig.~\ref{fig30}) \cite{JOHNZ}.

There are other things revealed by the doping dependence of the ARPES 
data as well.  
Both the peak energy and the hump energy increase strongly with 
underdoping (Fig.~\ref{fig27}), yet the ratio of their energies is 
roughly constant (3.5-4)
as would be expected for a strong-coupling superconductor 
\cite{JC99} (a result difficult to explain if their energy 
separation were simply due to bilayer splitting).  Moreover, the hump 
dispersion (Fig.~\ref{fig36})
increasingly begins to resemble that (Fig.~\ref{fig22})
expected of a spin 
density wave insulator as the doping is reduced \cite{JC99}.  That is, 
as far as the hump dispersion is concerned, the wavevector $(\pi,\pi)$ 
begins to look more and more like a reciprocal lattice vector.  This 
is not unexpected, since as the resonance mode energy goes soft with
underdoping, the 
material will be unstable to long range order at this wavevector.

Similar effects to the ARPES ones have been inferred from a generalized 
Drude analysis of optics data, where the gap in the optical 
scattering rate has been interpreted in a similar fashion 
\cite{JULES}.  On the other hand, the optical scattering rate 
resembles most closely the behavior of the ARPES self-energy at the 
node, rather than at the $(\pi,0)$ point \cite{ADAM} (not surprising,
since as $(\pi,0)$ is 
a saddle point, the velocity there is zero, and so it does not 
contribute to the in-plane optical response).

As mentioned previously, the normal state scattering rate at the node 
from ARPES resembles that expected for a marginal Fermi liquid 
\cite{OLSON,VALLA}.  What has been controversial, though, is how this 
scattering rate changes below $T_{c}$.
\begin{figure}
\centerline{\epsfxsize=1.0\textwidth{\epsfbox{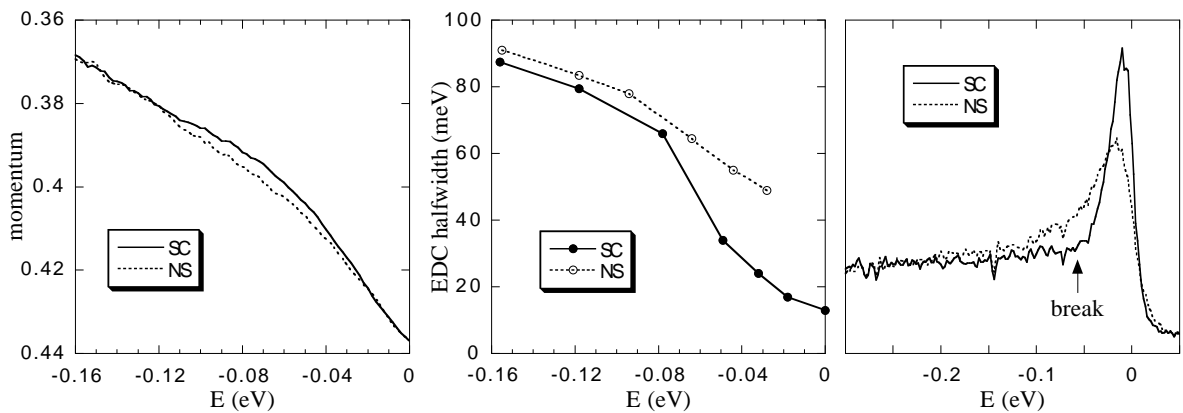}}}
\caption{Left:  Dispersion obtained from ARPES MDCs along the nodal 
direction for optimal doped (90K) Bi2212.  
The difference of the superconducting state (SC) as compared to the
normal state (NS) gives rise to a kink in the dispersion.  Middle:
The change in the EDC linewidth, with the drop in the 
scattering rate in the superconducting state connected to the 
dispersion kink by Kramers-Kronig relations.  Right:
EDC at the nodal point.  Note the break in the superconducting
case marking the
separation of the coherent peak from the incoherent part.
Adapted from Ref.~\cite{ADAM2}.}
\label{fig31}
\end{figure}
The latest results are consistent 
with a strong drop in the scattering rate below some threshold energy (middle
panel, Fig.~\ref{fig31})
\cite{ADAM}, though the expected superlinear behavior this implies 
with temperature has yet to be positively identified \cite{VALLA}.  By 
Kramers-Kronig, this drop implies a ``kink'' in the dispersion (left panel,
Fig.~\ref{fig31}).  For 
binding energies smaller than the ``kink'', the spectral peak is 
sharper and less dispersive, for larger energies, broader and more 
dispersive.  Surprisingly, the kink was not recognized at first
(it was later identified by Shen's group 
\cite{KINK}).  This kink is present throughout the zone 
\cite{KINK}, and occurs at the same energy as the  ``break'' in the ARPES 
lineshape at the Fermi surface separating the quasiparticle peak from 
the incoherent part (right panel, Fig.~\ref{fig31}) \cite{ADAM}.  
Moreover, it was later shown that this spectral 
break evolves into the spectral dip as the momentum is swept in the 
zone from the node to the $(\pi,0)$ point \cite{ADAM2}.  This led to 
the speculation that the ``kink'' effect was due to the resonance as
suggested earlier for  the spectral dip \cite{NORM97}.  This
was later confirmed by theoretical simulations \cite{ESCHRIG}.  
Strong support for this conjecture was offered by data from 
Johnson's group \cite{PETER01}, where the energy scale associated with 
the kink was found to track in doping with the neutron resonance energy.
\begin{figure}
\centerline{\epsfxsize=0.5\textwidth{\epsfbox{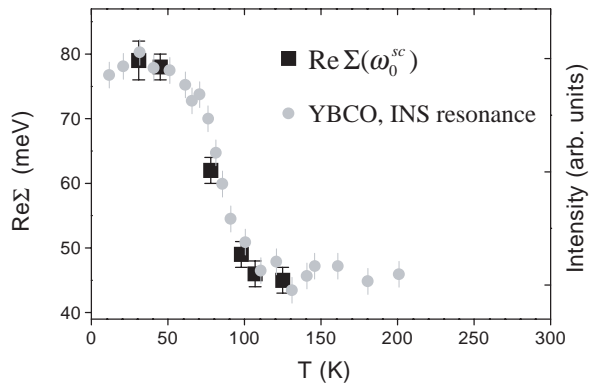}}}
\caption{Variation of the real part of the self-energy at the kink 
energy with temperature as determined from ARPES for Bi2212, compared to the 
intensity of the magnetic resonance for YBCO.  From Ref.~\cite{PETER01}.}
\label{fig32}
\end{figure}
More 
importantly, the change in the self-energy with temperature
associated with the kink has the same temperature dependence as
the resonance intensity (Fig.~\ref{fig32}) \cite{PETER01}.

This picture has been challenged by Shen's group \cite{LANZARA}.  
They observed that not only did the kink effect persist above $T_c$, 
it was universally present in all cuprates (Bi2212, Bi2201, LSCO) at 
roughly the same energy.  They argued that this implied the effect was 
due to a phonon, since the dynamic spin susceptibility of Bi2212 and 
LSCO look very different.  There is some attractiveness to this 
phonon picture, but one should recognize that (1) most of the 
``normal'' state data were actually taken in the pseudogap phase and 
(2) the constancy of the energy scale is somewhat surprising in a 
phonon model as well, since the kink energy, even at the node, should be the sum 
of the maximum superconducting gap energy plus the mode energy (phonon 
or otherwise).
Also, it is somewhat surprising that only a single 
phonon energy would appear in the data.  Still, the arguments 
being invoked in the Lanzara \etal paper \cite{LANZARA} are quite important, 
in that they address the fundamental issue of whether the 
many-body effects in the cuprates should be associated with phonons 
(as in classic superconductors) or with electron-electron 
interactions (as has been commonly assumed in the literature).

\subsection{Pseudogap Phase}
  
ARPES has revealed many unique features connected with the pseudogap phase, and 
has had a profound influence on our understanding of this unusual 
state of matter.

\begin{figure}
\centerline{\epsfxsize=0.5\textwidth{\epsfbox{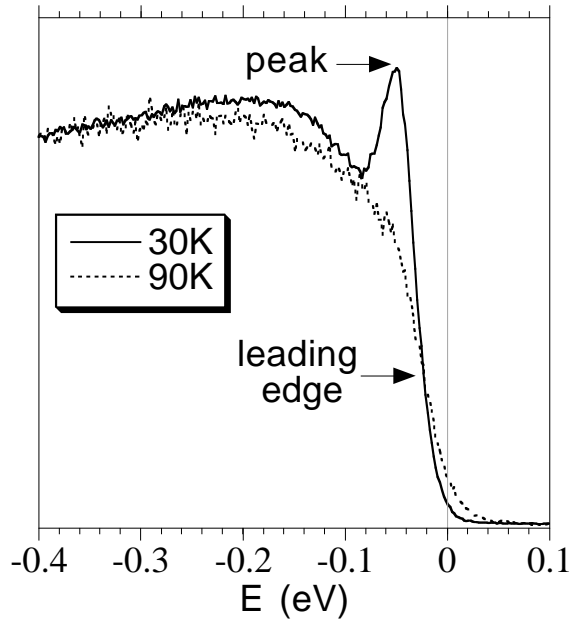}}}
\caption{ARPES spectrum at $(\pi,0)$ for an underdoped Bi2212 sample 
in the superconducting state (30K) and the pseudogap phase (90K).  The 
sharp peak in the superconducting state is replaced by a leading edge gap 
in the pseudogap phase.  
Data courtesy of A. Kaminski and J. C. Campuzano.}
\label{fig33}
\end{figure}

We start our discussion by considering states near the $(\pi,0)$ point 
of the zone for underdoped samples.  Upon heating above $T_{c}$, the 
sharp spectral peak disappears, but the leading edge of the spectrum
is still pulled
back from the chemical potential (leading edge gap, see Fig.~\ref{fig33})
\cite{LOESER,DING}.
This is quite unusual, in that the spectral function is completely
incoherent in nature, but the leading edge is still quite sharp.  As 
the temperature is raised, the leading edge gap appears to go away, 
with the leading edge becoming degenerate with the Fermi function at 
a temperature $T^{*}$ \cite{DING}, similar to $T^{*}$ inferred from NMR 
measurements of the spin gap.

One of the more surprising findings, though, was that this leading 
edge gap has an anisotropy in momentum space quite similar to the d-wave 
gap in the 
superconductor (Fig.~\ref{fig8}) \cite{LOESER,DING}.  This has been taken as strong support 
for those theories proposing that the pseuodgap involves pairs of 
some kind.
\begin{figure}
\centerline{\epsfxsize=0.8\textwidth{\epsfbox{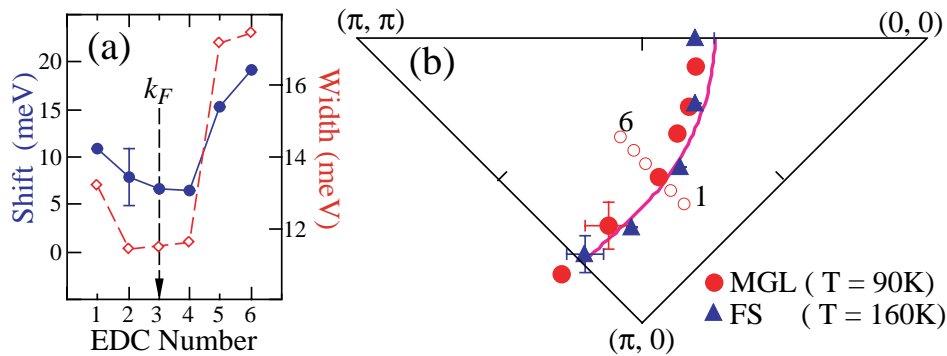}}}
\caption{Left panel shows the determination of the minimum gap locus 
along a cut (open circles of right panel)
in momentum space from the leading edge shift of ARPES 
EDCs.  The right panel shows that the minimum gap locus (MGL) of the 
pseudogap phase matches the normal state Fermi surface (FS).  From 
Ref.~\cite{DING97}.}
\label{fig34}
\end{figure}
In further support of this picture, it was observed that 
the minimum gap locus coincided with the normal state
Fermi surface (Fig.~\ref{fig34}) \cite{DING97}, 
as occurs for superconductors.  This set of points is obtained by 
taking various cuts in momentum space, and looking for that point along 
the cut where the leading edge gap is smallest.  This behavior can be 
contrasted with a spin density wave precursor, for instance, where 
the minimum gap locus would have a new symmetry defined by the 
magnetic Brillouin zone boundary running from $(\pi,0)$ to $(0,\pi)$ 
(see Fig.~\ref{fig22}).

There are, though, a number of unusual features of the data
which are not as easily
understood in terms of a precursor pairing scenario.  In 
particular, the pseudogap phase does not have a node like for a d-wave 
superconductor, instead, it possesses a ``Fermi arc'' 
(Fig.~\ref{fig8}) centered at the 
d-wave node \cite{MARSHALL}.  This arc expands in temperature, 
eventually recovering the full Fermi surface at $T^{*}$ \cite{NAT98}.

For states near the $(\pi,0)$ point, the pseuodgap appears to fill in 
with temperature rather than close \cite{NAT98}, much like what is observed 
in c-axis 
conductivity (Fig.~\ref{fig7}) \cite{HOMES}.
\begin{figure}
\centerline{\epsfxsize=0.5\textwidth{\epsfbox{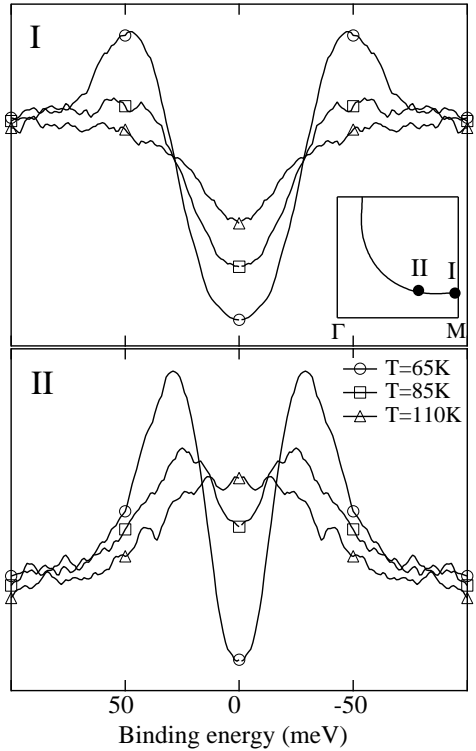}}}
\caption{Filling in of the spectral gap at the antinode 
(I, $(\pi,0)-(\pi,\pi)$ Fermi crossing) as compared to the closing of the 
spectral gap about halfway between the antinode and the node (II).  
Symmetrized ARPES data for underdoped (75K) Bi2212 from Ref.~\cite{NAT98}.}
\label{fig35}
\end{figure}
On the other hand, away from this region, 
the spectral gap clearly closes \cite{NAT98}.  This is most easily 
visualized (Fig.~\ref{fig35}) by ``symmetrizing'' the data (a way of removing the Fermi 
function from the data by assuming particle-hole symmetry in the spectral 
function).  This behavior has been further confirmed by data fitting 
\cite{PHEN98}, where the spectral gap parameter, $\Delta$, at the antinode 
($(\pi,0)-(\pi,\pi)$ Fermi crossing) is found to be relatively temperature 
independent.  The self-energy has the form 
$\Sigma=-i\Gamma_{1}+\Delta^{2}/(\omega+i\Gamma_{0})$.  $\Gamma_{1}$ is a 
crude approximation for the normal self-energy (this term becomes
strongly reduced in the superconducting state with the onset of 
coherence), whereas $\Gamma_{0}$ is the lifetime of the pair propagator, which 
is proportional to $T-T_{c}$.
Note the contrast with
Eliashberg theory, where $\Gamma_0$ would be equal to $\Gamma_1$.
An incoherent spectrum with a pseudogap is
formed when $\Gamma_0 \ll \Delta \ll \Gamma_1$.  The strong
temperature variation of $\Gamma_0$ leads to a filling in of the pseuodgap
($\Delta$ being roughly constant in $T$).
$T^{*}$ is then simply 
the temperature where $\Delta = \Gamma_{0}(T)$.
This behavior can be contrasted with that away from the $(\pi,0)$ 
region, where $\Delta$ closes with temperature in a BCS 
like fashion \cite{PHEN98}.

These findings seem to imply the possibility of two regions in the 
Brillouin
zone, a ``pseudogap'' region centered at the $(\pi,0)$ point and an 
``arc'' region centered at the d-wave node.  This picture would be in 
support of a competitive scenario, where the pseudogap and 
superconducting gap were different phenomena.  On the other hand, 
newer high resoultion data do not necessarily support the picture of two 
regions of the zone, rather it appears that the gap ``closing'' and 
gap ``filling in'' behaviors smoothly evolve into one another as a 
function of momentum.  In fact, there are several pair precursor 
calculations \cite{JAN,GIAN} which predict the presence of Fermi 
arcs.  In the strong-coupling RVB approaches, these Fermi arcs are 
also found, and are
due to fluctuations in the pseudogap regime between d-wave pairs and 
the staggered flux phase state, which are nearly degenerate in energy
\cite{PLEE}.  Arcs are also found in a one loop renormalization group 
treatment of interacting fermions in 2D \cite{FRS}, and in a high 
temperature expansion study of the 2D t-J model \cite{PUTIKKA}.

The resemblance of the arcs to one side of a hole pocket 
(Fig.~\ref{fig22}) has been 
noted by a number of authors \cite{MARSHALL}, and as such hole pockets 
are expected when doping a magnetic insulator, then a magnetic 
precursor scenario is a possibility.  On the other hand, there 
is no clear evidence from ARPES that the arc deviates from the large
Fermi surface and ``turns in" so that its normal would be parallel to the magnetic
zone boundary at the magnetic zone crossing as would be expected in such a scenario
(see Fig.~\ref{fig22}).  And
although ``shadow'' bands have been seen in ARPES \cite{AEBI} (the 
image of the main band translated by $Q=(\pi,\pi)$, which would thus form the 
back side of this pocket), their intensity seems to scale with 
$T_{c}$ \cite{DRESDEN}, and thus drops off as the 
doping is reduced, in complete contrast with the expected behavior if 
the shadows were due to magnetic correlations.

\begin{figure}
\centerline{\epsfxsize=0.4\textwidth{\epsfbox{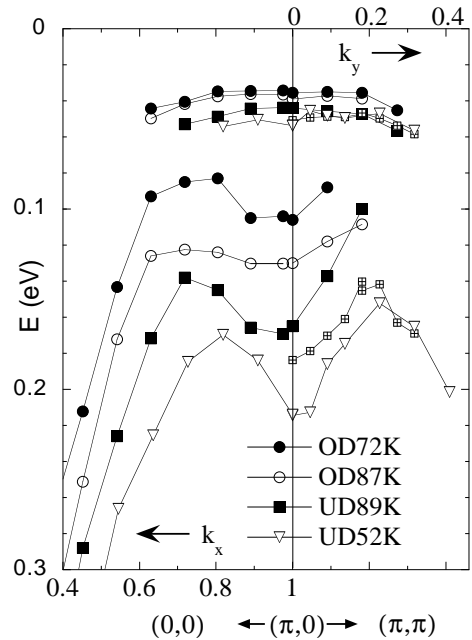}}}
\caption{Peak (top) and hump (bottom) dispersion from ARPES data of
Bi2212 as a function 
of doping 
(superconducting state).  These quantities become the leading edge 
(strong) pseudogap and the high energy (weak) pseudogap in the 
pseudogap phase.  The two dispersion directions shown become increasingly 
similar as the doping is reduced.  If magnetic long range order was
present, the directions would be equivalent.  From Ref.~\cite{JC99}.}
\label{fig36}
\end{figure}

All of the above discussion concerns the leading edge gap, also known 
as the strong pseudogap.  ARPES studies also find a higher energy 
pseudogap, known as the weak pseudogap.  The presence of the latter was 
evident from the earliest studies \cite{MARSHALL}, but it was not 
until later that the two effects were clearly differentiated 
\cite{JC99}.  In constrast to the leading edge gap, which appears to 
be a precursor to the d-wave superconducting gap, the high energy 
pseudogap behaves differently.  It is simply the continuation of the 
``hump'' from the superconducting state, and has a dispersion 
which increasingly resembles that of a spin density wave insulator as 
the doping is reduced (Fig.~\ref{fig36}) \cite{JC99}.
In essence, the sharp spectral 
peak in the superconducting state is
replaced by the leading edge gap, the spectral dip is filled in, and 
the high energy hump becomes the weak pseudogap.
This high energy gap is 
what is commonly observed by ARPES in LSCO and NCCO, as the actual 
superconducting gaps are difficult to see in these materials because 
of their small size.  This gap strongly increases with underdoping, 
and adiabatically connects to the Mott insulating gap of the undoped 
phase \cite{BIGBOB}.  The presence of these two gaps may resolve the 
precursor pair versus competitive scenario debate, in that the leading 
edge gap 
seems to be the precursor to the superconducting gap, whereas the 
high energy gap is the precursor to the magnetic insulating gap.  Of 
course, these two gaps are connected, in that their energies scale 
together with doping, with a ratio of 3.5-4 (Fig.~\ref{fig27}) \cite{JC99}.
This again demonstrates the intimate 
relation of magnetic and pairing correlations in the cuprates.

\begin{figure}
\centerline{\epsfxsize=0.8\textwidth{\epsfbox{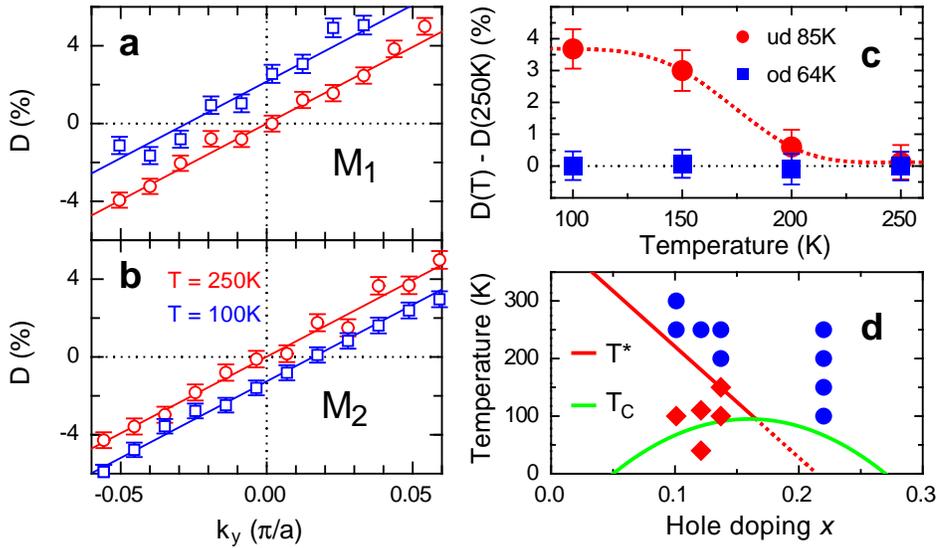}}}
\caption{Circularly polarized ARPES of underdoped Bi2212.  The intensity 
difference between left and right polarized light normalized to the 
average is plotted along $(\pi,0)-(\pi,\pi)$ in the 
left panels, with M1 (panel a) and M2 (panel b) indicating orthogonal 
($(\pi,0)$ and $(0,\pi)$) directions (circles for T=250K, squares for T=100K).  
$k_{y}=0$ is the mirror plane.
The shift with temperature at $k_y=0$
represents chiral symmetry breaking, and is plotted in panel c versus 
temperature for an 
overdoped (squares) and underdoped (circles) sample.  
Panel d shows that this shift (diamonds) only exists for temperatures 
below the pseudogap temperature, $T^{*}$ (circles indicate no 
shift).  From Ref.~\cite{NAT02}.}
\label{fig37}
\end{figure}

None of the spectroscopic data, though, support a picture where the 
pseudogap phase represents a phase with true long range order, as 
advocated by a number of theories, in particular those involving a 
quantum critical point near optimal doping.  On the other hand, a 
recent ARPES experiment does find evidence for broken symmetry in the 
pseudogap phase (Fig.~\ref{fig37}) 
\cite{NAT02}.  In these experiments, circularly polarized light is 
employed.  In general, the signal for left and right polarized light 
is different, but in a mirror plane, they should be equivalent.  This 
mirror plane effect is seen in overdoped samples along the 
$(0,0)-(\pi,0)$ ($\Gamma-M$) line.  But in underdoped 
samples, the signals are no longer equivalent in this direction in 
the pseudogap phase, rather, they become degenerate at some other k 
point shifted off this line.
Moreover, the size of this shift 
increases below $T^{*}$ as would be expected if an order parameter
developed (Fig.~\ref{fig37}c).  The 
implication is that time reversal symmetry (or chiral symmetry, 
depending upon interpretation), is being broken in the pseudogap phase.

The first worry about such an experiment is that the $(0,0)-(\pi,0)$ 
direction is technically not a mirror plane in the Bi2212 crystal 
structure (due to the orthorhombicity and superstructure).  On the 
other hand, it has been known for some time that the main band signal 
from ARPES in Bi2212 appears to obey dipole selection rules consistent 
with tetragonal symmetry \cite{DING96}, and such is the case in these 
measurements as well.  Of course, there could be some structural 
effect associated with $T^{*}$, but these authors did x-ray 
scattering on the samples, and found no evidence for a structural 
change below $T^{*}$ \cite{NAT02}.  Moreover, the effect of the pseudogap is to 
shift the overall intensity of the left and right signals relative to 
one another, as if chiral symmetry was being broken.

A similar effect has not been seen in the $(0,0)-(\pi,\pi)$ 
($\Gamma-Y$)
mirror plane, though this has not been studied as extensively yet.  
If this continues to hold, then it has definite implications.  
Simple ferromagnetism would cause an effect in both mirror planes.  In 
addition, most orbital current models (the d density wave state, for 
example) would predict an effect along $\Gamma-Y$ and not along 
$\Gamma-M$ (opposite to experiment).  One of the two orbital current 
patterns discussed by Varma (left panel, Fig.~\ref{fig20})
behaves the same way, but the other (right panel, Fig.~\ref{fig20}) has 
a signature similar to experiment \cite{VARMA02}.  So, it is indeed 
possible that the data represent an effect which can be attributed to 
orbital currents, but more experiments would certainly be 
desirable.  What this particular experiment illustrates, though, is the 
power
of photoemission in addressing fundamental issues connected with 
the cuprates.

\section{Inelastic Neutron Scattering}

The other momentum resolved probe in the cuprates is inelastic 
neutron scattering.  The part of the signal of interest here is the 
magnetic part, which is proportional to the imaginary part of the
spin-spin response function, $\chi({\bf q},\omega)$, times a Bose 
population factor.  For elastic scattering, one sees Bragg peaks 
associated with the magnetism if the material is magnetically ordered.
(Phonons and structure are measured by neutrons as well, but this takes us 
beyond the scope of this review).

The first result with neutrons was finding the antiferromagnetic order in the 
undoped phase \cite{INS}.  Magnetic moments of 2/3 $\mu_{B}$ per Cu 
site are 
found \cite{JOHN88}, the reduction from 1 being due to quantum 
fluctuations associated with the small spin (S=1/2) of the Cu ion.  
The ordering wavevector is  $Q=(\pi,\pi,\pi)$, which means that 
successive planes are antiferromagnetically coupled as well.

\begin{figure}
\centerline{\epsfxsize=0.5\textwidth{\epsfbox{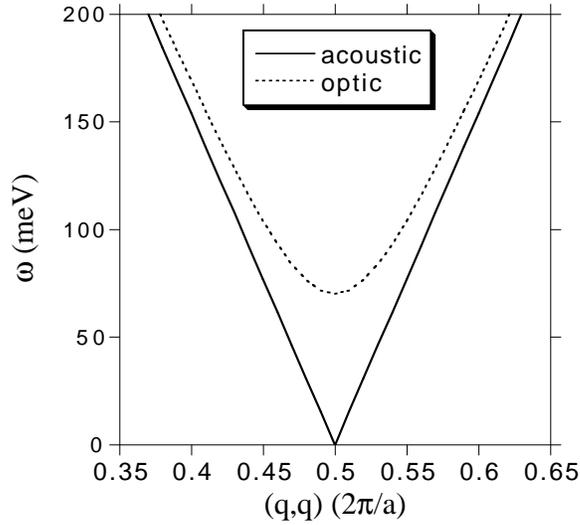}}}
\caption{Acoustic and optic branches of the spin wave dispersion in the 
magnetic phase of YBCO as revealed by inelastic neutron scattering.
Adapted from Ref.~\cite{REZNIK}.}
\label{fig38}
\end{figure}

For bilayer systems like YBCO, there are two branches of the spectrum
(Fig.~\ref{fig38}), an 
acoustic branch with form factor $\sin^{2}(Q_{z}d/2)$ and an optic branch 
with form factor $\cos^{2}(Q_{z}d/2)$,
where $d$ is the separation of the two CuO layers of the bilayer \cite{REZNIK}.
The acoustic branch has the classic spin-wave dispersion with respect 
to $(\pi,\pi)$, with the planar exchage energy $J_{\parallel} \sim 100$ meV
determined from the slope.  The optic branch has a gap of order 60 meV.  As the
optic gap is equal to $2\sqrt{J_{\parallel}J_{\perp}}$, where 
$J_{\perp}$ is the intrabilayer exchange, then $J_{\perp} \sim 10$ meV \cite{FONG00}.

\subsection{LSCO}

The first doping dependent studies were done on LSCO.  They revealed the 
presence of four incommensurate peaks (see Fig.~\ref{fig10}) at locations 
$2\pi(0.5\pm\delta,0.5)$ and $2\pi(0.5,0.5\pm\delta)$, with $\delta$ scaling 
with the doping, $x$ \cite{NS-LSCO}.
The original explanation for this incommensurability was related to 
the Fermi surface geometry.  As the doping increases, the 
Fermi surface hole volume expands, and the predicted 
incommensurability with doping from RPA calculations more or less agrees 
with experiment \cite{LEVIN}.

\begin{figure}
\centerline{\epsfxsize=0.5\textwidth{\epsfbox{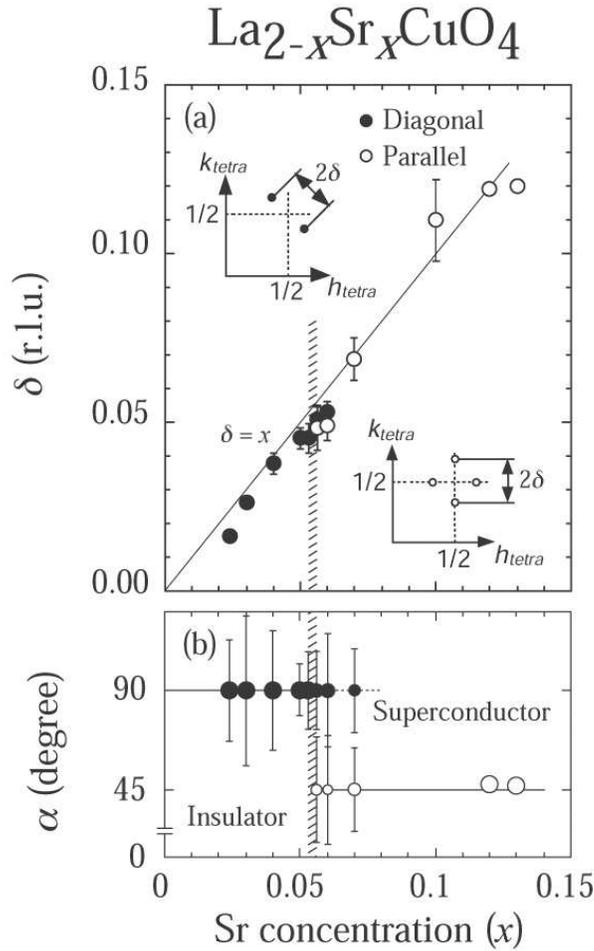}}}
\caption{Neutron scattering peaks versus doping for LSCO.  The spot 
pattern rotates by 45$^{\circ}$ at the spin glass/superconducting 
boundary.  $\delta$ is the incommenusability, and $\alpha$ the angle 
of the spots in momentum space relative to the (1/2,1/2) wavevector.  
From Ref.~\cite{YAMADA}.}
\label{fig39}
\end{figure}

This view changed, though, with the observation of elastic scattering 
peaks in the LTT (low temperature tetragonal)
phase of Nd doped LSCO \cite{JOHN95}.  The 
incommensurate elastic peaks were accompained by charge ordering 
peaks (see Fig.~\ref{fig10}) 
at $2\pi(\pm2\delta,0)$ and $2\pi(0,\pm2\delta)$.  Tranquada and co-workers 
interpreted this behavior as due to the formation of stripes of doped 
holes with commensurate antiferromagnetic domains between the 
stripes.  If the stripes act as antiphase domain walls, then the 
prediction is that the magnetic signal will be incommensurate, with 
$\delta$ proportional to $x$ (upper panel, Fig.~\ref{fig39}).
This picture received further support 
when it was observed in LSCO that 
for dopings smaller than those where superconductivity occured, only 
two spots were present, and they were rotated $45^{\circ}$ relative 
to the previous spots (bottom panel, Fig.~\ref{fig39}) \cite{YAMADA}.  This implies one 
dimensional behavior, consistent with stripes.

A similar incommensurate pattern was later seen in YBCO 
\cite{NS-YBCO}, and this pattern is also found to be 1D like in detwinned
samples (where the CuO chains are all aligned) \cite{YBCO1D}.
This again gives evidence for 
stripes, though the effect may have a more benign origin due to the
influence of the CuO chains.  At low dopings, charge ordering 
peaks are seen in YBCO as well \cite{MOOK02}.

\begin{figure}
\centerline{\epsfxsize=0.5\textwidth{\epsfbox{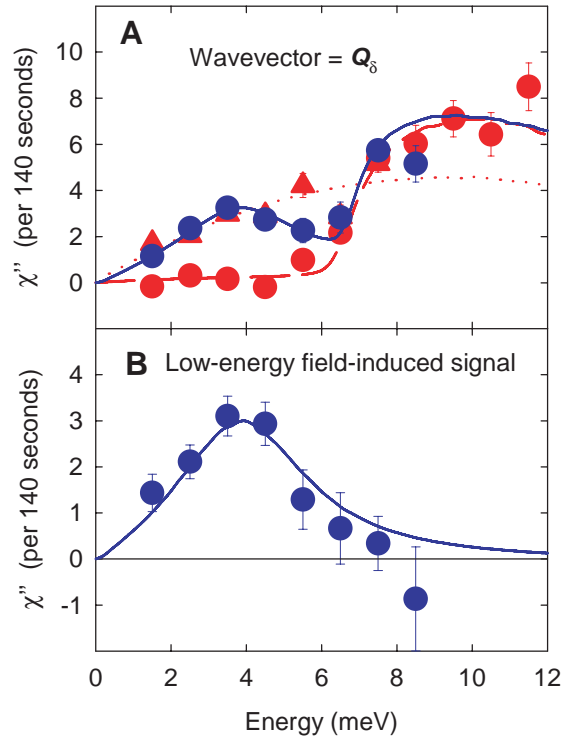}}}
\caption{Imaginary part of the dynamic susceptibility versus energy 
(panel A)
for the superconducting state of optimal doped (x=0.163) LSCO
at zero field (lower set of circles), at a field 
of 7.5 T (upper set of circles), and in the normal state (triangles).  The 
difference of zero and 7.5 T data are plotted in the panel B.  
From Ref.~\cite{LAKE2}.}
\label{fig40}
\end{figure}

The effect of superconductivity on LSCO is to lead to a sharpening of 
the incommensurate peaks, and the formation of a ``spin gap'' at low 
energies of $\sim$ 6 meV (Fig.~\ref{fig40}).  This spin gap is fairly isotropic in momentum space 
\cite{LAKE1}, which was taken as evidence against an
interpretation of it being the 2$\Delta$ continuum gap since 
$\Delta$ is anisotropic for a d-wave superconductor (in particular, there should be
a significant continuum gap at $(\pi,\pi)$).
This statement should be treated with care, though, since the only low 
frequency structure is at these incommensurate wavevectors, with the 
intensity at other wavevectors, like $(\pi,\pi)$,
due to overlap from these peaks given their finite width in momentum.

Subsequent experiments found that this spin gap filled in at modest 
values ($H \ll H_{c2}$) of the magnetic field (Fig.~\ref{fig40}) \cite{LAKE2}.
For more underdoped 
samples, magnetic ordering could even be induced by applying a field
\cite{LAKE3}.  The 
obvious idea is that some kind of SDW ordering is being stabilized by 
the vortices \cite{AROVAS,DSZ}.  But, the vortex density 
is quite 
low at the field values studied, which would imply that there is a
very large magnetic polarization cloud around the vortices.  This
is certainly consistent with the neutron results, in that the magnetic 
correlation length is quite long in underdoped LSCO samples.

On the other hand, one might interpret these results as a 
stabilization of stripe formation.  In this context, STM experiments 
on Bi2212 find a charge density wave pattern associated with the 
vortex cores \cite{HOFFMAN}.  The Fourier pattern was anisotropic (factor
of 3 intensity difference between orthogonal planar directions), which 
would argue for 1D behavior, although it is possible that this 
could be an extrinsic effect due to the STM tip.

\subsection{YBCO}

\begin{figure}
\centerline{\epsfxsize=1.0\textwidth{\epsfbox{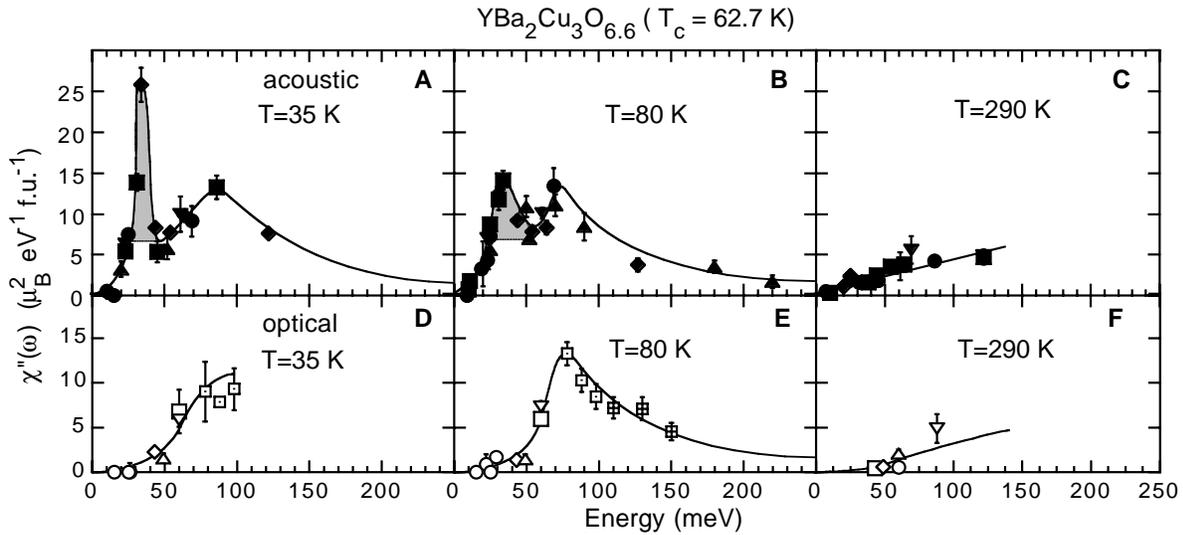}}}
\caption{Wavevector integrated dynamic susceptibility for underdoped 
YBCO at 35K (superconducting state), 80K (pseudogap phase), and 
290K (normal state) in the acoustic and optic channels.  The gray 
shaded area represents the resonance.  From Ref.\cite{DAI99}.}
\label{fig41}
\end{figure}

Perhaps the most dramatic effect associated with neutron scattering in 
the superconducting state is the formation of a sharp commensurate
resonance at 
about 40 meV in YBCO (Fig.~\ref{fig41}) \cite{ROSSAT}.
The magnetic nature of the 
resonance was confirmed by later polarized measurements 
\cite{MOOK93}.  Subsequently, the resonance was seen in Bi2212 
\cite{KEIMER}.  
The resonance has a form factor equivalent to the 
acoustic branch of the undoped material, and so is centered about 
the $(\pi,\pi,\pi)$ wavevector.  In particular, the ``optic branch'' 
has no resonance, and remains gapped as in the insulator (Fig.~\ref{fig41})
\cite{FONG00}.  These observations led to speculations that the 
resonance might be a bilayer effect, since is it not seen in LSCO, 
but a new experiment has identified the resonance in single layer 
Tl2201 \cite{TL2201}.  Several 
theories concerning the resonance were discussed in Section 2, in 
particular the controversy concerning whether it represents a 
particle-hole or particle-particle collective mode.

The resonance energy scales with doping like 
$5T_{c}$ \cite{FONG00}, 
and its intensity has a variation with temperature much like that of
the superconducting order parameter \cite{FONG96}.  Based on previous theoretical work 
\cite{WHITE}, Demler and Zhang made the provocative suggestion
that these results implied an 
equivalence between the exchange energy difference between the normal 
and superconducting state and the resonance weight \cite{DZNAT}; that 
is, the superconducting condensation energy was related to the 
formation of the resonance.
\begin{figure}
\centerline{\epsfxsize=0.5\textwidth{\epsfbox{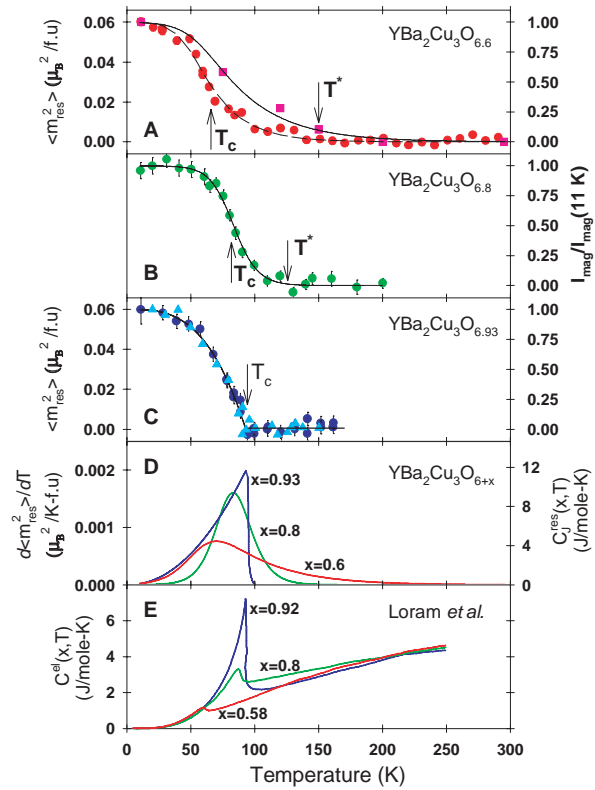}}}
\caption{Neutron data for YBCO.  Panels A-C are the temperature 
variation of the resonance peak intensity for samples at three different 
dopings.  Panel D is the temperature derivative of the intensity data, in 
comparison to specific heat data in panel E.  From Ref.~\cite{DAI99}.}
\label{fig42}
\end{figure}
Their suggestion was tested by
measurements on YBCO, which demonstrated a 
similarity of the specific heat anomaly at $T_{c}$ and the temperature 
derivative of the resonance weight (Fig.~\ref{fig42}) \cite{DAI99}.  As magnetic fields 
along the c-axis are known to suppress the specific heat anomaly, this 
motivated experiments which looked at the field dependence of the 
resonance.  It was found that a field applied 
perpendicular to the planes did lead to a strong suppression of the 
resonance in an underdoped YBCO sample \cite{DAI00}, as opposed to a 
parallel field which did not \cite{BOURGES}.  To understand this result,
we note that STM measurements find in underdoped Bi2212 samples that the 
pseudogap phase is present in the vortex cores, with no coherence
peaks \cite{CORE}.  This
would imply that the resonance, which is a property of the coherent
superconducting state, is suppressed in the cores.  On the other 
hand, an analysis of the neutron data indicates that the effective core radius 
needed to explain the observed suppression of the resonance is significantly larger 
than the known superconducting correlation length \cite{RESVOR}.  This 
again demonstrates that for YBCO, like for LSCO, the 
cores polarize the surrounding medium.  This leads to a suppression of the
resonance out to a magnetic correlation length about the cores.  In that sense, 
it should be remarked that the YBCO sample studied in the field 
dependent experiment exhibits an anomalously large magnetic 
correlation length (28 $\AA$) as determined from the resonance width
in momentum space \cite{DAI01}.

Perhaps the biggest controversy surrounding the resonance is whether it is 
responsible for structure seen in other spectroscopic probes, like 
ARPES and tunneling \cite{NORM97}.  The smallness of the resonance 
weight (a few percent of the local moment sum rule) argues against 
this \cite{KEE}.  On the other hand, phase space considerations can be used 
as a counterargument \cite{GANG5}.  The resonance weight is small 
because it is localized in momentum space around $(\pi,\pi)$.  But as 
electronic states near $(\pi,0)$ are connected by these wavevectors,
there is no problem for these states to 
be strongly affected by the resonance despite its overall small 
weight.  These arguments can be extended to states in other regions 
of the zone as well \cite{ESCHRIG}.

One reason this controversy has arisen is the suggestion by certain 
researchers that the resonance interpretation of ARPES and tunneling
implies that the 
resonance alone is responsible for pairing.  Actually, the 
formation of the resonance, as well as the profound changes in the 
ARPES lineshape, is a consequence of the onset of superconductivity, 
rather than the cause of it.  Such ``feedback'' effects are 
unavoidable in any theory where the pairing is not due to 
phonons \cite{AC}.  The classic example is the stabilization of the A phase in 
$^{3}He$ relative to the B phase due to the feedback of pairing on 
the spin fluctuations \cite{BA}.

\begin{figure}
\centerline{\epsfxsize=0.8\textwidth{\epsfbox{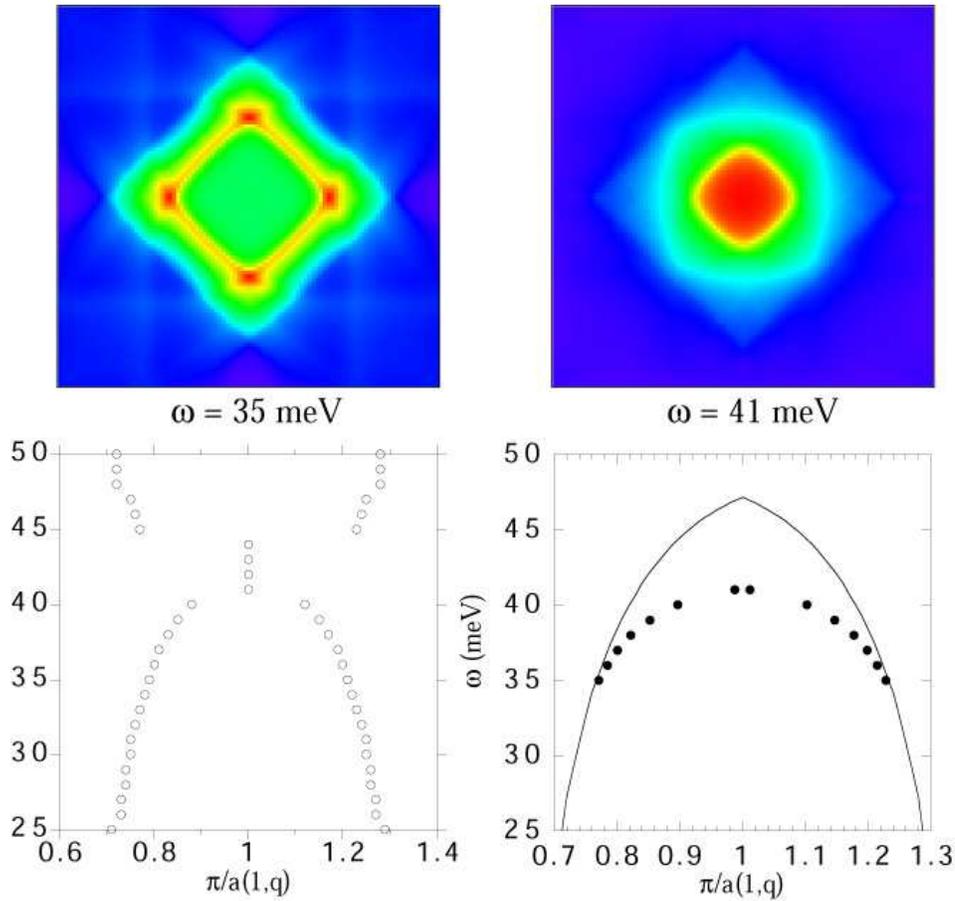}}}
\caption{RPA calculation of the dynamic susceptibility for YBCO in the 
superconducting state based 
on ARPES dispersions.  Momentum dependence of the intensity is 
plotted at resonance in top right panel, and below resonance in top
left panel.  Note the commensurate pattern at resonance compared to 
the incommensurate (diamond shaped) pattern below resonance.  
Resulting $\omega(q)$ dispersion relation (hourglass shaped)
is plotted in the bottom left panel.  (The maximum superconducting gap 
was 29 meV and the superexchange energy 110 meV.)  This is
further illustrated in the bottom right panel, where the pole of the RPA response
function is plotted (circles), with the solid line representing the edge of the
continuum.
Adapted from Refs.~\cite{NORM00} and \cite{NORMAN2}.}
\label{fig43}
\end{figure}

Below the resonance energy in YBCO, the magnetic response becomes 
incommensurate (top left panel, Fig.~\ref{fig43}) \cite{NS-YBCO}.
Detailed studies of the $\omega({\bf 
q})$ dispersion relation find an ``hourglass'' shape (bottom left panel,
Fig.~\ref{fig43}) \cite{ARAI}, with 
incommensurate ``sidebranches'' appearing both above and below the 
resonance energy, though the upper branch is damped (it lies 
above the two particle continuum).  This sidebranch behavior can be 
understood from 
linear response calculations for a d-wave superconductor 
(Fig.~\ref{fig18}) 
\cite{FLORA,KAO,NORM00}.  Under certain conditions, the 
lower incommensurate branch is a collective mode
pulled below the two particle continuum, with the commensurate
resonance at the top of 
this lower branch (bottom right panel, Fig.~\ref{fig43})  \cite{FLORA,NORMAN2}.
This is consistent with an interpretation of recent 
data on slightly underdoped YBCO \cite{BOURGES00}, but differs from an 
interpretation of more heavily underdoped samples, where the resonance 
and sidebranches appear to represent separate effects \cite{MOOK02}.  
In the latter case, the incommensurate sidebranches have been interpreted as 
due to antiphase domain stripes, with the commensurate resonance 
presumably due in phase domain stripes.
It should be remarked that the 
condition to get pole-like behavior for the lower incommensurate 
branch is difficult in linear response calculations, and requires 
reduced curvature of the
Fermi surface around the nodes, so that the constant energy contours of the Dirac cones 
discussed in Section 3 (Fig.~\ref{fig26}) are flat instead of banana 
shaped \cite{BLEE,NORM00,NORMAN2}.  For instance, the ARPES Fermi 
surface of Bi2212 does not indicate such flat contours, 
and the prediction would be that the incommensurate effects in
Bi2212 are weak and 
unconnected to the resonance \cite{NORM00,NORMAN2}, as inferred 
experimentally in heavily underdoped YBCO \cite{MOOK02}.  Unfortunately, 
neutron scattering experiments on Bi2212 are difficult due to the 
small crystals available, and ARPES results for YBCO are controversial 
because of surface related problems.  Still, such studies would 
be useful to test these ideas.

As discussed by Batista and 
co-workers \cite{BALATSKY}, differentiation between these various 
interpretations is not as straightforward as it might seem.
\begin{figure}
\centerline{\epsfxsize=0.4\textwidth{\epsfbox{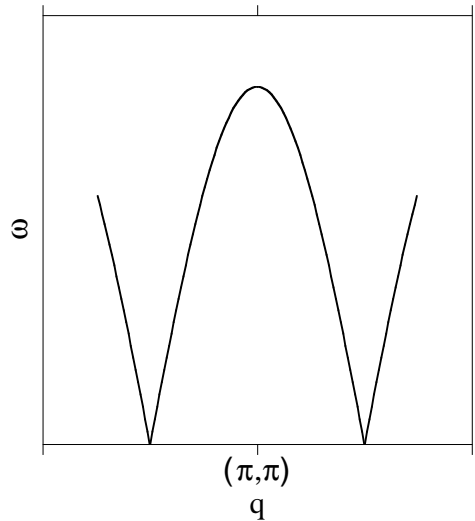}}}
\caption{Proposed $\omega(q)$ dispersion relation for the dynamic 
susceptibility of YBCO based on overlapping spin waves from the AF 
domains between stripes.  Adapted from Ref.~\cite{BALATSKY}.}
\label{fig44}
\end{figure}
The 
picture they offer is that the lower incommensurate branch is 
analogous to the spin wave dispersion in an incommensurate antiferromagnet, 
with the resonance being where the two spin wave branches from +Q and -Q 
intersect (Fig.~\ref{fig44}).  Actually, the dispersion they plot follows the two 
particle continuum gap of the linear response calculations.  In these 
RPA calculations, the incommensurate branch either follows the 
continuum gap (and thus is not a pole), or is pulled beneath (becoming 
a pole), depending on how flat the constant energy contours of the Dirac cones are
\cite{FLORA,NORMAN2}.  Presumably, this physics is not unrelated to a 
stripes interpretation, where quasi 1D behavior would cause the same 
effects as the flat contours.

Finally, although a lot has been made about the differences between 
the dynamic susceptibilities of YBCO and LSCO, it should be noted that 
the most recent data on LSCO find the presence of energy dispersion in 
the incommensurate response (like in YBCO), with commensurate excitations
present beyond 20 meV \cite{HIRAKA}.  Thus, it may be that 
the central difference between these two materials is that the bulk 
of the spectral weight sits in the lower incommensurate branch for 
LSCO, but in the commensurate resonance for YBCO.  This difference
is probably 
due to a variety of factors:  the smaller energy gap in LSCO, 
differences in the Fermi surface topology, and the stronger 
tendency for disorder and inhomogeneous behavior (stripes) in LSCO.  In this 
respect, Bi2212 is probably intermediate between YBCO and LSCO, so 
more neutron studies of Bi2212 would be desirable.

\section{Optical Conductivity}

Optics measures a two-particle response function as well, the 
current-current correlation function.  Because of the tiny momentum 
associated with the light used, optics measures the zero $q$ limit of 
this function.  
This can be represented as a particle-hole bubble, like in the 
previous section, but with the spin operators replaced by current 
operators at the vertices (the Kubo bubble).

\begin{figure}
\centerline{\epsfxsize=0.5\textwidth{\epsfbox{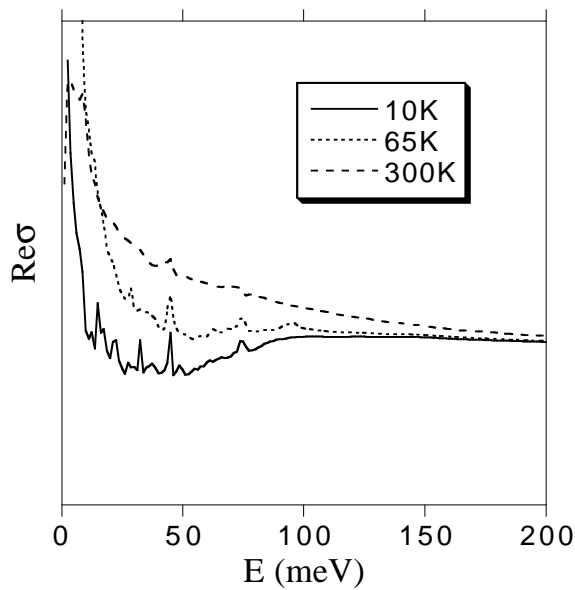}}}
\caption{Real part of planar infrared conductivity of underdoped YBCO 
at three temperatures corresponding to the superconducting state (10K), 
pseudogap phase (65K), and normal state (300K).  Note gap-like 
depression which develops at around 50 meV.  From Ref.~\cite{PUCH}.}
\label{fig45}
\end{figure}

\begin{figure}
\centerline{\epsfxsize=0.5\textwidth{\epsfbox{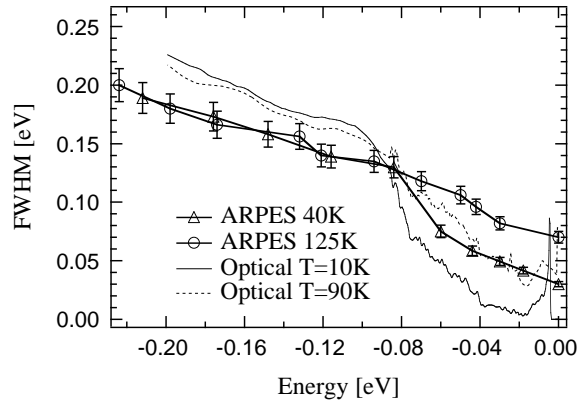}}}
\caption{Comparison of optical scattering rate, $1/\tau$, from 
Ref.~\cite{PUCH} to ARPES nodal linewidth in the superconducting 
state and pseudogap phase of optimal doped Bi2212.  From 
Ref.~\cite{ADAM}.}
\label{fig46}
\end{figure}

We start with the planar response.  The normal state is characterized 
by a broad, Drude-like response centered at $\omega=0$ (Fig.~\ref{fig45}).  
The Drude 
tail, though, has an anomalous behavior.  The data are best 
appreciated by representing the optical response in a generalized 
Drude form \cite{ZACK,PUCH}
\begin{equation}
\sigma(\omega) = \frac{1}{4\pi}
\frac{\omega_{pl}^{2}}{1/\tau(\omega)-i\omega[1+\lambda(\omega)]}
\end{equation}
where $\omega_{pl}$ (the plasma frequency) is given by the sum rule
\begin{equation}
\int_{0}^{\infty} d\omega Re\sigma(\omega) = \omega_{pl}^{2}/8
\end{equation}
(the last integral is usually cut-off at 1 eV or so to avoid interband 
contributions).  In this form, $1/\tau$ is the scattering rate, and 
$1+\lambda$ the mass renormalization.  The two terms are related by 
a Kramers-Kronig tranform.  Such an analysis reveals that the 
scattering rate has the form $a+b\omega$ (Fig.~\ref{fig46}).  The $b$ term is what is
expected for a marginal Fermi liquid, but the $a$ term is not,
since it does not appear to be proportional to T.
Abrahams and Varma attribute it to scattering from 
off planar impurities \cite{AV}, but more likely, it is a signature 
of the non-Fermi liquid nature of the normal state.  In particular, there is some
evidence from ARPES that the $a$ term is associated with the pseudogap,
since it has a similar momentum anisotropy.

When superconductivity sets in, $Re\sigma$ develops a depression at 
an energy close to the value $2\Delta$ expected for a superconductor 
\cite{COOPER,ZACK,JOE2,ROTTER,PUCH}, but a true gap never fully 
develops.  Instead, a narrower Drude peak is present at energies
below this depression, 
representing uncondensed carriers.  For a superconductor, one 
expects the presence of a $\delta$ function at $\omega=0$ due to the 
dissipationless response of the superfluid, and 
this is indeed seen as a $1/\omega$ term in $Im\sigma$ at low 
frequencies.

\begin{figure}
\centerline{\epsfxsize=0.5\textwidth{\epsfbox{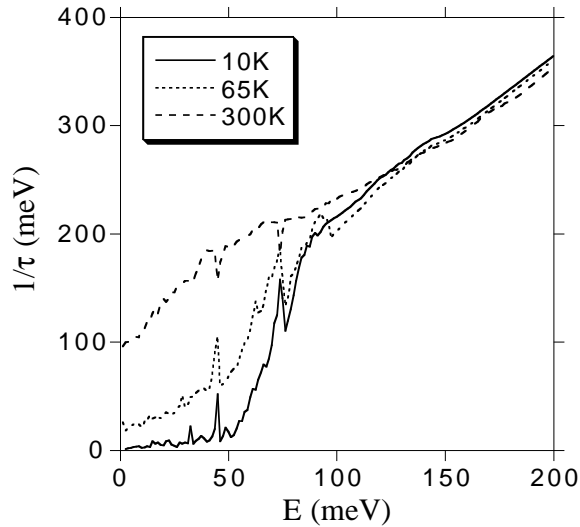}}}
\caption{Variation of optical scattering rate with energy for 
underdoped YBCO (10K - superconducting state, 65K - pseudogap phase, 
300K - normal state).  From Ref.~\cite{PUCH}.}
\label{fig47}
\end{figure}

A generalized Drude analysis in the superconducting state \cite{PUCH} 
reveals that $1/\tau$ becomes strongly gapped below some threshold 
energy (Figs.~\ref{fig46} and \ref{fig47}), and the behavior seen is 
very similar to that inferred from ARPES along the nodal direction (Fig.~\ref{fig46})
\cite{ADAM}.
The low frequency limit is most easily seen from 
microwave \cite{BONN} and thermal conductivity \cite{ONG} measurements,
which find a strong collapse with temperature (see Fig.~\ref{fig5})
of the scattering rate 
below $T_{c}$, with a residual low temperature
mean free path of order microns for 
clean samples of YBCO.  It should be remembered that for 
electron-electron scattering, only Umklapp processes 
contribute to the electromagnetic response, whereas normal processes 
contribute to the ARPES and thermal response as well.  This has been 
used to quantitatively account for differences between the microwave 
and thermal conductivity scattering rates \cite{DUFFY}.  We should 
note that although the strong collapse of the scattering rate
with temperature below $T_c$ is consistent 
with ARPES results near $(\pi,0)$ \cite{PHEN98}, the ARPES results along 
the nodal direction are still controversial, where a linear T behavior below $T_c$
has been claimed by one group (Fig.~\ref{fig23}) \cite{VALLA} but not by others.  All ARPES 
results are on Bi2212, which is more disordered than YBCO, and in fact 
terahertz ($\omega \sim 1$ meV) measurements on Bi2212 claim a linear T scattering rate of 
the residual carriers as well \cite{JOE3}.  The size of this rate
is difficult to compare to 
ARPES, given the fact that the ARPES resolution width greatly exceeds 
the frequency of this measurement.

The other interesting point is that for underdoped samples, a partial 
scattering rate gap persists in the pseudogap phase (Fig.~\ref{fig47}), likely due to the
gapping of electronic states near $(\pi,0)$ \cite{PUCH}.  In 
fact, the optics data have a smooth evolution through $T_{c}$ for 
underdoped samples.  Even a finite frequency signature of the 
superfluid response persists above $T_c$, as revealed by terahertz
measurements of the electrodynamic response \cite{JOE}.  These data
have been successfully modeled assuming the superconducting transition is of
the Kosterlitz-Thouless type.

All of these behaviors would be difficult to attribute to phonons, in 
that the latter are not gapped in the superconducting state.  There is 
a drop in the electron-phonon scattering rate, since the electrons 
themselves are gapped, but the effect is not as dramatic as in the 
electron-electron case.
\begin{figure}
\centerline{\epsfxsize=0.4\textwidth{\epsfbox{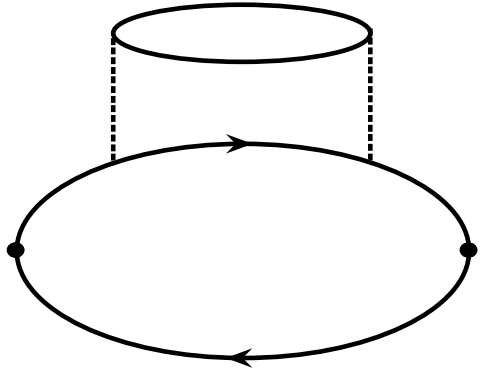}}}
\caption{Kubo bubble for optical conductivity.  Dots represents 
current vertices.  The attached part at the top is the lowest order 
self-energy insertion to the bubble due to electron-electron scattering.}
\label{fig48}
\end{figure}
A Feynman diagram analysis of the 
Kubo bubble (Fig.~\ref{fig48}) would give a scattering rate drop setting in below 4$\Delta$
(analogous to the 3$\Delta$ gap in the single particle scattering 
rate (Fig.~\ref{fig5}) \cite{NOZ}) for the electron-electron scattering case (in the electron-phonon case,
the self-energy insertion in Fig.~\ref{fig48} would be replaced by a phonon).
The actual optics data, though, show an abrupt 
onset of the drop at a frequency smaller than this.  This implies 
collective effects, and in fact several researchers 
\cite{MUNZAR,JULES,AC3} 
have used the same explanation for the optics data as invoked previously 
to explain the ARPES data \cite{NORM97,NDING}.  In particular, the 
scattering rate onset corresponds to $2\Delta + \Omega_{res}$, 
where $\Omega_{res}$ is the spin resonance energy
seen by inelastic neutron scattering.

The c-axis response, though, is quite different (see Fig.~\ref{fig7}).
Most studies have 
been done on YBCO, since the electronic part of the c-axis 
conductivity is small, and non-trivial to separate from the phonons 
present in the data, particularly for Bi2212 which is more 2D like.  The normal state 
response is non-Drude like in nature, $Re\sigma$ being more or less flat in 
frequency \cite{HOMES}.  A hard gap begins to open up in this spectrum 
below $T_{c}$ at a
frequency near $2\Delta$ \cite{HOMES}.  Beyond this energy, a peak is present 
in the data whose origin is controversial.  It has been identified as 
the optic plasmon of the bilayer \cite{DIRK}, but also appears to be related to 
the neutron resonance as well \cite{TOM}.

Recent developments in optics have focused on the issue of the 
condensation energy, which brings us to the topic of the next section.

\section{Condensation Energy}

The condensation energy is defined as the energy difference between 
the normal and superconducting states at T=0 (at finite temperature, 
the entropy must be taken into account).  Since normal state properties 
are temperature dependent, then this requires an extrapolation 
to infer a hypothetical zero temperature normal state.  Most 
estimates of the condensation energy have been made from specific heat 
data, which have been recently used to suggest a quantum critical point 
just beyond optimal doping based in part on the doping dependence of 
the inferred condensation energy \cite{LORAM}, which has a maximum near
the suggested critical point (middle panel, Fig.~\ref{fig9}).

We begin our story with a piece of pre-BCS history.  In 1956, Chester 
published a paper demonstrating where the condensation energy was 
coming from in conventional superconductors \cite{CHESTER}.  What he 
chose to study was the full Hamiltonian of the solid, which is 
composed of the kinetic energies of the electrons and ions,
and the electron-electron, electron-ion, and ion-ion interactions.  
Exploiting the dependence of $T_{c}$ on ion mass (isotope effect)
and the virial 
theorem, he was able to demonstrate that: (1) the potential energy of 
the electrons is reduced in the superconducting state, (2) the kinetic 
energy of the electrons is increased, and (3) the ion kinetic energy is 
reduced.  For the classic value of the isotope coefficient (1/2), the 
potential and kinetic energy changes of the electrons actually 
cancel, which means that the entire condensation energy is 
equivalent to the lowering of the ion kinetic energy.

This is not what occurs in BCS theory, of course.  The reason is that 
it is an effective theory gotten by projecting the full 
Hamiltonian onto the low energy subspace.  In such a theory, the ion 
kinetic energy term is absorbed into the definition of the potential 
energy, and so superconductivity is due to a lowering of the 
effective potential energy of the low energy electrons.  Note that 
the virial theorem does not apply to the projected Hamiltonian, and so 
cannot be used as a guide.

\begin{figure}
\centerline{\epsfxsize=0.5\textwidth{\epsfbox{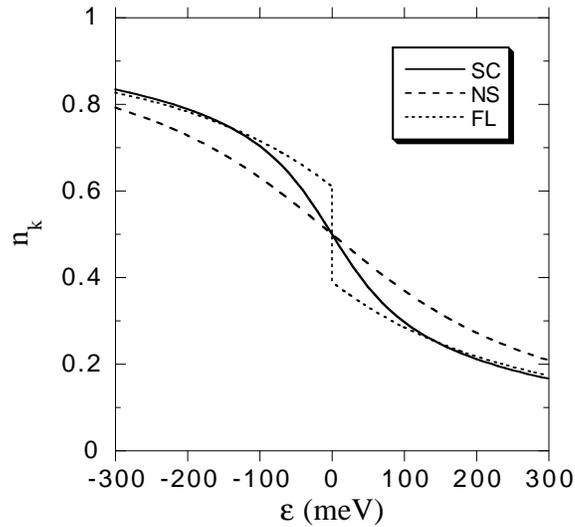}}}
\caption{Model calculations of the momentum distribution function 
in the normal state (NS) and 
the superconducting state (SC) compared to a hypothetical Fermi 
liquid normal state (FL).  From Ref.~\cite{GANG4}.}
\label{fig49}
\end{figure}

As for the kinetic energy, in BCS theory, the normal state is taken to 
be that of non-interacting electrons.  Because of this, the effective 
kinetic energy of the electrons is exactly diagonalized.  That is, 
the momentum distribution function is equal to 1 for $k < k_{F}$ and 
0 for $k > k_{F}$.  In the superconducting state, particle-hole 
mixing occurs, leading to $n_{k} = v_{k}^{2} = \frac{1}{2}
(1-\epsilon_{k}/\sqrt{\epsilon_{k}^{2}+\Delta_{k}^{2}})$ (see 
Fig.~\ref{fig12}).  So, the 
``smearing'' of $n_{k}$ (Fig.~\ref{fig49}) leads to an increase of the kinetic energy.

The potential and kinetic energy differences are straightforward to 
calculate in BCS theory, and this exercise can be found in Tinkham's book 
\cite{TINKHAM}.  Although each term is ultraviolet divergent (and 
thus cut-off at the Debye energy), the sum is convergent, equal to 
$\frac{1}{2} N \Delta^{2}$ where $N$ is the density of states.
Of course, the normal state of the cuprates is not a Fermi liquid (at 
least for optimal and underdoped samples), so BCS considerations 
could be misleading.

In Section 4, we discussed one attempt to get at a piece of the 
condensation energy.  It was noted by Scalapino and White \cite{WHITE} 
that the difference of the exchange energy between the normal and 
superconducting states could be obtained from an integral involving 
the difference of $Im\chi$ between the normal and 
superconducting states.  Demler and Zhang \cite{DZNAT} noted that for 
optimal doped YBCO, the normal state $Im\chi$ is small, and so 
an estimate could be made in that case by simply integrating the 
neutron resonance (see Figs.~\ref{fig41} and \ref{fig42}).
The resulting energy change (the exchange energy 
is lowered in the superconducting state) is in excess of the known 
condensation energy, and they suggested this was compensated by an 
expected increase in the kinetic energy in the superconducting state.
Although the latter is true in BCS theory, is it necessarily true in 
general?  Recent optics data give evidence to the contrary, as we now
discuss.

\subsection{Optics}

\begin{figure}
\centerline{\epsfxsize=0.5\textwidth{\epsfbox{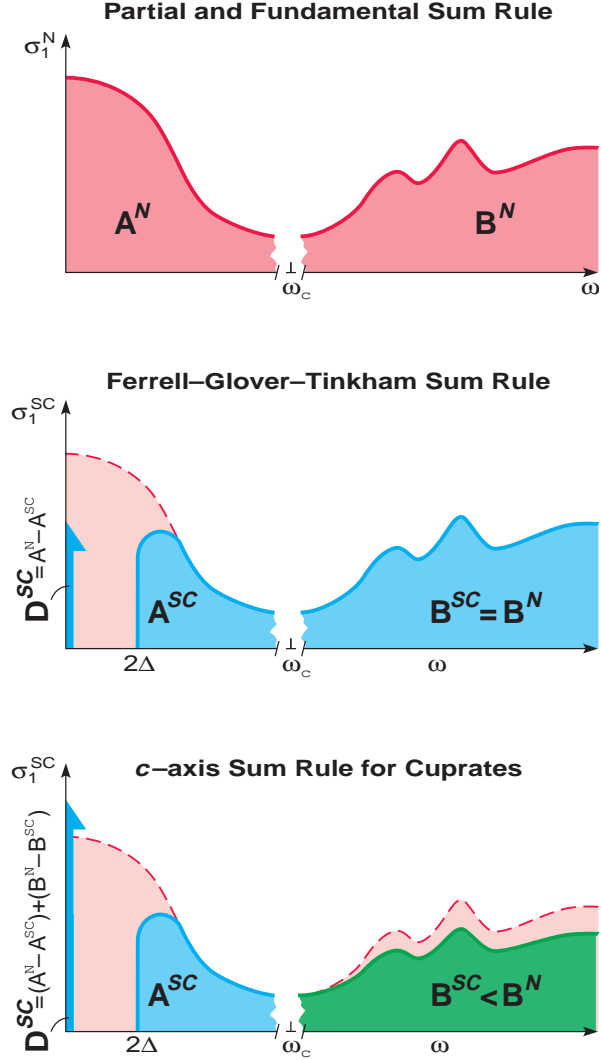}}}
\caption{Top panel:  schematic of optical conductivity (real part) in the 
normal state (A is low energy, B high energy).  Middle panel:  
Comparison of normal state to superconducting state if the single band 
sum rule is preserved.  Gapped weight appears as a $\delta$ function 
at zero energy.  Bottom panel:  When a sum rule ``violation'' is present, 
more weight appears in the $\delta$ function than is gapped out.  
This implies that the remaining weight must come from high energies to 
satisfy the full sum rule.  From Ref.~\cite{KLEIN}.}
\label{fig50}
\end{figure}

Anderson has noted that several unusual properties of the cuprates 
could be understood if spectral weight from high energies were 
transfered to low energies when going from the normal to the 
superconducting state \cite{PWA90,ANDERSON}.  Under such conditions, one 
might anticipate the kinetic energy of the electrons to actually be 
lowered in the superconducting state rather than raised as in BCS 
theory.  This was also suggested by Hirsch, who predicted a 
``violation'' of the optical sum rule due to this effect \cite{HIRSCH}.
That is, in a conventional superconductor, the weight appearing in the 
zero frequency condensate peak comes entirely from the weight removed 
by the 
$2\Delta$ gap in the optical response (a result known as the 
Tinkham-Ferrell-Glover sum rule \cite{TINKHAM}).  But a ``violation'' 
would indicate that more weight was present in the condensate peak 
than expected (Fig.~\ref{fig50}) \cite{KLEIN}.

To understand this, it is necessary to find a relation between the 
optics and kinetic energy.  The optical sum rule of interest is the 
same one discussed in Section 5
\begin{equation}
\int_{0}^{\infty} d\omega Re\sigma_{jj}(\omega) = \omega_{pl}^{2}/8
\end{equation}
This integral is well known when integrated over all energy bands, and 
is simply proportional to the bare 
carrier density, $n$, over the bare electron mass, $m$.
By charge conservation ($n$ fixed), 
this integral is always conserved.

On the other hand, what is of 
interest here is the ``single band'' sum rule
\begin{equation}
{\hat P} \int_{0}^{\infty} d\omega Re\sigma_{jj}(\omega) =
\frac{\pi e^{2} a^{2}}{2\hbar^{2}V} E_{K}
\end{equation}
where
\begin{equation}
E_{K} = \frac{2}{a^{2}N} \sum_{k}
\frac{\partial^{2} \epsilon_{k}}{\partial k_{j}^{2}} n_{k}
\end{equation}
and ${\hat P}$ projects onto the single band subspace.  In these 
expressions,
$V$ is the unit cell volume, $a$ the in-plane lattice constant,
$N$ the number of $k$ vectors, $\epsilon_{k}$ the dispersion 
defined from the kinetic energy part of the effective single band 
Hamiltonian, and $n_{k}$ the momentum distribution function.  This was 
first derived by Kubo \cite{KUBO}.   On the other hand, the kinetic 
energy for this band is
\begin{equation}
E_{kin} = \frac{2}{N} \sum_{k} \epsilon_{k} n_{k}
\end{equation}
Thus, the optical integral is of similar form, but not identical, to 
the kinetic energy \cite{MNCP}, except for a near neighbor tight binding 
form for $\epsilon_{k}$, where $E_{K} = -E_{kin}$.
In practice, the optical integral must 
be cut off at some energy so that interband terms are not included 
(typically 1 eV in the cuprates).

There have been a number of studies which show anomalous changes in 
the optical response between normal and superconducting states at
energies beyond 1 eV in the cuprates \cite{FUGOL,LITTLE,RUBHAUSEN}.
But matters came to the forefront 
when Basov and co-workers demonstrated an explicit violation of the 
single band sum rule for the c-axis optical response of underdoped 
cuprates \cite{BASOV}.  Since a near neighbor tight binding model 
should be adequate to describe the hopping along the c-axis, this 
finding gives direct evidence that the c-axis kinetic energy is 
lowered in the superconducting state, as suggested by Anderson and 
co-workers \cite{INTLAY,ANDERSON} and Hirsch as well
\cite{HIRSCH}.  There has been an alternate interpretation of this 
observation, though, from 
Ioffe and Millis \cite{LEV}.  They claim that a ``violation'' is 
possible for the c-axis response if the normal state reference is 
taken to be a pseudogap phase involving pairs without long range 
phase order.

Regardless, the c-axis kinetic energy is 
so small, the energy savings inferred is far below what is needed to 
account for the actual condensation energy of the cuprates, which is
about 3K per CuO plane in optimal doped YBCO \cite{LORSH}.
On the other hand, the in-plane kinetic energy is quite large, of 
order 1 eV.  Therefore, if a similar relative violation of the size seen 
for the c-axis occurs for the in-plane response, then the energy 
savings could be enough to account for the condensation energy.

\begin{figure}
\centerline{\epsfxsize=0.5\textwidth{\epsfbox{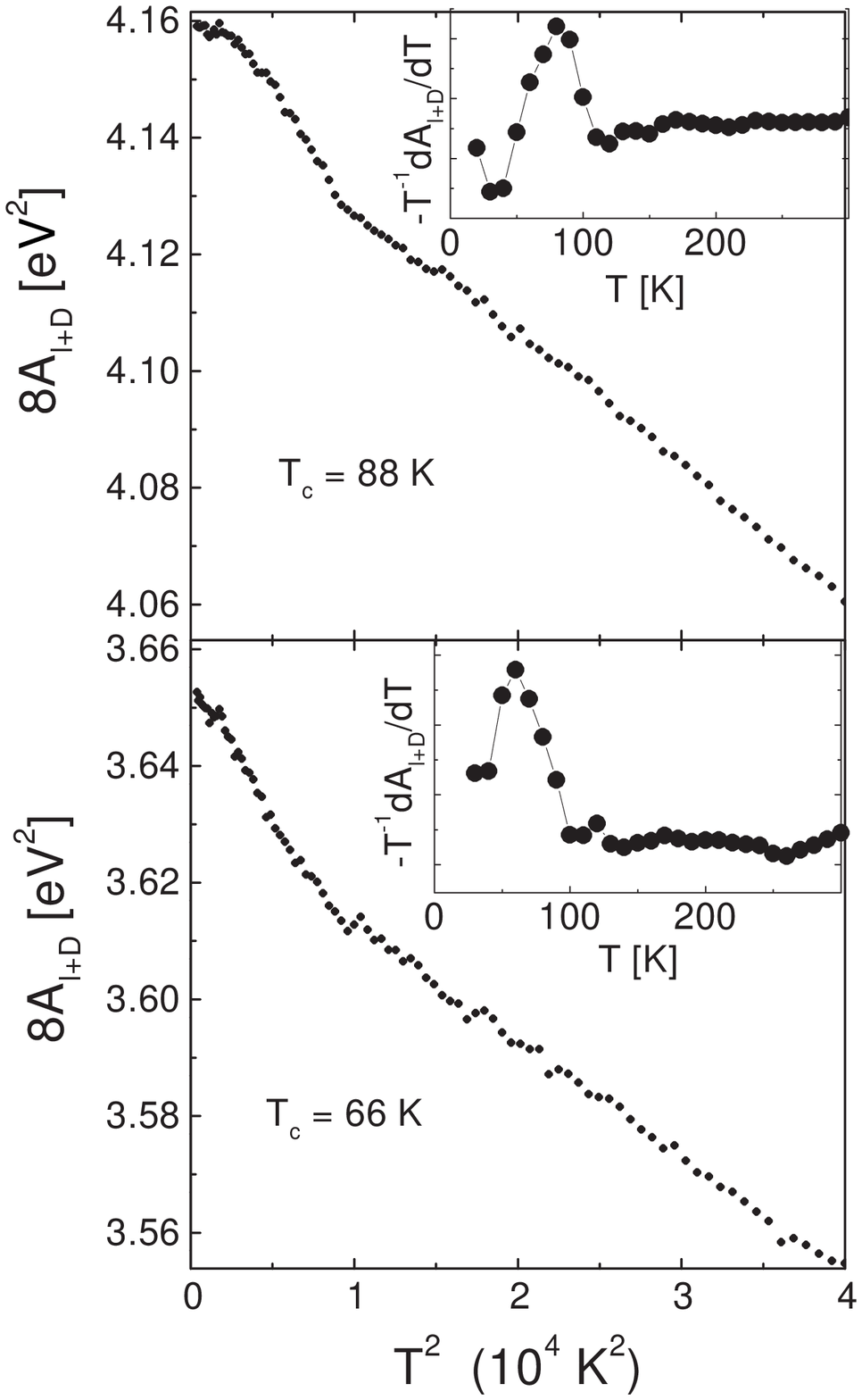}}
\epsfxsize=0.5\textwidth{\epsfbox{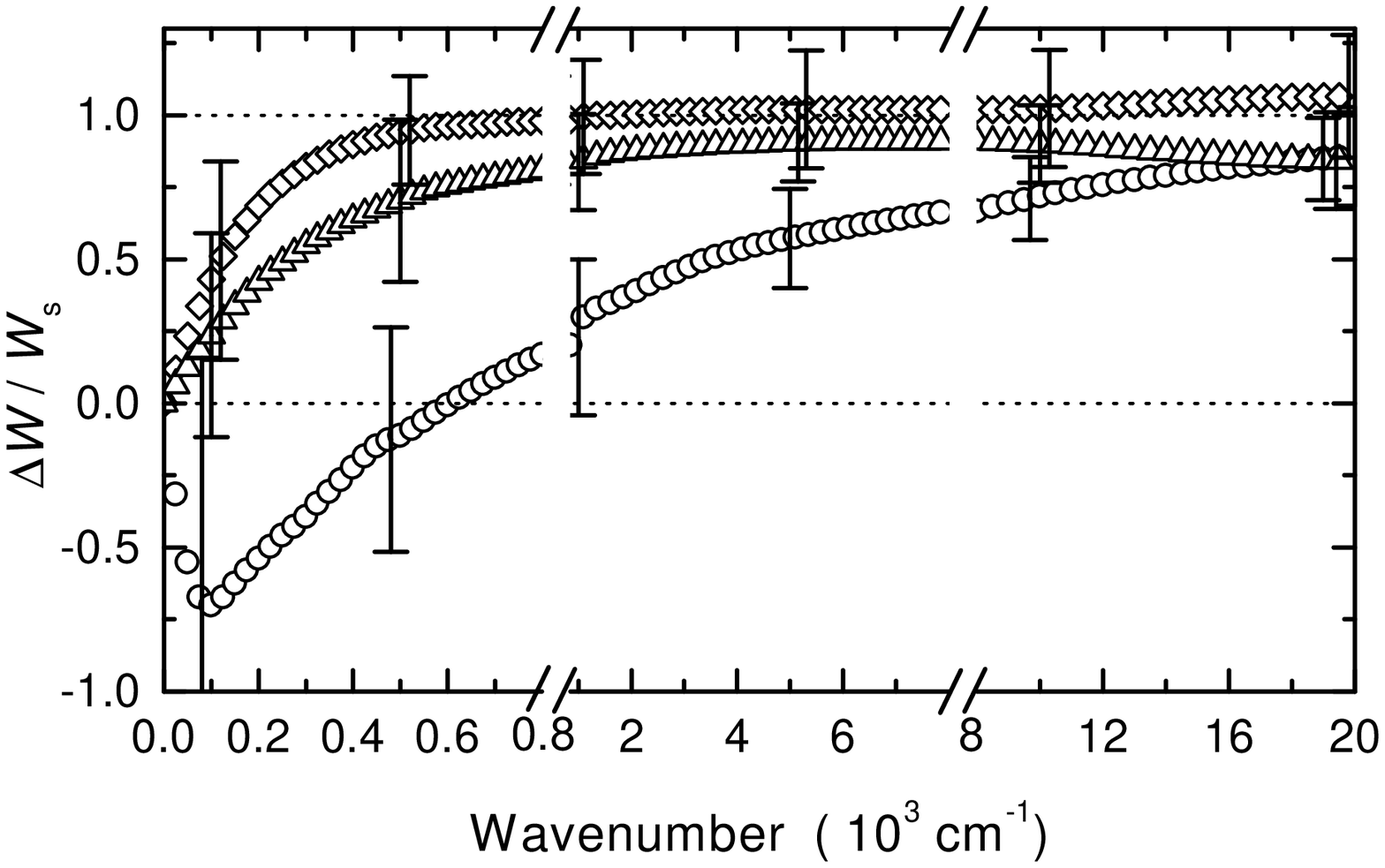}}}
\caption{Left:  Variation of the optical spectral weight (integrated to 1.25 eV)
with temperature
for an optimal (top left) and an underdoped (bottom left)
sample of Bi2212.  Anomalous rise below $T_c$
implies kinetic energy lowering in the 
superconducting state.  From 
Ref.\cite{DVMSC}.  Right:  Optical spectral weight difference between a 
temperature above $T_{c}$ and 10K integrated to the energy 
plotted on the x axis normalized to the weight of the superconducting 
condensate for various samples of Bi2212 (diamonds for overdoped, 
triangles for optimal doped, and circles for underdoped). Spectral 
weight balance (sum rule) would correspond to a value of 1.
From Ref.~\cite{NICOLE}.}
\label{fig51}
\end{figure}

Motivated by this, two recent experiments on optimal and underdoped 
Bi2212 found evidence for a change in the planar optical integral between 
normal and superconducting states large enough to account for the 
condensation energy (Fig.~\ref{fig51}) \cite{DVMSC,NICOLE} (though an earlier study of
underdoped YBCO did not \cite{JOE2}).
No such violation was found 
for an overdoped Bi2212 sample \cite{NICOLE}.  In both experiments, more 
spectral weight showed up in the zero frequency condensate peak than can 
be accounted for by the loss of finite frequency weight up to about 1 eV.
Integrating up 
to about 2 eV, though, spectral weight balance is found.  This confirms 
the earlier speculations by Anderson \cite{PWA90,ANDERSON} of transfer of 
spectral weight from high to low energies.

\begin{figure}
\centerline{\epsfxsize=0.8\textwidth{\epsfbox{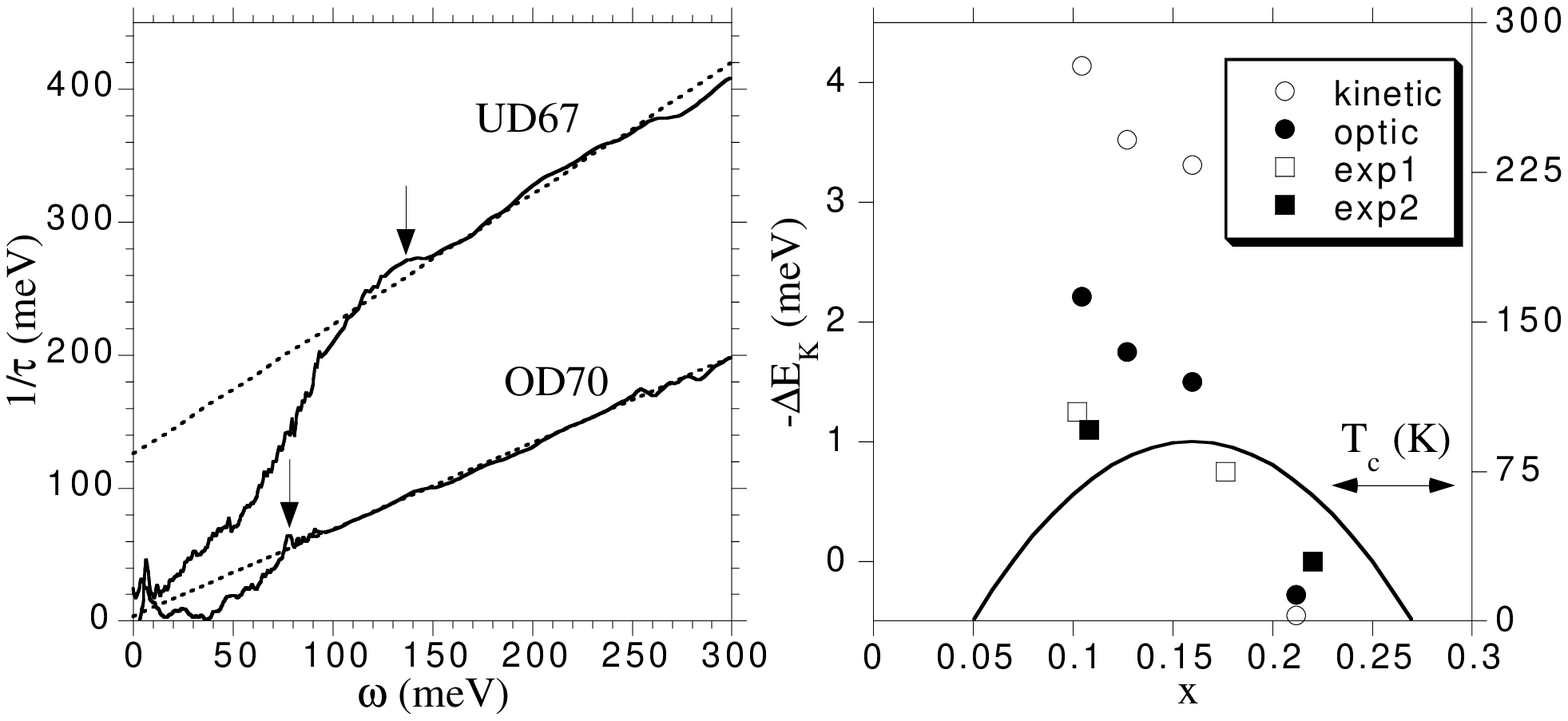}}}
\caption{Left: $1/\tau$ vs $\omega$ in the superconducting state
as extracted from optics for an overdoped (OD70)
and an underdoped (UD67) sample of Bi2212 \cite{PUCH}.  Dotted lines are 
$a+b\omega$ fits to normal state data, arrows the locations of the scattering rate
gaps.  Note the near zero value of $a$ in the OD case,
as contrasted to the large value in the UD case.
Right: Calculated change in the kinetic energy (open circles) and 
optical integral (full circles) versus hole doping, x, with parameters determined
from optics (left panel)  and ARPES.
The open squares are the data of 
Ref.~\cite{DVMSC} and the full squares that of Ref.~\cite{NICOLE}.  
Adapted from Ref.~\cite{MNCP}.}
\label{fig52}
\end{figure}

Using the observed form of
the scattering rate in the normal state,
$a_{k} + b\omega$ \cite{AV}, and assuming both of these terms are 
gapped below some threshold energy in the superconducting state
(left panel, Fig.~\ref{fig52}),
the optical integral 
difference was calculated by Norman and P\'{e}pin \cite{MNCP} (right
panel, Fig.~\ref{fig52}) from 
an $\epsilon_{k}$ extracted from ARPES data (more properly, band structure
values should be used for $\epsilon_k$).  They 
find that such a calculation gives a good estimate of the optical 
integral change, including its doping dependence.
In these 
calculations, the sum rule violation is coming from the $a_{k}$ term (that 
is, if the normal state were a pure marginal Fermi liquid, there would 
be no sum rule violation).  This $a_k$ term, which as stated earlier seems to
be associated with the pseudogap, has a strong dependence on
doping (left panel, Fig.~\ref{fig52}), which explains the large doping
dependence of the sum rule violation.
The same calculations find that the 
actual kinetic energy change is about twice that indicated by the 
optical integral, the difference due to the fact that the 
inverse mass tensor is not simply the negative of $\epsilon_{k}$ as in a 
near neighbor tight binding model of the energy dispersion.  In summary, the 
origin of the sum rule ``violation'' and kinetic energy lowering
can be traced to the formation of 
quasiparticle peaks in the superconducting state due to the opening of 
a scattering rate gap.

\subsection{ARPES}

These results can be generalized by considering the entire free 
energy.  If only two particle interactions are involved, it is easily 
shown that the full condensation energy can be written as \cite{GANG4}
\begin{equation}
E_{cond} = \sum_{k} \int_{-\infty}^{\infty} d\omega (\omega + 
\epsilon_{k}) f(\omega) [A_{N}(k,\omega)-A_{S}(k,\omega)]
\end{equation}
where $A_{N}$ is the normal state single particle spectral function, 
and $A_{S}$ that of the superconducting state.  This expression is 
the sum of two terms, a kinetic energy term (where $\omega+\epsilon_{k}$ 
is replaced by $2\epsilon_{k}$) and a potential energy term (where 
$\omega+\epsilon_{k}$ is replaced by $\omega-\epsilon_{k}$).

This expression can be reduced further by performing some of the 
integrals and sums
\begin{equation}
E_{cond} = \sum_{k} \epsilon_{k} [n_{N}(k) - n_{S}(k)]
+\int_{-\infty}^{\infty} d\omega \omega f(\omega) 
[N_{N}(\omega)-N_{S}(\omega)]
\end{equation}
where $n(k)$ is the momentum distribution function and $N(\omega)$ the density 
of states.  The first term is related to (but not the same as)
the optical integral we just discussed, the second term 
could be obtained from tunneling spectroscopy.  Both terms, though, 
can in principle be obtained from angle resolved photoemission, since 
only the occupied states enter.  In practice, resolution, 
normalization, and matrix element effects will be a limiting factor 
\cite{GANG4}.

The above expressions, though, represent a simple conceptual formalism 
for tackling the condensation energy issue which avoids the problem 
of considering complicated two particle correlation functions.  This 
was illustrated by Norman \etal \cite{GANG4}, who evaluated these 
expressions using the ``mode'' model of Norman and Ding 
(Fig.~\ref{fig29}) \cite{NDING} 
for fitting ARPES spectra (the same model was employed by Ioffe and 
Millis \cite{LEV} to analyze the c-axis optical sum rule violation, 
and Norman and P\'{e}pin \cite{MNCP} to analyze the in-plane one).  What 
these authors found was that for small normal state scattering rates, 
a result similar to BCS theory occurs, which presumably applies on 
the overdoped side of the phase diagram.  For scattering rates much 
larger than the superconducting energy gap, though, a result opposite 
to BCS theory was found, in that the kinetic energy was lowered and 
the potential energy raised in the superconducting state.

The kinetic energy result is rather straightforward to understand 
(Fig.~\ref{fig49}).  
There is the BCS effect of particle-hole mixing which raises the 
kinetic energy.  Opposed to this is the formation of coherent 
quasiparticle states from the incoherent normal state, which acts to 
lower the kinetic energy.  For a small scattering rate, the 
particle-hole mixing effect wins out, and the kinetic energy is 
raised, but for larger scattering rate, the quasiparticle formation 
effect wins out, and the kinetic energy is lowered.

\begin{figure}
\centerline{\epsfxsize=0.8\textwidth{\epsfbox{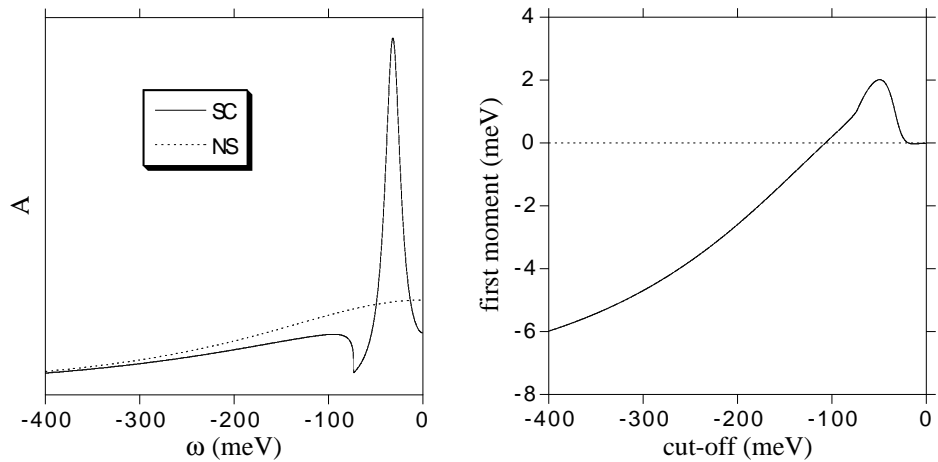}}}
\caption{Left panel:  calculated spectral function in the normal state (NS) 
and superconducting state (SC).  Right panel:  resulting difference in 
first moments of the spectrum
versus cut-off energy.  Note positive contribution (potential energy 
decrease in the SC state) from spectral peak, negative contribution 
(potential energy increase in the SC state) from the difference in the 
spectral tails.
From Ref.~\cite{GANG4}.}
\label{fig53}
\end{figure}

The potential energy part is also straightforward to understand (Fig.~\ref{fig53}).
There is a competition between the formation of a spectral gap (which 
lowers the potential energy as in BCS theory) and spectral weight 
transfer from high to low energy to form the coherent peak 
(which raises the potential energy).  For small scattering 
rate, the spectral gap effect wins out and the potential energy is 
lowered, whereas for large scattering rate, the weight transfer 
effect wins out and the potential energy is raised.

What this implies for the phase diagram is that on the overdoped side, 
one expects more or less BCS like physics.  But on the underdoped 
side, one expects dramatic departures.  In particular, the potential 
energy is lowered across the pseudogap $T^{*}$ line due to the formation 
of a spectral gap, then the kinetic energy lowered across the 
superconducting $T_{c}$ line due to the onset of coherence.

We would like to end with a ``cautionary'' remark.  What one calls 
kinetic or potential energy depends on what effective Hamiltonian is 
being employed.  The definition of this changes, for instance, if one 
goes from the three band Hubbard model, to the single band Hubbard 
model, to the t-J model.  The purpose of going through the above 
exercise is to demonstrate that one's preconceptions based on BCS 
theory could well 
be wrong in pairing models driven by electron-electron interactions, 
particularly if the normal state reference is non Fermi liquid like.

On that note, we would like to bring this review to a close.

\ack

This work was supported by the U.S. Dept.~of Energy, Office of Science, under
contract W-31-109-ENG-38.  This review is based on a series of lectures given
at the SPhT in the fall of 2001 by MRN.  We would like to thank the
staff of the SPhT, in particular Dr. J. P. Blaizot and Dr. G. Misguich, for
making these lectures possible.

\end{document}